\documentclass[final,5p,times,twocolumn]{elsarticle}

\usepackage{amssymb}

\newcommand{\commentout}[1]{}
\newcommand{\onethirdwidth}{0.31\linewidth}

\usepackage{subfigure}
\usepackage{multirow}
\usepackage{makecell}

\usepackage{color}
\usepackage[normalem]{ulem}
\newcommand{\revone}[2]{#2}
\newcommand{\revtwo}[2]{#2}
\newcommand{\revthree}[2]{#2}
\newcommand{\cutthree}[1]{}

\journal{Computer Networks}

\begin{document}

\begin{frontmatter}



\title{QUIC Throughput and Fairness over Dual Connectivity\tnoteref{label1}}
\tnotetext[label1]{A shorter preliminary version of this work appeared in the IEEE MASCOTS 2020 workshop~\cite{HLK+20}.}


\author{David Hasselquist}
\author{Christoffer Lindström }
\author{Nikita Korzhitskii}
\author{Niklas Carlsson}
\author{Andrei Gurtov\\~\\ Link\"oping University, Sweden}





\begin{abstract}
Dual Connectivity (DC) is an important lower-layer feature accelerating the transition from 4G to 5G that also is expected to play an important role in standalone 5G
\revone{}{radio networks}.  
However, even though the packet reordering introduced by DC can significantly impact the performance of upper-layer protocols, no prior work has studied the impact of DC on QUIC.  In this paper, we present the first such performance study. Using a series of throughput and fairness experiments, we show how QUIC is affected by different DC parameters, network conditions, and whether the DC implementation aims to improve throughput or reliability. 
\revone{}{Results for two QUIC implementations (aioquic, ngtcp2) and
two congestion control algorithms (NewReno, CUBIC) are presented under both static and highly time-varying network conditions.}
Our findings provide 
\revone{}{network operators with}
\revone{insights into}{insights and understanding into}
the impacts of splitting QUIC traffic in a DC environment. With reasonably selected DC parameters and increased UDP receive buffers, QUIC over DC performs similarly to TCP over DC and achieves optimal fairness under symmetric link conditions when DC is not used for packet duplication.
\revone{}{The insights can help network operators provide modern users with better end-to-end service when deploying DC.}
\end{abstract}



\begin{keyword}
QUIC \sep Dual Connectivity \sep Throughput \sep Fairness \sep Transport Protocol \sep Multipath
\end{keyword}

\end{frontmatter}


\section{Introduction}
\label{sec:introduction}

The end-to-end performance depends on the interactions between protocols in different network layers. As new features are introduced on the lower layers, it is therefore important to understand the impact that such features and their parameters have on the upper layer protocols~\cite{ABS2007}. One such feature is Dual Connectivity (DC).  DC was introduced in 4G, gained popularity with the introduction of 5G, and currently plays an integral role in accelerating the generational transition from 4G to 5G~\cite{smrwf15}.

With DC, users can transmit and receive data from two base stations concurrently. This allows users to use both 4G and 5G 
\revone{}{radio} 
networks in parallel, simplifying the above-mentioned generational transition. However, it has also been argued that DC should be a part of future 5G solutions needed to meet the requirements of Ultra-reliable and Low-Latency Communications (URLLC)~\cite{MLLPB18,3gpp15}. Combined with its increased usage, this has made DC an important 5G feature.

Like \revone{multi-path}{multipath} transport protocols~\cite{CB17,WRGH11,IAS06}, DC can be used to combine WiFi with 4G and 5G solutions. Furthermore, like these protocols, DC can be used to achieve improved throughput (by sending different data over different paths), to increase reliability (by transmitting the same data over the different paths), or both. However, in contrast to the transport-layer multipath solutions, DC is performed within the link layer of the network stack and is therefore in practice invisible to transport layer protocols such as TCP and QUIC. This is an important observation since DC may introduce jitter or reordering of packets that can significantly impact TCP and QUIC performance.

In parallel with the transitioning of different
\revone{}{cellular}
network generations, Google recently introduced QUIC as a next generation transport-layer solution aimed at addressing some shortcomings with TCP~\cite{LRW+17}.
\revone{}{In contrast to TCP,}
QUIC is implemented in the user-space, on top of UDP, and provides much improved stream multiplexing compared to TCP. This is important to speed up web connections in the presence of packet losses and/or modern HTTP/2 traffic. Initial research shows that QUIC allows performance improvements over TCP in several cases while providing an easy way to achieve fast incremental deployment~\cite{LRW+17}. Popular services that already today use QUIC include Google search services, Chrome, Chromium, \revone{YouTube}{YouTube,} and Facebook~\cite{LRW+17,ietf-106-singapore-quic}. 

Due to the increasing use and popularity of both QUIC and DC, combined with the continuous rollout of 5G networks using DC, it is important to understand how QUIC performs over DC under different network conditions, and the impact that different DC parameters have on QUIC performance.

In this paper, we present the first performance evaluation of QUIC over DC. First, a testbed is set up to simulate DC. The testbed captures QUIC and TCP performance under a wide range of network behaviors (based on bandwidth, delay, and loss conditions) and the impact of different DC parameters.  Second, using a series of throughput and fairness experiments, we show how QUIC is affected by different DC parameters, network conditions, and whether the DC implementation aims to improve throughput or reliability. For our throughput evaluation, we primarily compare the throughput of QUIC over DC with that of TCP over DC, and for our fairness comparisons we compare the throughput (and calculate a fairness index) of competing flows when using QUIC over DC. 
\revone{We also present results using different QUIC implementations (aioquic, ngtcp2) and congestion control algorithms (NewReno, CUBIC)
\revone{}{and include network conditions with both static and highly time-varying bandwidths}.}{We also present results using different QUIC implementations (aioquic, ngtcp2), congestion control algorithms (NewReno, CUBIC), and for networks with both static and highly time-varying bandwidths.}
Our findings provide insights into the impact that DC and its parameters have on QUIC performance.
\revone{For example, we}{Furthermore, we} 
show the value of increasing the UDP receive buffers when running QUIC over DC, that QUIC over DC can achieve similar throughput as TCP over DC, and that QUIC over DC can achieve optimal fairness under symmetric link conditions, except if DC duplicates packets to increase reliability.

\revtwo{}{Overall, the experimental findings presented in this paper provide an important step towards understanding the interplay between DC and QUIC under different conditions and scenarios.}
\revtwo{}{Again, while much prior work have studied either DC or QUIC, we are the first to consider these in combination.}

\revone{{\bf Outline:}}{The remainder of the paper is organized as follows.}
Sections~\ref{sec:dc} and~\ref{sec:related-work} introduce DC and 
present related works, respectively. The following sections present our methodology 
\revtwo{(Section~\ref{sec:method}), performance 
\revone{}{evaluation} results (Section~\ref{sec:results}), and}{(Section~\ref{sec:method}) and performance 
evaluation results (Section~\ref{sec:results}), including a summary of our key findings, before we present our}
conclusions (Section~\ref{sec:conclusion}).

\section{Dual Connectivity}\label{sec:dc}

DC, 
\revone{}{also} 
sometimes called inter-node radio resource aggregation, is a multi-connectivity technique introduced in release 12 of the third-generation partnership project (3GPP)~\cite{3gpp12}. The aim was to increase reliability, performance, and signaling due to frequent handovers in scenarios where macro and micro cells are connected with a non-ideal backhaul (X2) link. DC tries to achieve this by splitting the traffic over multiple paths.

\begin{figure}[t]
\centering
\includegraphics[width=0.325\textwidth]{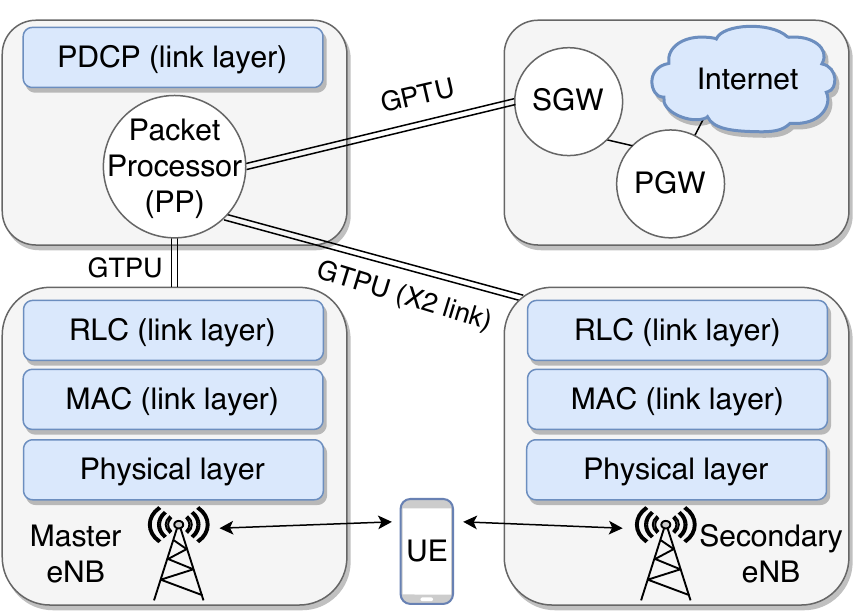}
\caption{Dual connectivity overview.}
\label{fig:dc}
\vspace{-8pt}
\end{figure}

Figure~\ref{fig:dc} shows an overview of DC in a Radio Access Network (RAN) environment. With DC, a User Equipment (UE) connects to two different 
\revone{}{network nodes, also known as}
Evolved Node Bs (eNBs)~\cite{3gpp15}.
One of the 
\revone{}{network} 
nodes will serve as Master eNB (MeNB), and the other one will serve as Secondary eNB (SeNB). Each of the MeNB and SeNB contains a separate Radio Link Control (RLC) and  Media Access Control (MAC) layer, while sharing the same Packet Data Convergence Protocol (PDCP) layer.

DC is similar to carrier aggregation~\cite{RRM+16}, but is performed in the PDCP layer instead of the MAC layer. Carrier aggregation uses the same scheduler for the separate connections and requires an ideal X2 link. The split connections are therefore often transmitted from the same node. In contrast, DC uses two separate 
\revone{}{packet}
schedulers together with a non-ideal X2 link, and packets are often originating from two different nodes.

PDCP is a sublayer located inside the link layer, just below the network layer and above RLC and MAC. The main tasks of PDCP are header compression and decompression, ciphering, integrity protection, transfer of data, sequence numbering, reordering and in-order delivery~\cite{3gpp16-etsi5g}. The PDCP layer can be broken out into a unit called a Packet Processor (PP), which connects to Serving Gateway (SGW), MeNB and SeNB using a GTPU-tunnel. SGW is connected to the Packet Data Network Gateway (PGW), which 
\revone{}{in turn}
connects to the public internet. The PP can also be a part of MeNB. In this case, MeNB splits the traffic and the link between MeNB and SeNB becomes the X2 link. In both scenarios, the traffic is split in the PDCP layer.

\begin{figure*}
\centering
\begin{minipage}{.45\textwidth}
    \subfigure[\revthree{}{Conceptual overview of the experimental setup}\label{fig:model-abstract}]{
        \includegraphics[trim = 0mm -4mm 0mm 0mm, height=0.63\textwidth]{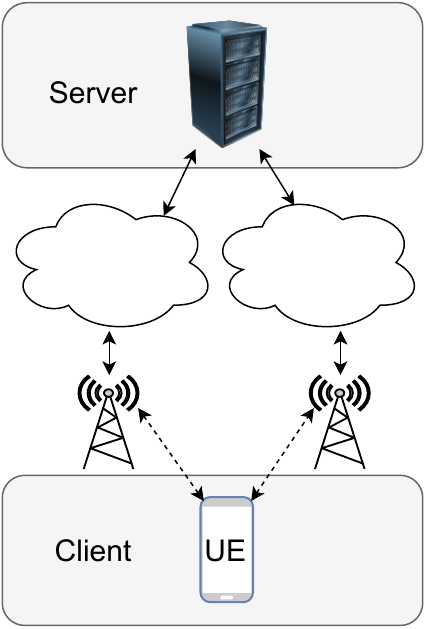}}
    \end{minipage}
    \hspace{-110pt}
    \begin{minipage}{.45\textwidth}
    \subfigure[Throughput\label{fig:model-basic}]{
        \includegraphics[height=0.83\textwidth]{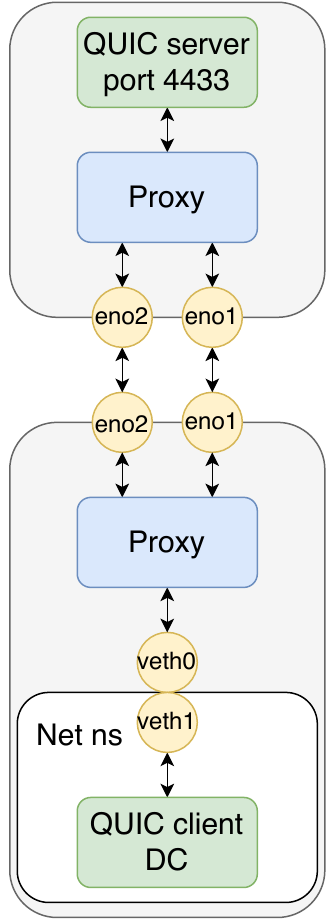}}
    \hspace{20pt}
    \subfigure[Fairness\label{fig:model-extended}]{
        \includegraphics[height=0.83\textwidth]{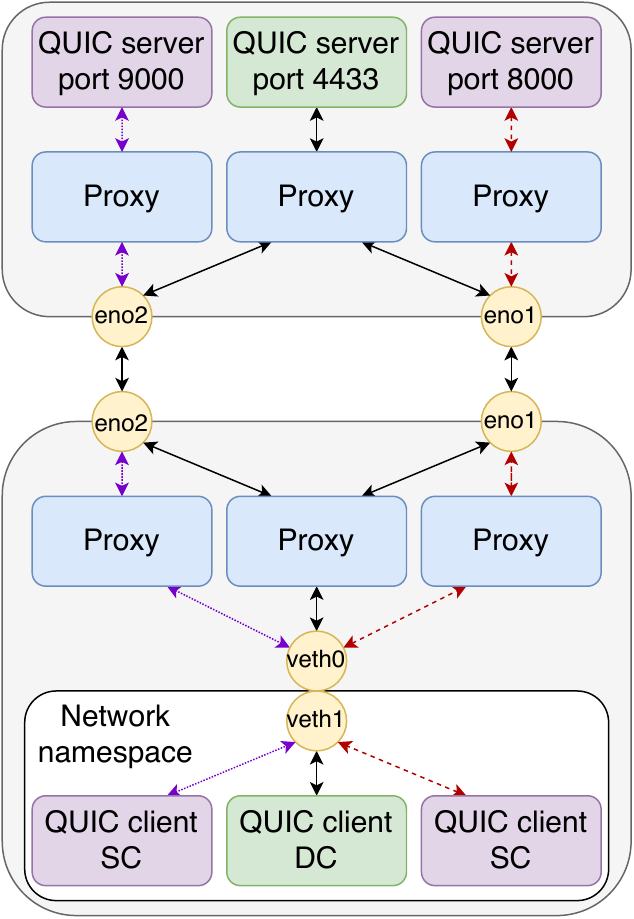}}
\end{minipage}
    \vspace{-2pt}
    \caption{Testbed for throughput and fairness \revtwo{}{tests.}}
    \label{fig:model}
    \vspace{-2pt}
\end{figure*}

\section{Related Work}
\label{sec:related-work}

{\bf Dual connectivity:} Unlike TCP, QUIC is relatively new, and there are few studies of it in specific scenarios such as DC. As QUIC shares similarities with TCP, we can obtain initial insights from research about DC that uses TCP as the transport protocol. Polese et al.~\cite{PMRZ17} study the performance of TCP when using DC to perform mobile handovers for
\revone{an}{a}
UE and compare the performance with different single connection mobility management solutions. They show that DC can improve TCP goodput by quickly moving the traffic from one of the two DC links to the other.

Other studies have focused on specializations of DC;
e.g., LTE-WLAN Aggregation (LWA)~\cite{JKYLKY17,KLG16}, which allows for network traffic over LTE and WLAN. Jin et al.~\cite{JKYLKY17} 
\revone{}{propose LTE-W and}
show that splitting TCP over LTE and WiFi 
\revone{}{links} 
at the PDCP layer can achieve similar throughput and better fairness than MP-TCP; demonstrating the value of lower-layer traffic splitting. Khadraoui et al.~\cite{KLG16} investigate the effect of PDCP reordering when using TCP in LWA over heterogeneous links. Their results show that PDCP reordering can have adverse effects on TCP throughput, and that in some cases it is better to use only one link. 
\revtwo{}{Others have presented performance optimizations for DC~\cite{wu2018optimal, wu2018optimal-b, sharma2019performance, he2021delay} but have not considered QUIC over DC.}
\revtwo{}{For example, Wu et al.~\cite{wu2018optimal} present an optimized DC traffic scheme for offloading the uplink of mobile users. Considering various tradeoffs, they 
achieve significant performance benefits compared to when using a fixed bandwidth allocation or scheduling scheme. At a high level, our study differs as we focus on QUIC and experimentally study its performance when using DC.}
While some works have looked at TCP with DC, no prior work has studied 
\revone{the QUIC performance over DC.}{the performance of QUIC over DC.}

{\bf Upper-layer multipathing:} 
Multipathing is similar to DC but performed higher up in the network stack. Most such solutions are implemented in the transport layer, e.g., SCTP~\cite{IAS06} and MP-TCP~\cite{WRGH11}, but some are implemented in the network layer~\cite{GuPo09}. Here, we focus on QUIC-based solutions. De Coninck and Bonaventure~\cite{CB17} implement Multipath QUIC (MP-QUIC) based on quic-go and lessons learned from MP-TCP, and show that serving QUIC over multiple paths is beneficial. Mogensen et al.~\cite{MMMKPL19} expands MP-QUIC to Selective Redundant MP-QUIC (SR-MPQUIC). Their solution modifies the congestion control algorithm, the scheduler, and the stream framer. SR-MPQUIC reduces latencies and improves reliability for priority data at a small increase in bandwidth usage and latencies for background data. The results show the importance of proper packet scheduling and the value of packet duplication.
While additional cross-layer communication would be required to benefit DC, QUIC 
\revone{}{(and HTTP/3)}
also includes some unique attributes to assist packet/flow scheduling~\cite{RHB18}.

{\bf Fairness:} Fairness can be difficult to judge when there are multiple paths with different amount of resources. 
\revone{}{As such, we take a look at studies on multipathing fairness.}
Becke et al.~\cite{BDAR12} study the fairness of different congestion control algorithms in multipath scenarios, 
\revone{}{targeting TCP-friendliness and}
focusing on two fairness types: link-centric 
\revone{}{fairness (where each flow share the link capacity equally),}
and network-centric flow fairness
\revone{}{(where fairness of flows is considered in the whole network with several paths)}.
Raiciu et al.~\cite{RPBGWH2010} study how MP-TCP can replace single connections and load balancing in data centers
\revone{}{with different network topologies}.
For specific topologies, MP-TCP significantly improved fairness and provided throughput closer to optimal compared to single connectivity using random load balancing. To judge 
\revone{}{the}
fairness,
they and many others~\cite{RPBGWH2010,WRGH11,ZL08} evaluate multipathing using Jain's fairness index (JFI)~\cite{JCH84}. Similar to these works, we use JFI here.

\begin{table*}[t]
  \caption{Hardware and operating systems \revtwo{}{used in tests.}}
  \vspace{4pt}
  \label{tab:specs}
  \centering
  {\small
  \begin{tabular}{|l|l|l|}
     \hline
     \textbf{Component} & \textbf{Client} & \textbf{Server} \\
     \hline
     OS  & Ubuntu 18.04.3 LTS  & Ubuntu 18.04.3 LTS \\                     
     \hline
     Kernel & Linux 4.15.0-74-lowlatency & Linux 5.3.0-26-generic \\
     \hline
     Processor 1 \& 2~ & Intel(R) Xeon(R) \scriptsize{CPU E5-2690 v3~} & Intel(R) Xeon(R) \scriptsize{CPU E5-2667 v3~} \\ 
     \hline
     eno1 \& eno2 & 82599ES 10-Gigabit SFI/SFP+ & 82599ES 10-Gigabit SFI/SFP+ \\
\hline
  \end{tabular}}
\end{table*}

\section{Methodology}
\label{sec:method}

\subsection{Dual connectivity testbed}

\revthree{Figure~\ref{fig:model} shows an overview of the testbed used for studying QUIC performance over DC.  
\revone{We}{At a high level, we}
used one machine to capture client-side behavior and performance, and one machine to capture server- and network-side effects. The two machines were connected via two network interface pairs, each supporting 10 Gbps full duplex. 
\revone{Hardware}{The hardware}
specifications are given in Table~\ref{tab:specs}. We next describe the configurations used for our throughput and fairness tests, respectively, and our proxy implementation used to simulate DC and PDCP.}{To evaluate the QUIC performance over DC, we used two machines in our evaluation framework: one machine to capture client-side behavior and performance, and one machine to capture server- and network-side effects. The two machines were connected via two network interface pairs (eno1 and eno2), each supporting 10 Gbps full duplex. To simulate different network conditions, tc in Linux was used to add extra delay, jitter, loss, and bandwidth limitations for each of the two links.} 

\revthree{}{Depending on the specific experiment, we ran one or three QUIC/TCP clients on the client side and one or three QUIC/TCP servers on the server side. Figure~\ref{fig:model} shows an overview of the testbed, starting with a high-level conceptual overview (Figure~\ref{fig:model-abstract}) and then two sub-figures (Figures~\ref{fig:model-basic} and~\ref{fig:model-extended}) that captures the specific designs used for our throughput and fairness experiments. Furthermore, to simulate the functionality of DC and PDCP, two proxies were implemented for each QUIC client-server pair: one on the client side and one on the server side. Finally, each QUIC client was launched inside a network namespace (kernel functionality in Linux). This is to avoid the QUIC clients binding to a random interface and to ensure that we can control which interface each DC connection eventually is sent over. Here, two virtual interfaces were created to forward data to and from the namespace. We also note that there is no need for a network namespace on the server side, as the server will respond over the incoming interface, and DC is only studied on the downlink (server to client).}

\revthree{}{Table~\ref{tab:specs} specifies the hardware used in our experiments. We next describe the specific configurations used for our throughput and fairness tests, respectively, as well as our proxy implementation used to simulate DC and PDCP.}

{\bf Throughput test configuration:}
For our throughput tests (Figure~\ref{fig:model-basic}), we used one client, one server, and studied the performance impact of DC parameters and the network conditions between them. In our baseline tests, both the QUIC server and QUIC client used aioquic~\cite{aioquic}. When 
\revthree{running comparison tests}{comparing} 
with TCP, we used a Hypercorn server using HTTP/2 over TLS 1.3, and the client used curl to make HTTP/2 requests. As baseline, both TCP and QUIC used NewReno for congestion control.
\cutthree{Since traffic splitting with DC is implemented in the link layer, QUIC (and TCP) are unaware that the traffic is sent over multiple paths, and therefore do not need to be modified. However, as DC was introduced for radio technology, the link layer functions 
\revone{}{and structure} 
differ from the Ethernet links used here. To simulate the functionality of DC and PDCP, two proxies were implemented: one at the client and one at the server.}
\cutthree{The QUIC client was launched inside a network namespace. Two virtual interfaces were created to forward data to and from the namespace. The server side does not require a network namespace as DC is only studied on the 
\revtwo{downlink.}{downlink (to client).}
To simulate different network conditions, tc in Linux was used to add extra delay, jitter, loss, and bandwidth limitations.}

{\bf Fairness test configuration:}
For our fairness tests (Figure~\ref{fig:model-extended}), we used three clients and three servers. One end-to-end connection was performing DC, while the other two used single connectivity (SC) over interface eno1 and eno2, respectively. The server with port 8000 was operating only on eno1, while the server on port 9000 only on eno2. The QUIC server on port 4433 used DC and operated on both interfaces. 
\cutthree{Each server and client were equipped with its own proxy, simulating the PDCP functionality for each connection independently.}

{\bf Proxy-based implementation:}
\revthree{To}{Since traffic splitting with DC is implemented in the link layer, QUIC (and TCP) are unaware that the traffic is sent over multiple paths, and therefore do not need to be modified. However, as DC was introduced for radio technology, the link layer functions and structure differ from the Ethernet links used here. To simulate the functionality of DC and PDCP, two proxies were implemented: one at the client and one at the server. First, to}
capture the PDCP functionality, packets originating from the server are caught by iptables OUTPUT chain and delivered to a NFQUEUE, before being read by the server proxy. The server proxy then adds a 2-byte PDCP sequence number to each packet and routes the packets to the client over two interfaces. When running DC, the server proxy alternates between the two interfaces.

At the client proxy, packets are caught in the PREROUTING chain and delivered to NFQUEUE. The client proxy can then read from the queue, perform PDCP convergence of the two streams, do PDCP reordering, and remove the sequence numbers that were added by the server proxy.

If a packet is received in order, it is immediately forwarded to the client. However, if a packet is out of order, it is kept until the missing packets are processed or until a PDCP timer of 200~ms is reached. If the timer is reached, all packets before the missing packet and all consecutive packets after the missing packets are delivered. The reordering algorithm follows the PDCP standard described in 3GPP~\cite{3gpp16-etsi5g} and the testbed was developed in close consultation with Ericsson. Our proxy adds around 1~ms to the total RTT, assuming that the packets arrive in order. Without PDCP, large reordering occurs, resulting in QUIC having a very low throughput.

\subsection{Performance testing}

To understand how DC affects QUIC, a series of tests are performed that captures the impact of different DC parameters and network conditions. In our experiments, we vary one parameter at a time, starting with a default configuration, while keeping the others constant (as per the default configuration). In the throughput tests, the client downloads a 100MB file, and in the fairness tests each client downloads a 1GB file and we measure the clients' performance for the first three minutes of the download. For each test configuration, we run ten tests and calculate both average and standard deviation values for the metrics of interest
\revone{}{(i.e., throughput and JFI)}.
\revthree{}{When discussing JFI, we also stress that there is no global JFI threshold that can be considered as a threshold for what is ``fair” and ``not fair". Rather than suggesting any threshold, we instead focus on relative comparisons as a parameter changes or between the JFI of two alternative configurations operating under otherwise similar conditions.}

{\bf DC parameters and default configurations:}
The primary DC parameters we varied were the DC batch size and DC batch split. These parameters determine how many packets are sent over each interface before the server proxy switches to the other interface. For example, with a DC batch size of 100 and a DC ratio of 9:1 (90\% eno1 and 10\% eno2), the proxy would send 90 packets over eno1, before switching over to send 10 packets over eno2. In our default experiments, the default DC batch size and DC ratio was configured to 100 and 1:1, respectively.

{\bf Network emulation parameters and default configurations:}
To capture different network conditions, we primarily varied the bandwidths, delays, and loss rates of the links. For both the bandwidths and delays, we present experiments both where we vary the average values and where we vary the ratio between the two links. In the case we vary one of the ratios, we keep the average value of that metric constant. For example, a bandwidth ratio of 3:1 corresponds to 30~Mbps and 10~Mbps for the downlink interfaces eno1 and eno2, respectively. In our default experiments, each link operates at 20~Mbps and has normally distributed per-packet delays with a mean of 10~ms and a 
\revtwo{standard deviation}{coefficient of variation} 
of 10\%.

{\bf QUIC and TCP configurations/versions:}
Throughput tests for QUIC are performed both with the default UDP receive socket buffer size and a larger receive buffer size. The larger size is used to give a fair comparison to TCP, as the kernel performs buffer autotuning for TCP~\cite{PKB13}. When studying fairness, QUIC with modified buffer size is used and the fairness is calculated using JFI.
\revone{}{Furthermore, we use Qvis and Qlog~\cite{MaRQ18, MaPQ20} to debug and study QUIC at a more detailed level.}

In our default scenarios we use aioquic with NewReno. However, as discussed by McMillan and Zuck~\cite{McZu19}, the QUIC RFC is ambiguous, open for interpretation, and differences between QUIC implementations following the 
\revone{}{same} 
RFC have been demonstrated using specification testing. 
\revone{}{Marx et al.~\cite{MaHQ20} also highlight this diversity by comparing 15 different QUIC implementations 
and showing that they are highly heterogeneous.}
We therefore repeated our experiments with both another QUIC implementation (ngtcp2~\cite{ngtcp2}) and congestion control algorithm (CUBIC).

\begin{figure}[t]
    \vspace{-8pt}
    \centering
    \includegraphics[trim = 0mm 4mm 0mm 0mm, width=0.4\textwidth]{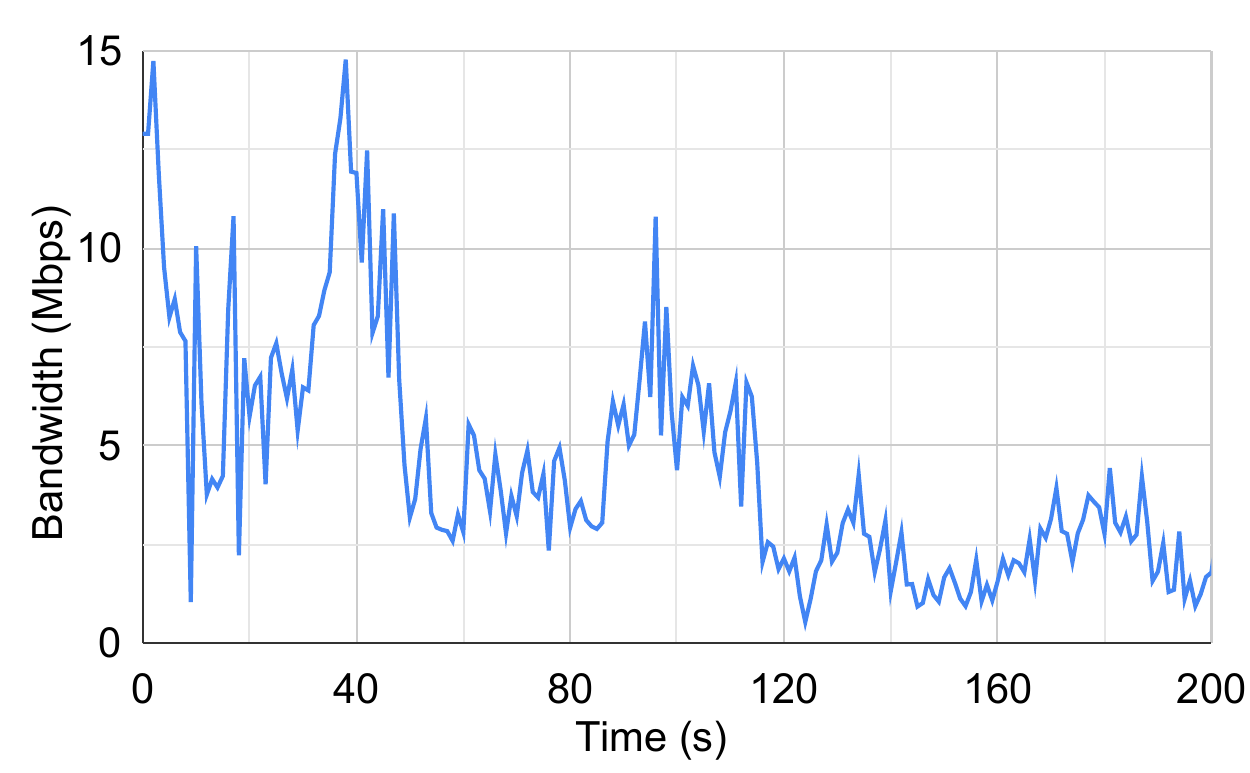}
    \vspace{-4pt}
    \caption{\revone{}{Example LTE bandwidth trace~\cite{RQZS18}}}
    \label{fig:bw-traces}
\end{figure}

\begin{table}[t]
  \vspace{-6pt}
  \caption{\revtwo{}{Average throughput and RTT with and without PDCP proxy.  Standard deviations within parentheses.}}
  \vspace{3pt}
  \label{tab:results-proxy-eval}
  \centering
  {\small
    \begin{tabular}{|c|c|c|}
    \cline{2-3}
    \multicolumn{1}{c|}{~} & \textbf{Without proxy} & \textbf{With proxy} \\ \hline
    \textbf{RTT (ms)} & 56.01 (0.62) & 58.07 (0.85) \\ \hline
    \textbf{Throughput (Mbps)} & 18.28 (0.17) & 17.00 (0.30) \\ \hline
    \end{tabular}
  }
\vspace{-4pt}
\end{table}

{\bf Trace-based evaluation:}
Finally, 
\revone{}{to capture a use case with varying bandwidth,}
experiments were repeated using a real LTE bandwidth trace collected by Raca et al.~\cite{RQZS18}. The specific bandwidth trace (\textit{Static~A\_2018.02.12\_16.14.02})
\revone{}{is used for both downlinks and}
has an average throughput of 4.5~Mbps for the first 200 seconds.
\revone{}{Figure~\ref{fig:bw-traces} shows the sampled bandwidth trace used as a function of time.} 

\subsection{\revtwo{}{Dual connectivity overhead}}
\revtwo{}{To evaluate the overhead invoked by our testbed simulating DC, we perform 10 independent measurements using our default configuration and capture both the bandwidth and RTT with and without using our PDCP proxies. When running without a proxy, packets are being passed directly to the interfaces, skipping the NFQUEUE. Table~\ref{tab:results-proxy-eval} shows the evaluation results with their standard deviations. When using PDCP proxy, the average RTT increases with 2~ms. This corresponds to 1~ms per proxy, as packets are passed through both a client and server proxy. The increase of RTT causes the throughput to drop from 18.28~Mbps to 17.00~Mbps. This shows that only small performance overheads are added from our DC simulations.
}


\begin{table*}[t]
\caption{\revtwo{}{Performance evaluation results overview.}}
\vspace{6pt}
\label{tab:results-overview}
\centering
{\small
\begin{tabular}{|l|l|l|l|l|l|}
\hline
\textbf{Class} & \textbf{Parameter} & \textbf{QUIC impl.} & \textbf{CC algorithm} & \textbf{Throughput Fig.} & \textbf{Fairness Fig.} \\ \hline
\multirow{2}{*}{\makecell[l]{DC parameters \\ 
(Section~\ref{ssec:results-dc-param})}}                   & DC batch size                                      & \multirow{8}{*}{aioquic} & \multirow{8}{*}{NewReno} & \ref{fig:A-1} & \ref{fig:B-1}             \\ \cline{2-2} \cline{5-6} 
                                                          & DC batch split                                     &                          &                          & \ref{fig:D-1} & \ref{fig:E-1}             \\ \cline{1-2} \cline{5-6} 
\multirow{5}{*}{\makecell[l]{Network conditions \\ 
(Section~\ref{ssec:results-network-conditions})}}         & BW ratio                                           &                          &                          & \ref{fig:D-2} & \ref{fig:E-2}             \\ \cline{2-2} \cline{5-6} 
                                                          & BW \& DC ratio                                     &                          &                          & \ref{fig:D-3} & \ref{fig:E-3}             \\ \cline{2-2} \cline{5-6} 
                                                          & Low delay ratio                                    &                          &                          & \ref{fig:F-1} & \ref{fig:G-1}             \\ \cline{2-2} \cline{5-6} 
                                                          & High delay ratio                                   &                          &                          & \ref{fig:F-2} & \ref{fig:G-2}             \\ \cline{2-2} \cline{5-6} 
                                                          & Loss rates                                         &                          &                          & \ref{fig:F-3} & \ref{fig:G-3}             \\ \cline{1-2} \cline{5-6} 
Duplicate packets 
(Section~\ref{ssec:results-duplicate-packets})            & Loss rates                                         &                          &                          & \ref{fig:H-1} & \ref{fig:H-3}             \\ \hline
\multirow{16}{*}{\makecell[l]{QUIC implementation \\
and congestion control \\ algorithm 
(Section~\ref{ssec:results-cc-versions})}}                & \multirow{2}{*}{DC batch size}                     & \multirow{16}{*}{ngtcp2} & NewReno                  & \multirow{2}{*}{\ref{fig:I-1}} & \ref{fig:I-2}   \\ \cline{4-4} \cline{6-6} 
                                                          &                                                    &                          & CUBIC                    &                                & \ref{fig:I-3}   \\ \cline{2-2} \cline{4-6} 
                                                          & \multirow{2}{*}{DC batch split}                    &                          & NewReno                  & \multirow{2}{*}{\ref{fig:J-1}} & \ref{fig:K-1}   \\ \cline{4-4} \cline{6-6}  
                                                          &                                                    &                          & CUBIC                    &                                & \ref{fig:KC-1}  \\ \cline{2-2} \cline{4-6} 
                                                          & \multirow{2}{*}{High delay ratio}                  &                          & NewReno                  & \multirow{2}{*}{\ref{fig:J-2}} & \ref{fig:K-2}   \\ \cline{4-4} \cline{6-6}  
                                                          &                                                    &                          & CUBIC                    &                                & \ref{fig:KC-2}  \\ \cline{2-2} \cline{4-6} 
                                                          & \multirow{2}{*}{Loss rates}                        &                          & NewReno                  & \multirow{2}{*}{\ref{fig:J-3}} & \ref{fig:K-3}   \\ \cline{4-4} \cline{6-6}   
                                                          &                                                    &                          & CUBIC                    &                                & \ref{fig:KC-3}  \\ \cline{2-2} \cline{4-6} 
                                                          & \multirow{2}{*}{BW ratio}                          &                          & NewReno                  & \multirow{2}{*}{\ref{fig:X-1}} & \ref{fig:X-2}   \\ \cline{4-4} \cline{6-6}   
                                                          &                                                    &                          & CUBIC                    &                                & \ref{fig:X-3}   \\ \cline{2-2} \cline{4-6} 
                                                          & \multirow{2}{*}{BW \& DC ratio}                    &                          & NewReno                  & \multirow{2}{*}{\ref{fig:Y-1}} & \ref{fig:Y-2}   \\ \cline{4-4} \cline{6-6}   
                                                          &                                                    &                          & CUBIC                    &                                & \ref{fig:Y-3}   \\ \cline{2-2} \cline{4-6} 
                                                          & \multirow{2}{*}{Low delay ratio}                   &                          & NewReno                  & \multirow{2}{*}{\ref{fig:Z-1}} & \ref{fig:Z-2}   \\ \cline{4-4} \cline{6-6}   
                                                          &                                                    &                          & CUBIC                    &                                & \ref{fig:Z-3}   \\ \cline{2-2} \cline{4-6} 
                                                          & \multirow{2}{*}{\makecell[l]{Loss rates with \\ 
                                                                                    duplicate packets}}        &                          & NewReno                  & \multirow{2}{*}{\ref{fig:Q-1}} & \ref{fig:Q-2}   \\ \cline{4-4} \cline{6-6}   
                                                          &                                                    &                          & CUBIC                    &                                & \ref{fig:Q-3}   \\ \hline
\multirow{12}{*}{\makecell[l]{Bandwidth variability \\
(Section~\ref{ssec:results-bw-variability})}}             & \multirow{2}{*}{DC batch size}                     & aioquic                  & NewReno                  & \ref{fig:L-1} & \multirow{12}{*}{N/A$^*$}            \\ \cline{3-5} 
                                                          &                                                    & ngtcp2                   & NewReno + CUBIC          & \ref{fig:L-4} &              \\ \cline{2-5} 
                                                          & \multirow{2}{*}{High delay ratio}                  & aioquic                  & NewReno                  & \ref{fig:L-2} &              \\ \cline{3-5} 
                                                          &                                                    & ngtcp2                   & NewReno + CUBIC          & \ref{fig:L-5} &              \\ \cline{2-5} 
                                                          & \multirow{2}{*}{Loss rates}                        & aioquic                  & NewReno                  & \ref{fig:L-3} &              \\ \cline{3-5} 
                                                          &                                                    & ngtcp2                   & NewReno + CUBIC          & \ref{fig:L-6} &              \\ \cline{2-5} 
                                                          & \multirow{2}{*}{DC batch split}                    & aioquic                  & NewReno                  & \ref{fig:U-1} &              \\ \cline{3-5} 
                                                          &                                                    & ngtcp2                   & NewReno + CUBIC          & \ref{fig:V-1} &              \\ \cline{2-5} 
                                                          & \multirow{2}{*}{Low delay ratio}                   & aioquic                  & NewReno                  & \ref{fig:U-2} &              \\ \cline{3-5} 
                                                          &                                                    & ngtcp2                   & NewReno + CUBIC          & \ref{fig:V-2} &              \\ \cline{2-5} 
                                                          & \multirow{2}{*}{\makecell[l]{Loss rates with \\
                                                                                    duplicate packets}}        & aioquic                  & NewReno                  & \ref{fig:U-3} &              \\ \cline{3-5} 
                                                          &                                                    & ngtcp2                   & NewReno + CUBIC          & \ref{fig:V-3} &              \\ \hline
\multicolumn{6}{l}{
\footnotesize{\makecell*[l]{ $^*$ After validating that the throughput results held with variable bandwidth conditions, it was decided to not run a full set of corresponding fairness tests.}}
}
\end{tabular}
}
\vspace{-9pt}
\end{table*}


\begin{figure*}[t]
    \subfigure[\revtwo{}{Impact of} DC batch size\label{fig:A-1}]{
        \includegraphics[trim = 0mm 4mm 0mm 0mm, width=\onethirdwidth]{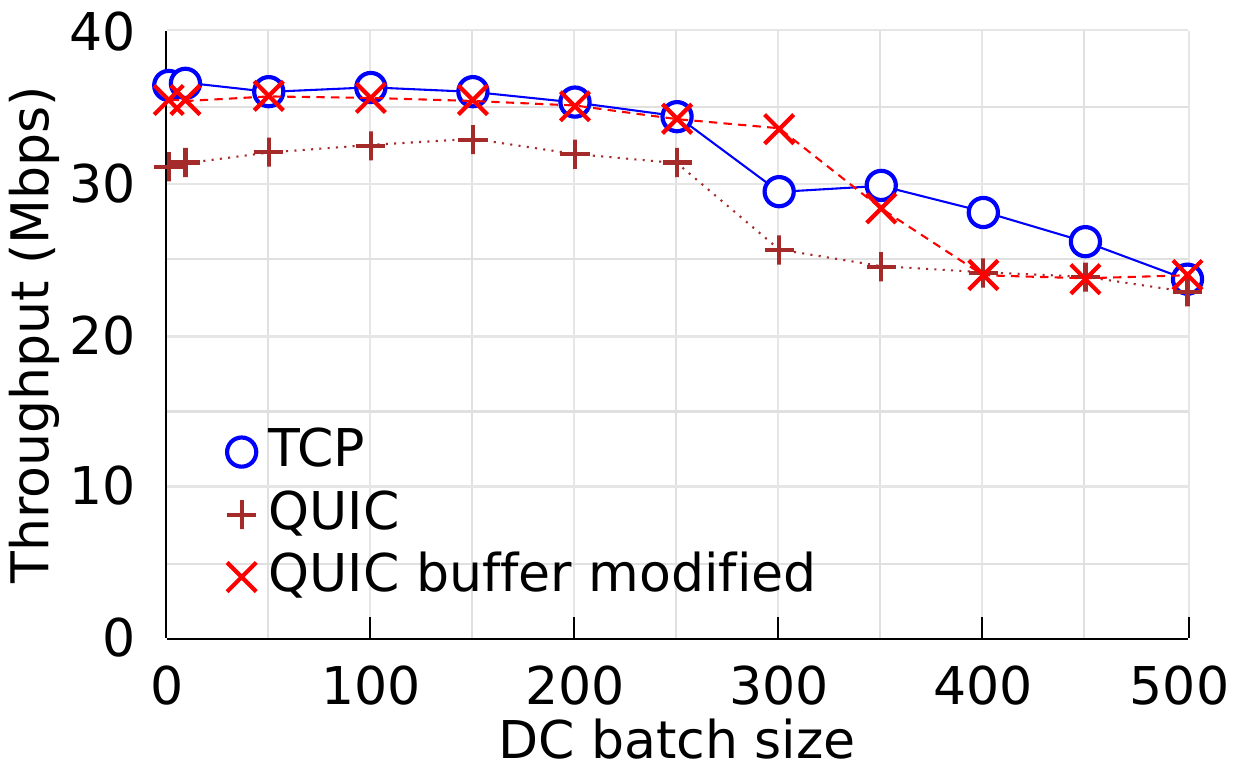}}
    \hspace*{\fill}
    \subfigure[Link utilization with batch size of 50\label{fig:A-2}]{
        \includegraphics[trim = 0mm 4mm 0mm 0mm, width=\onethirdwidth]{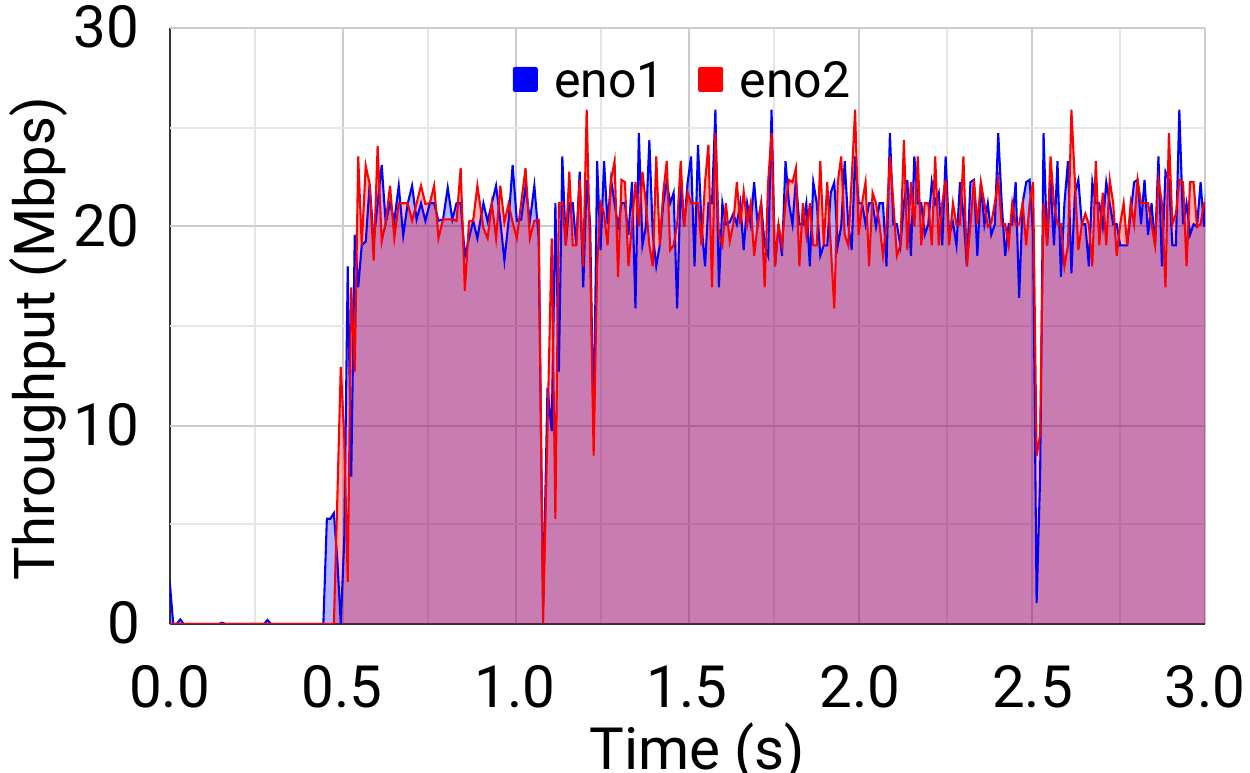}}
    \hspace*{\fill}
    \subfigure[Link utilization with batch size of 500\label{fig:A-3}]{
        \includegraphics[trim = 0mm 4mm 0mm 0mm, width=\onethirdwidth]{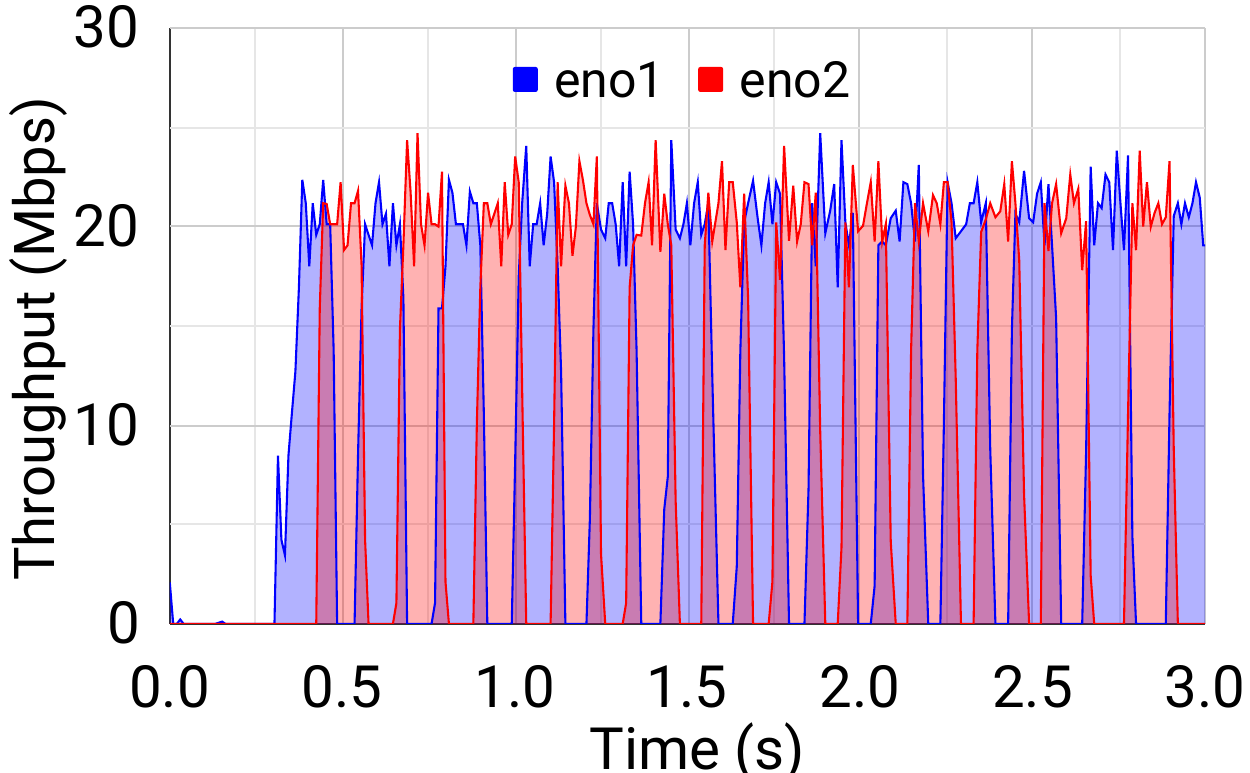}}
        \vspace{-7pt}
    \caption{Throughput and link utilization for different DC batch sizes}
    \label{fig:A}
    \vspace{-6pt}
\end{figure*}

\begin{figure*}[t]
    \begin{minipage}{0.66\linewidth}
    \subfigure[\revtwo{}{Impact of} DC batch size\label{fig:B-1}]{
        \includegraphics[trim = 0mm 4mm 0mm 0mm, width=0.47\linewidth]{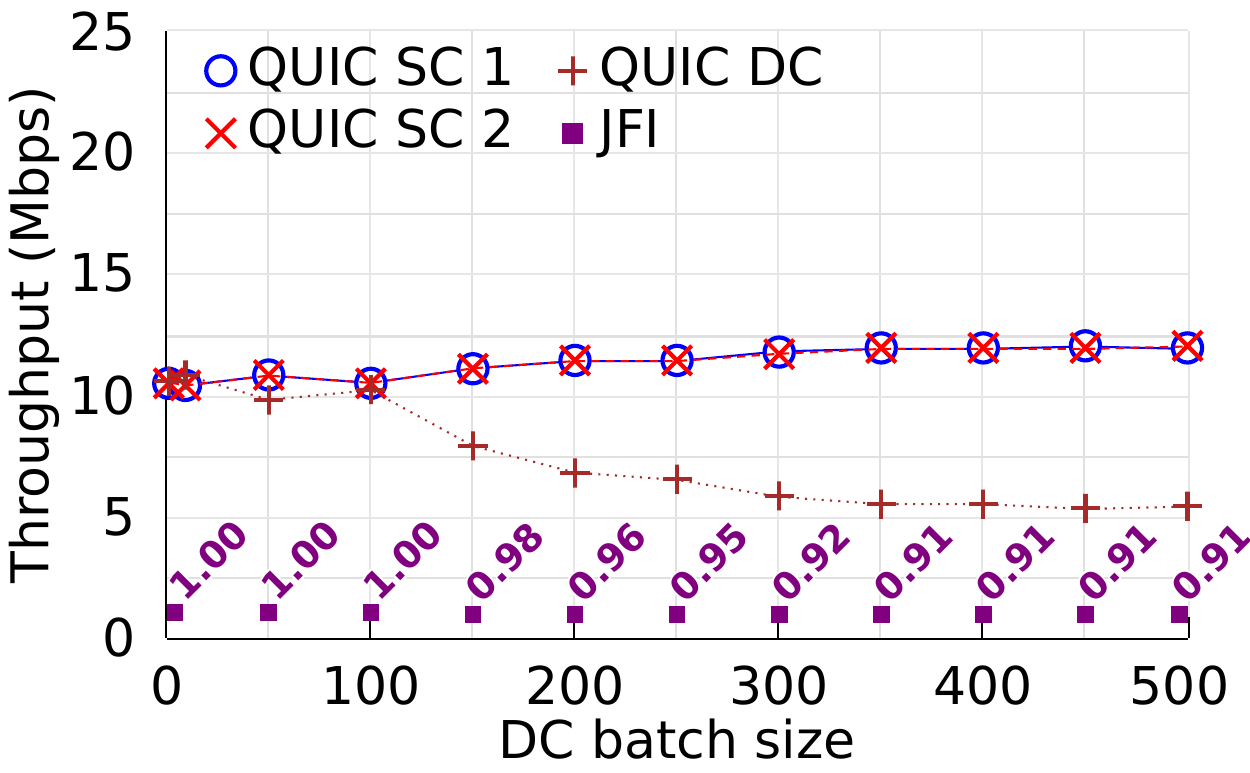}}
    \hfill
    \subfigure[DC fairness example \label{fig:B-2}]{
        \includegraphics[trim = 0mm 4mm 0mm 0mm, width=0.47\linewidth]{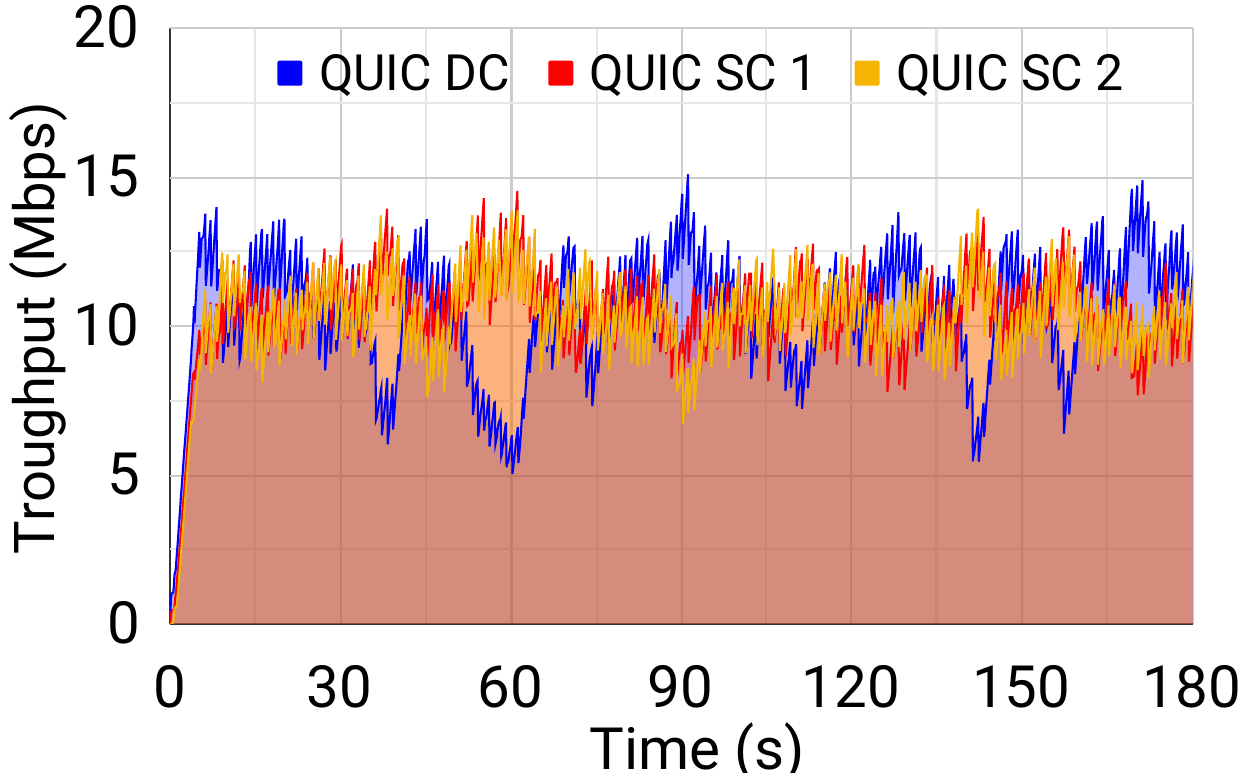}}
    \vspace{-9pt}
    \caption{Fairness for different DC batch sizes}
    \label{fig:B}
    \end{minipage}
    \hfill
    \begin{minipage}{0.305\linewidth}
    \vspace{4pt} 
        \includegraphics[trim = 0mm 4mm 0mm 0mm, width=\linewidth]{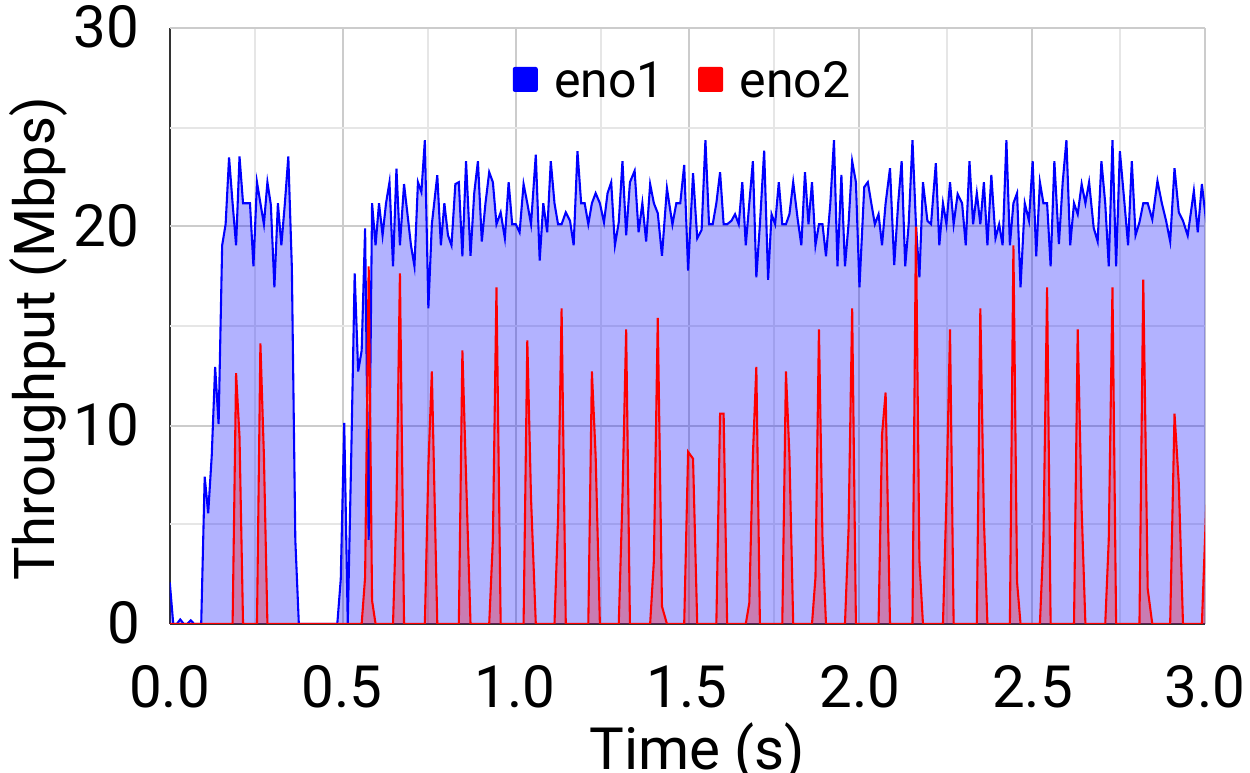}
    \caption{Link utilization example with unbalanced DC ratio of 9:1}
    \label{fig:C}
    \end{minipage}
\end{figure*}

\section{Evaluation Results}
\label{sec:results}

\revtwo{}{
In this section, we present our performance evaluation results under different scenarios and configuration.
First, we show the impact of DC parameters that network operators control (Section~\ref{ssec:results-dc-param}). Then, we present our experiments under different network conditions (Section~\ref{ssec:results-network-conditions}), before showing the impact of using DC to duplicate packets and improve reliability (Section~\ref{ssec:results-duplicate-packets}). Next, we repeat our experiments using another QUIC implementation and congestion control algorithm (Section~\ref{ssec:results-cc-versions}), and over a variable bandwidth scenario (Section~\ref{ssec:results-bw-variability}). 
Finally, we give a summary of the main observations from our results (Section~\ref{ssec:results-summary}).
Table~\ref{tab:results-overview} summarizes our experiments and provides an overview of the presented figures in this section.
}

\begin{figure*}[!t]
    \subfigure[DC batch split\label{fig:D-1}]{\includegraphics[trim = 0mm 4mm 0mm 0mm, width=\onethirdwidth]{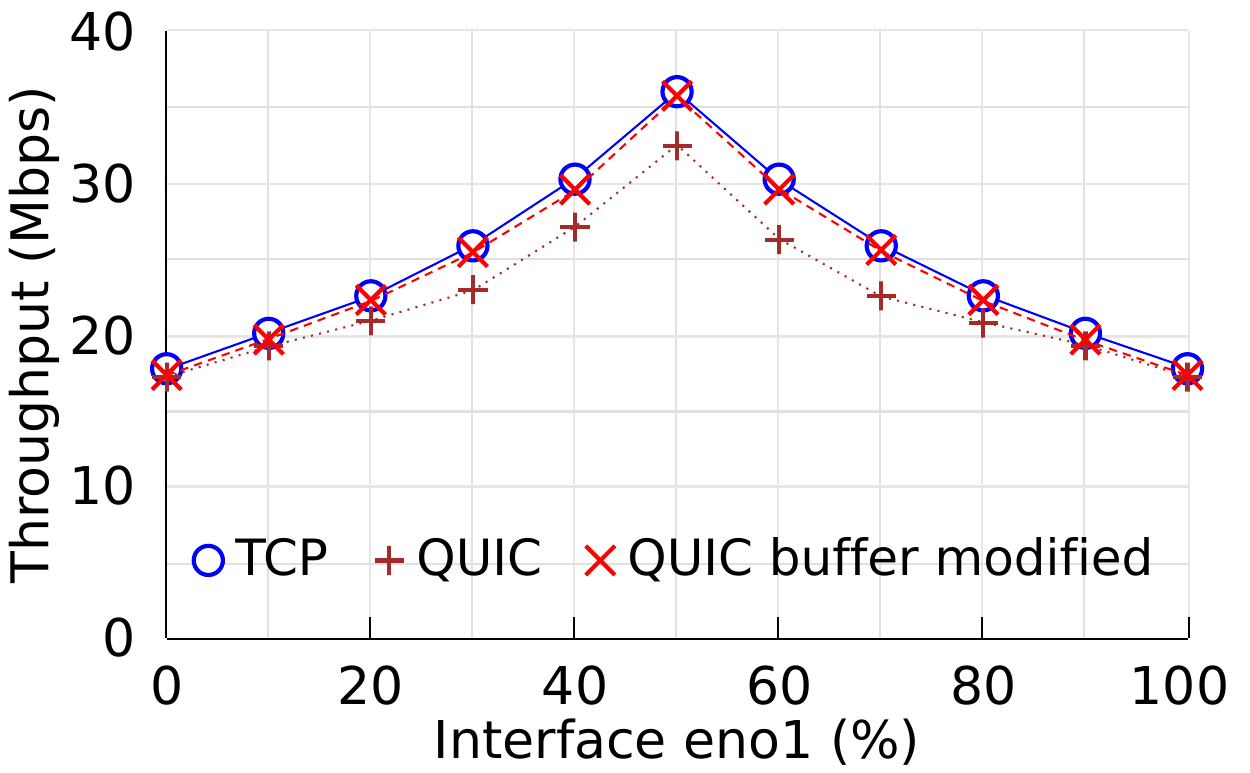}}
    \hspace*{\fill}
    \subfigure[BW ratio\label{fig:D-2}]{\includegraphics[trim = 0mm 4mm 0mm 0mm, width=\onethirdwidth]{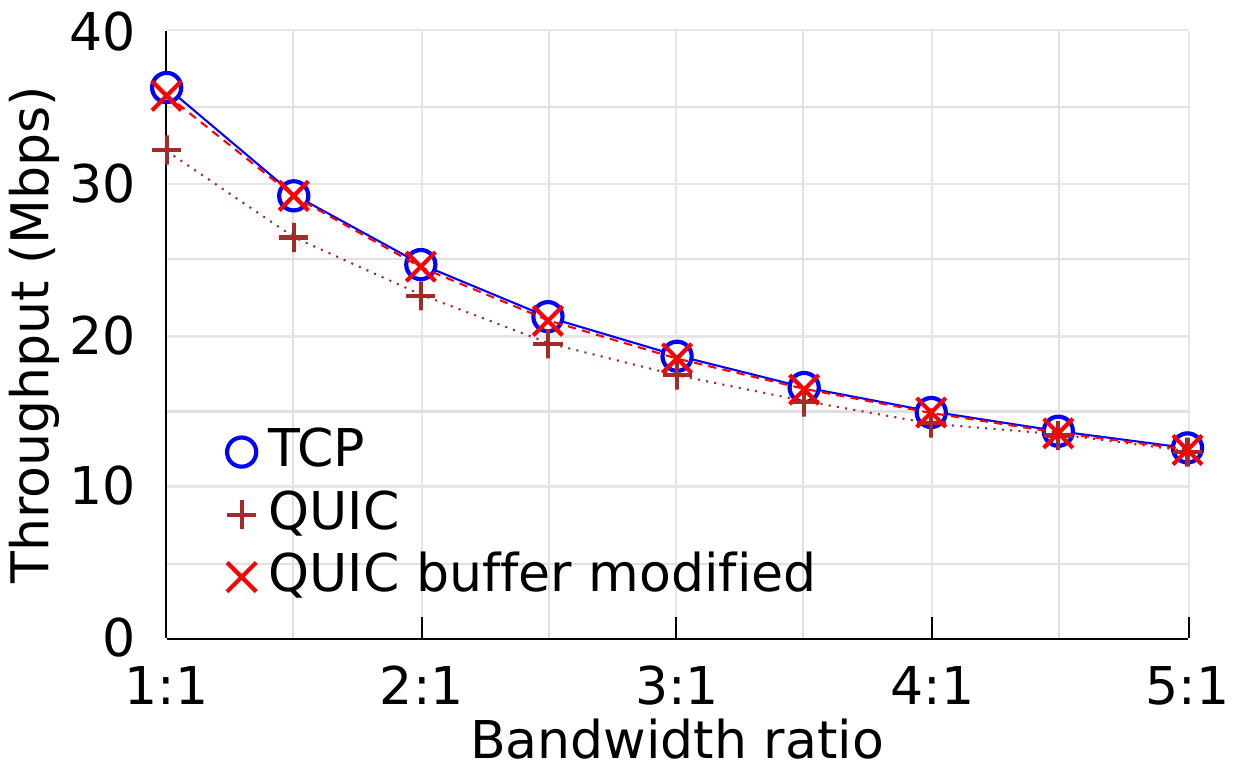}}
    \hspace*{\fill}
    \subfigure[BW \& DC ratio\label{fig:D-3}]{\includegraphics[trim = 0mm 4mm 0mm 0mm, width=\onethirdwidth]{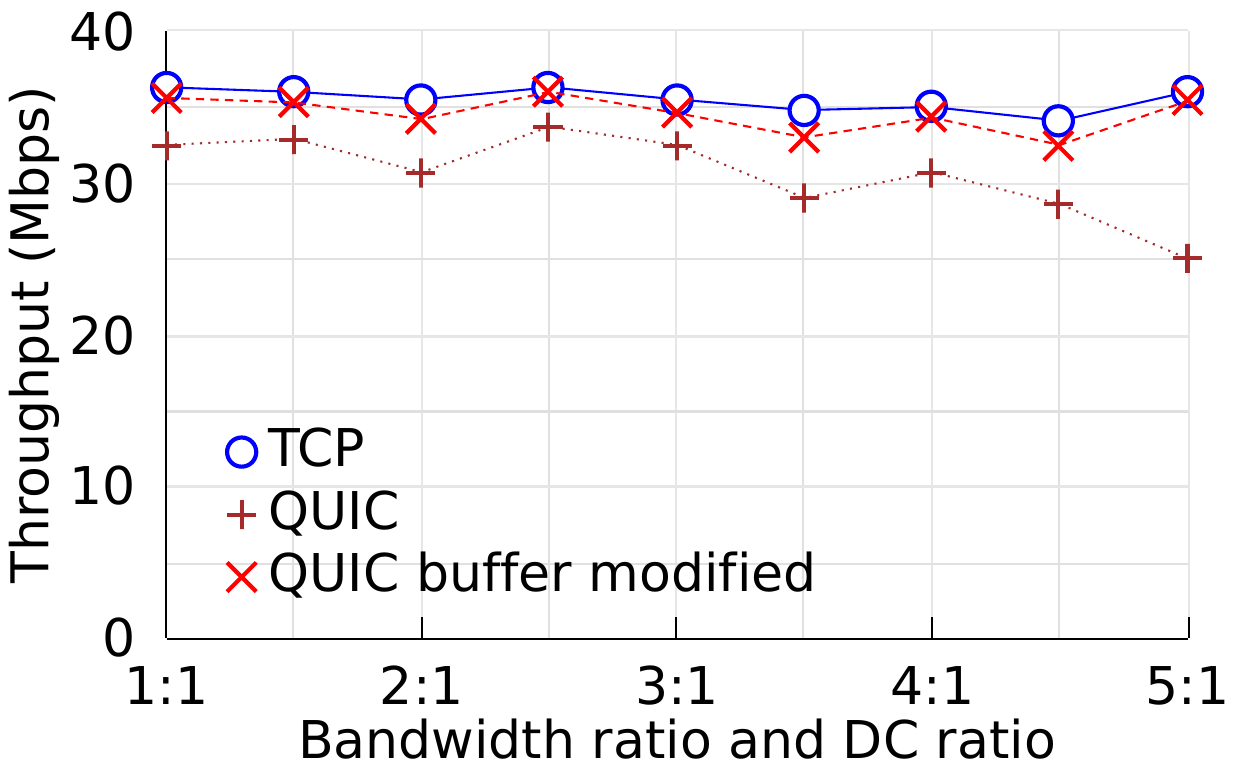}}
    \vspace{-9pt}
    \caption{Throughput for different batch splits and bandwidth ratios}
    \label{fig:D}
\vspace{-4pt}
\end{figure*}

\begin{figure*}[!t]
    \subfigure[DC batch split\label{fig:E-1}]{\includegraphics[trim = 0mm 4mm 0mm 0mm, width=\onethirdwidth]{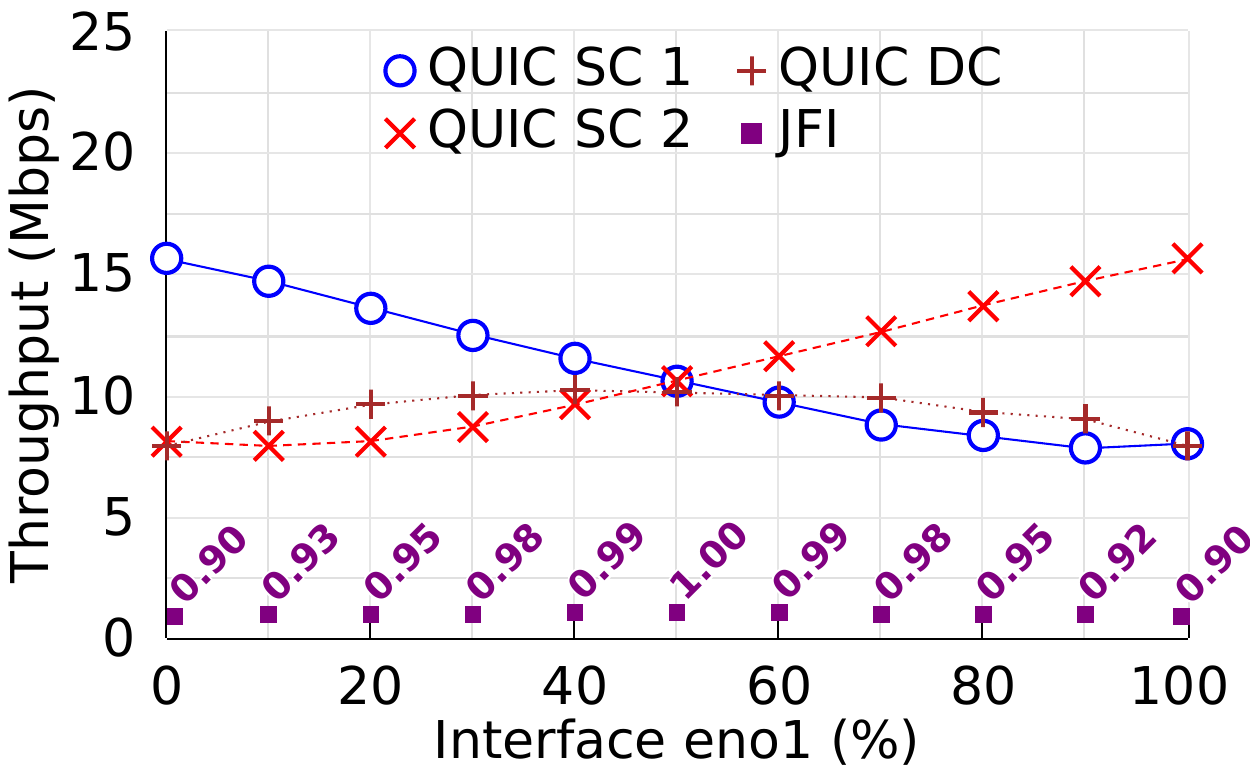}}
    \hspace*{\fill}
    \subfigure[BW ratio\label{fig:E-2}]{\includegraphics[trim = 0mm 4mm 0mm 0mm, width=\onethirdwidth]{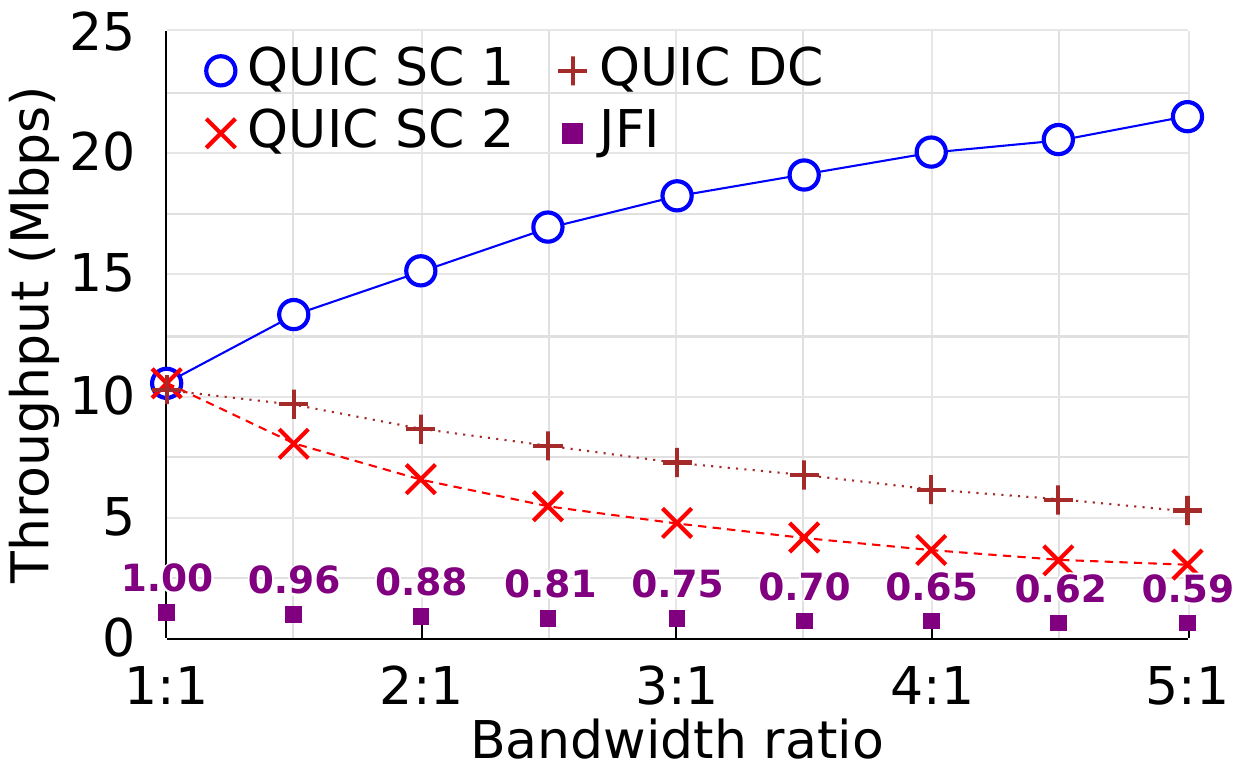}}
    \hspace*{\fill}
    \subfigure[BW \& DC ratio\label{fig:E-3}]{\includegraphics[trim = 0mm 4mm 0mm 0mm, width=\onethirdwidth]{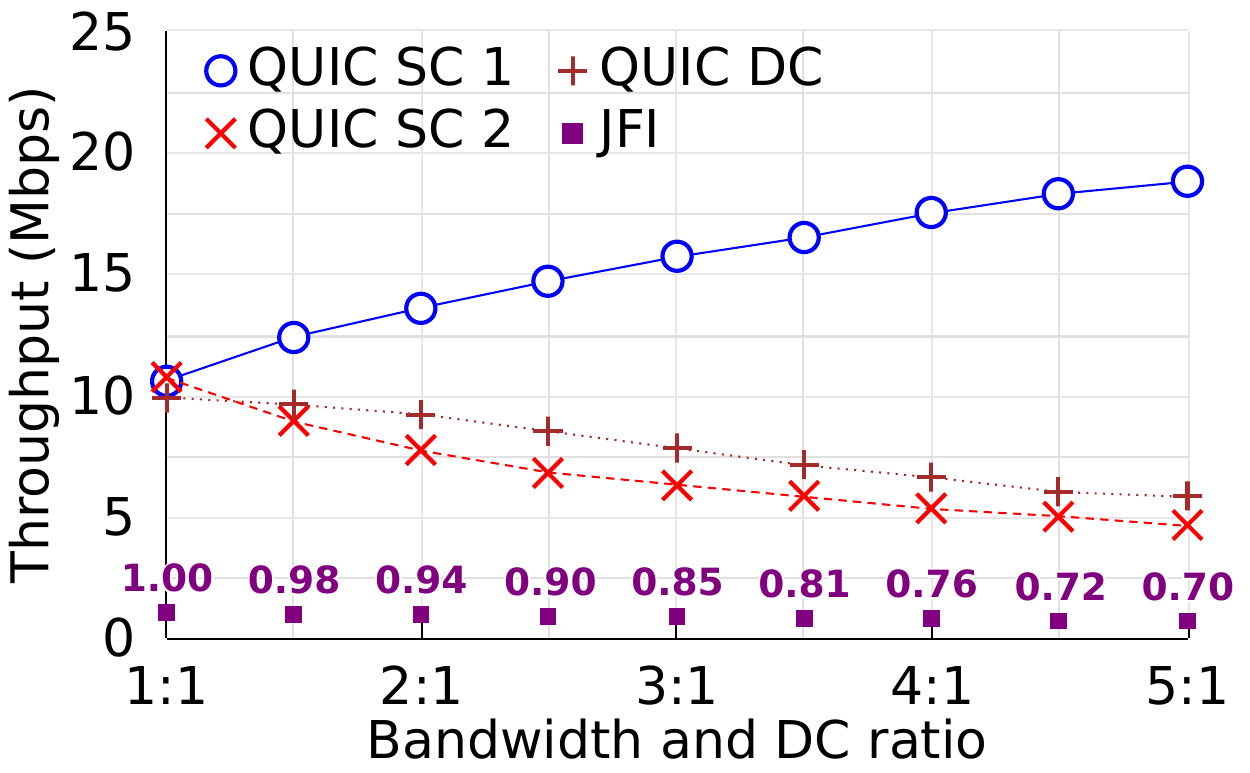}}
    \vspace{-9pt}
    \caption{Fairness for different batch splits and bandwidth ratios}
    \label{fig:E}
\vspace{-4pt}
\end{figure*}

\subsection{Dual connectivity parameters}\label{ssec:results-dc-param}

{\bf DC batch size:}
When using DC, network operators must select a good DC batch size for each connection.  To illustrate the impact of this choice on QUIC performance, Figure~\ref{fig:A-1} shows the throughput as a function of the DC batch size. (We omitted the standard deviations from all figures, as
\revone{the they}{they}
are small; e.g., well within 1~Mbps in more than 90\% of the cases.) In general, large DC batch sizes result in lower throughput.  One reason for this is reduced link utilization. For example, Figures~\ref{fig:A-2} and~\ref{fig:A-3} show the link utilization of the two links when using a DC batch size of 50 and 500, respectively. With a large DC batch size, we see significant periods during which one of the links is underutilized as almost all packets are being forwarded over the other interface. With smaller DC batch sizes, both links can better be used concurrently. However, there is also a penalty to using too small batch sizes, as this increases the number of re-order events. The best batch sizes are instead typically in the mid-range (e.g., around 100-150), with the sweet spot depending on the protocol being used. Finally, we note that QUIC with modified buffers
\revone{perform}{performs}
similar to TCP for much of the parameter range.

Figure~\ref{fig:B-1} shows summary results for our fairness tests with varying DC batch sizes.  Here, we measure fairness using Jain's fairness index (JFI), shown using purple text, as averaged over 10 full runs. When discussing fairness, it is important to note that the relative throughput of the competing clients can vary significantly over time.  This is illustrated in Figure~\ref{fig:B-2} where we show example throughput for the three competing clients over a 3-minute long experiment with the default settings.
\revthree{}{Here, it is important to note that the 3-minute time frame (duration of x-axis) is 60 times greater than the total duration shown for the example traces shown in Figures~\ref{fig:A-2} and~\ref{fig:A-3}.  We observe that short-term variations in the fairness persist up to time scales of about 30 seconds.  By using 3-minute traces and reporting averages over 10 runs, such temporal unfairness are filtered out.}
\revone{}{As an additional reference point, 
Appendix~A 
presents example traces with pairs of competing SC clients: (i) QUIC vs. QUIC, (ii) QUIC vs. TCP, and (iii) TCP vs. TCP.}

Similar to the throughput, the fairness is negatively affected by large DC batch sizes. In fact, the user using DC observe a significant throughput reduction with batch sizes of 150 and above. Here, SC clients can monopolize the links during the DC client's off periods, while DC is always sharing the link it is currently sending to at every point of time. This allows SC clients to increase their
\revone{cwnd}{congestion window}
further than the DC client and to use a larger bandwidth share.

{\bf DC batch split:}
\revone{Operators}{Network operators}
also control the DC ratio. This parameter determines the split over the two links. 
\revtwo{}{In contrast to 
Figures~\ref{fig:A-2} and~\ref{fig:A-3},}
Figure~\ref{fig:C} illustrates an unbalanced example with a DC batch split of 9:1, in which 90\% of the packets are sent over the main interface (eno1), and Figure~\ref{fig:D-1} shows the throughput as a function of the percent of packets sent over eno1. As per our default case, both links have the same network conditions. 
\revthree{In this case,}{Compared with the utilization examples of balanced DC ratio (Figures~\ref{fig:A-2} and~\ref{fig:A-3}), we note that the less loaded link almost never reaches its full capacity; not even during burst periods.  This results in sub-optimal link utilization in which only one of the links are well-utilized (close to 100\%).  This case corresponds to the 10\% case (and by symmetry the 90\% case) shown in Figure~\ref{fig:D-1}. More generally,}
the throughput peaks when using a 50/50 split, and decreases as a convex function as the split becomes more uneven. 
\revthree{This decrease is caused by poor link utilization of the less loaded link (eno2 in Figure~\ref{fig:C}), but also demonstrates the value of DC.}{The decrease in throughput caused by poor link utilization of the less loaded link (eno2 in Figure~\ref{fig:C}) highlights the value of careful batch split selection to best achieve the full potential of DC.}


\begin{figure*}[!t]
    \subfigure[Low delay ratio\label{fig:F-1}]{\includegraphics[trim = 0mm 4mm 0mm 0mm, width=\onethirdwidth]{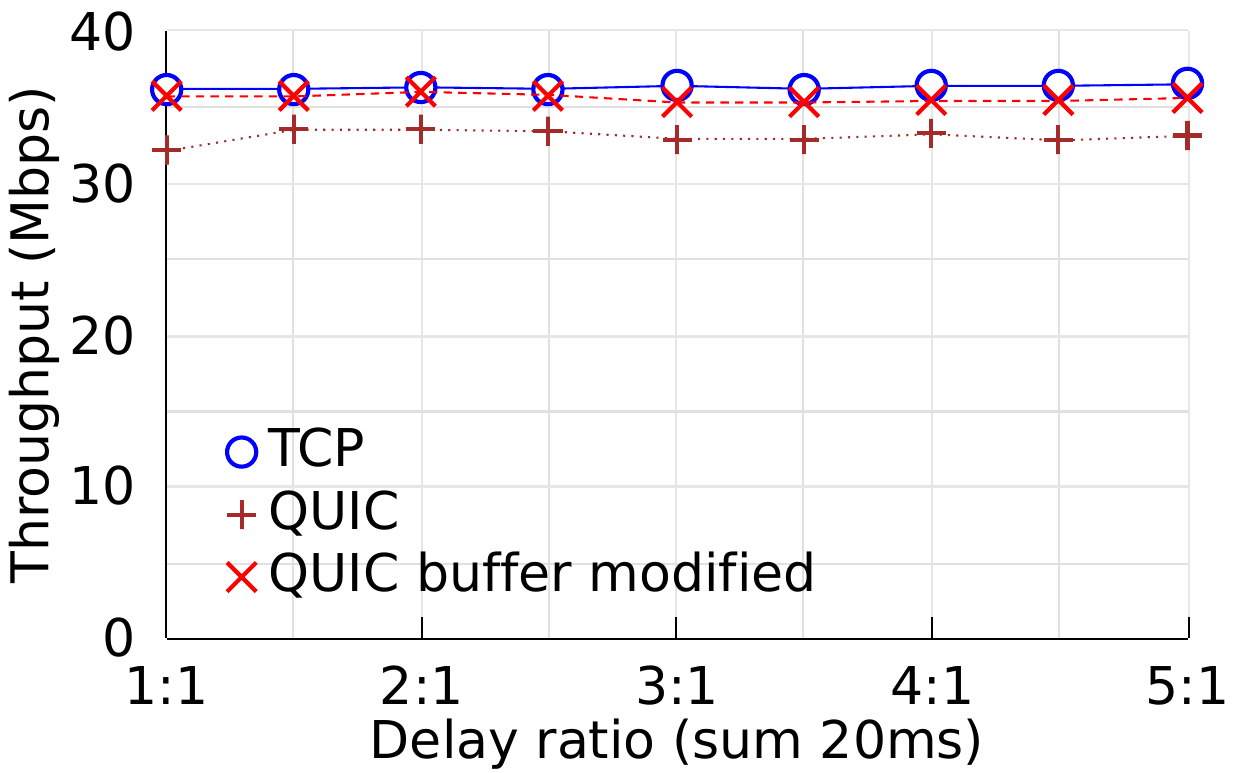}}
    \hspace*{\fill}
    \subfigure[High delay ratio\label{fig:F-2}]{\includegraphics[trim = 0mm 4mm 0mm 0mm, width=\onethirdwidth]{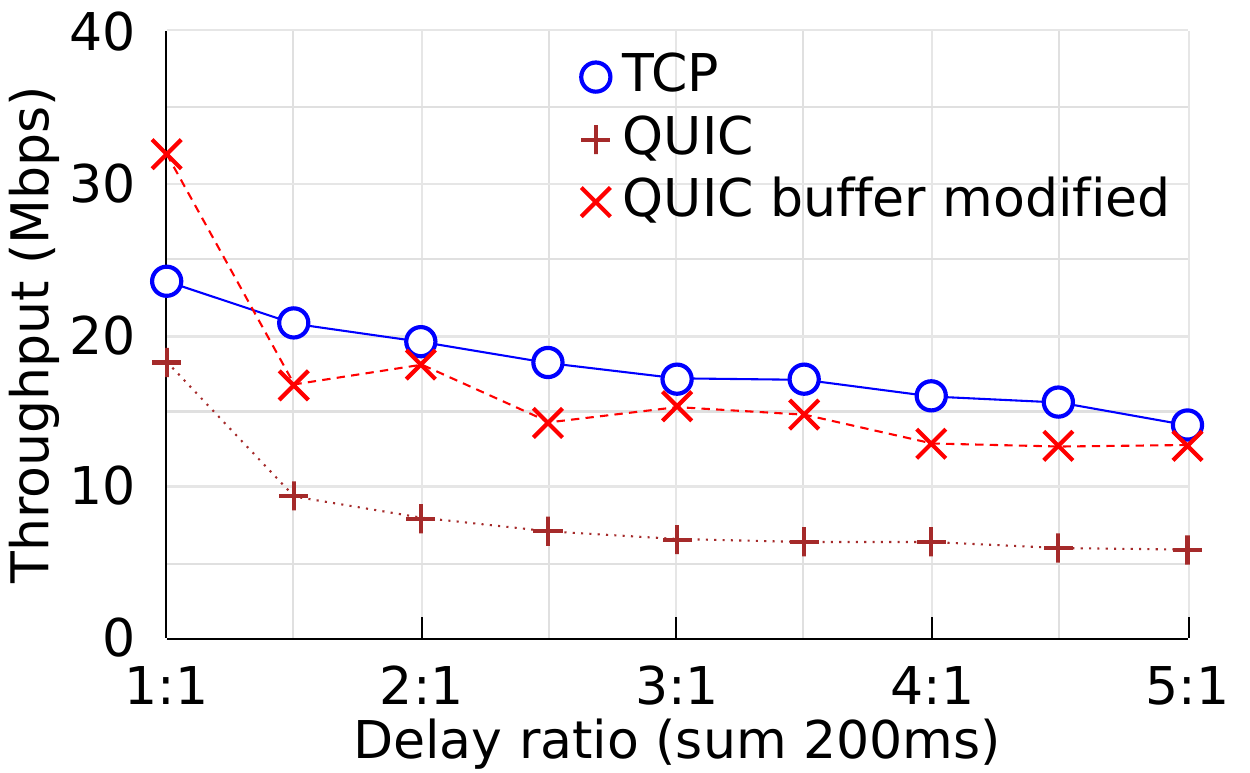}}
    \hspace*{\fill}
    \subfigure[Random loss\label{fig:F-3}]{\includegraphics[trim = 0mm 4mm 0mm 0mm, width=\onethirdwidth]{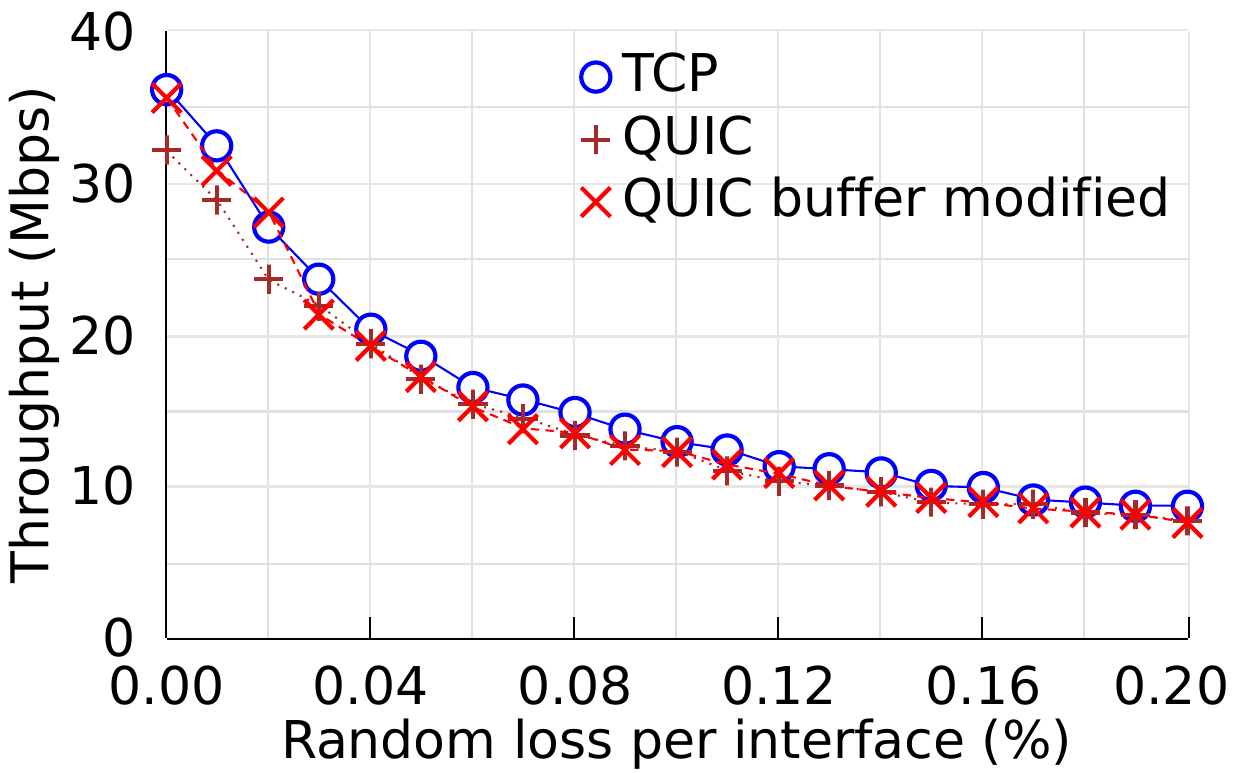}}
    \vspace{-9pt}
    \caption{Throughput for different delay and loss ratios}
    \label{fig:F}
%
    \subfigure[Low delay ratio\label{fig:G-1}]{\includegraphics[trim = 0mm 4mm 0mm 0mm, width=\onethirdwidth]{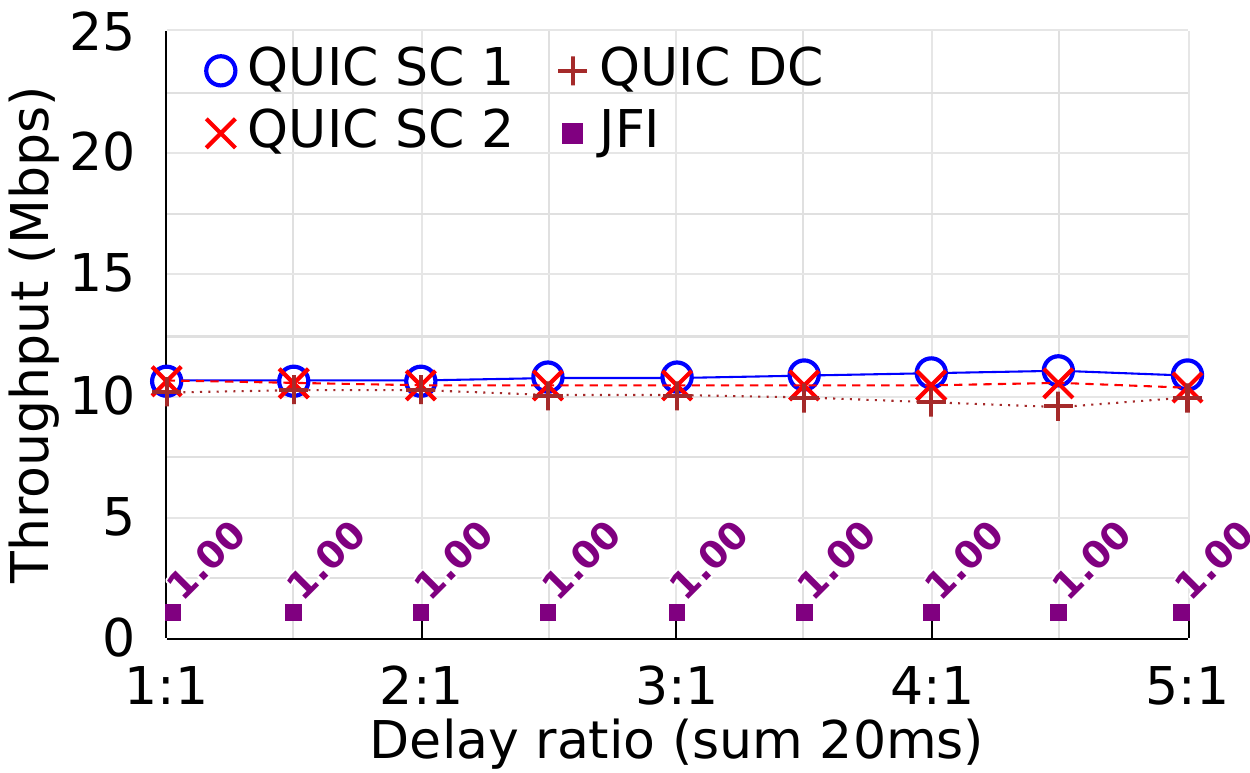}}
    \hspace*{\fill}
    \subfigure[High delay ratio\label{fig:G-2}]{\includegraphics[trim = 0mm 4mm 0mm 0mm, width=\onethirdwidth]{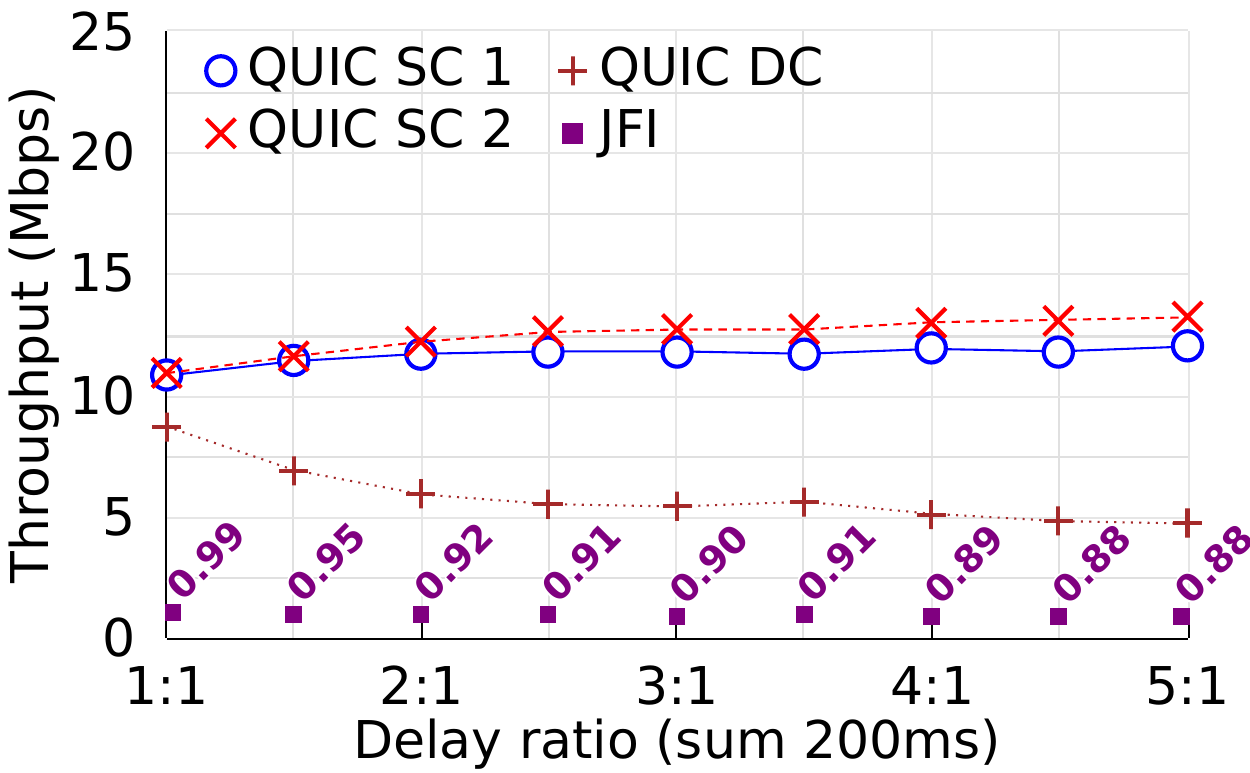}}
    \hspace*{\fill}
    \subfigure[Random loss\label{fig:G-3}]{\includegraphics[trim = 0mm 4mm 0mm 0mm, width=\onethirdwidth]{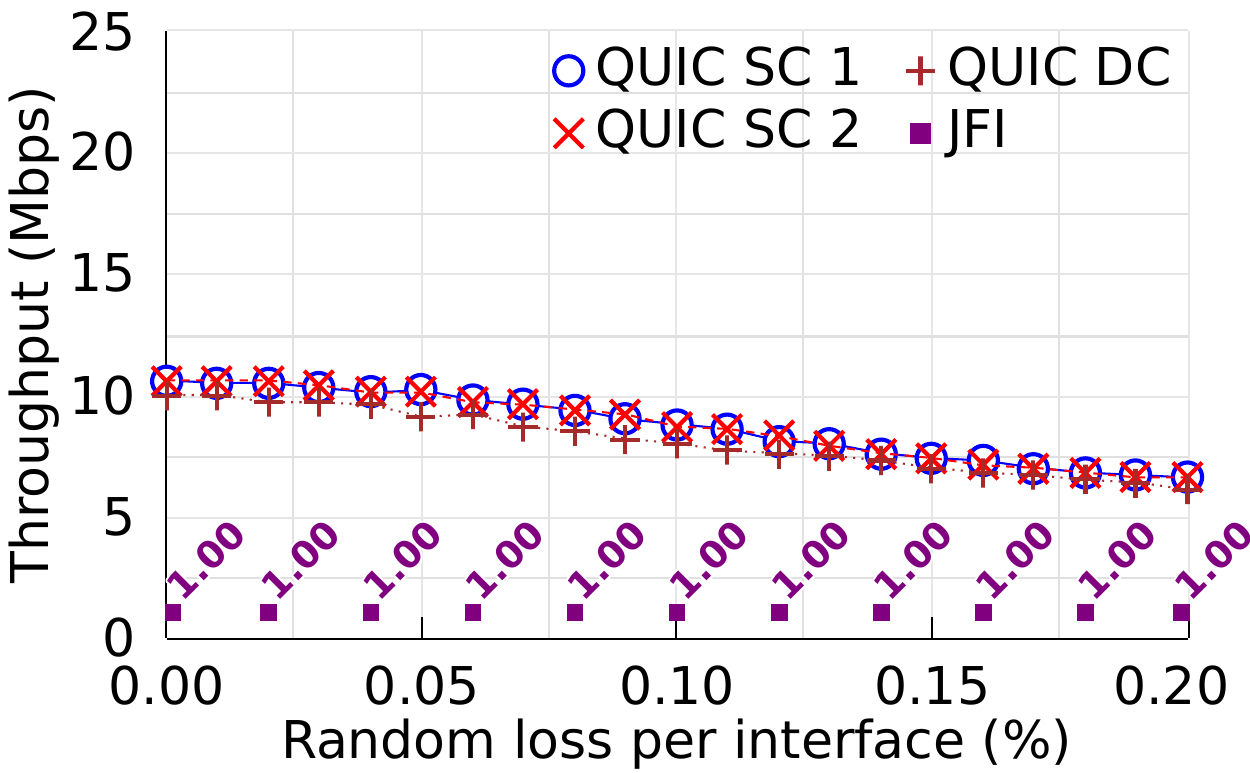}}
    \vspace{-9pt}
    \caption{Fairness for different delay and loss ratios}
    \label{fig:G}
\vspace{-4pt}
\end{figure*}

Figure~\ref{fig:E-1} shows our corresponding fairness results. When the ratio is significantly skewed (e.g., below 20\% or above 80\%), the throughput of the SC with the higher throughput increase/decrease at roughly the same rate as the DC's throughput increase/decrease, whereas the other SC 
\revone{}{with lower throughput}
has fairly constant throughput over these skewed splits. In this region, the DC compete (almost) fairly only over the more utilized interface. As the DC split becomes more even, the overall fairness improves, with optimal fairness and all connections having roughly equal throughput when perfectly balanced.

\begin{figure*}[t]
\centering
    \subfigure[Throughput\label{fig:H-1}]{\includegraphics[trim = -12mm 4mm -12mm 0mm, width=0.38\linewidth]{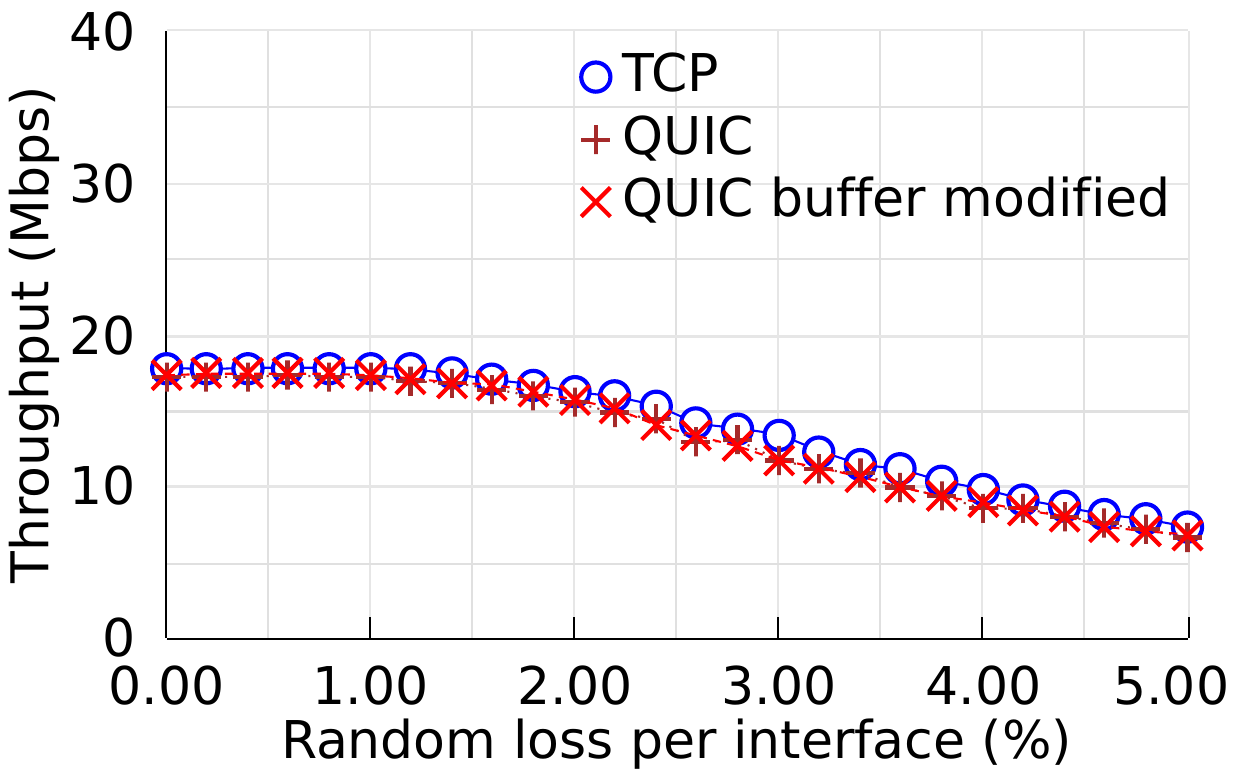}}
    \hspace{8pt}
    \subfigure[Goodput fairness\label{fig:H-3}]{\includegraphics[trim = -12mm 4mm -12mm 0mm, width=0.38\linewidth]{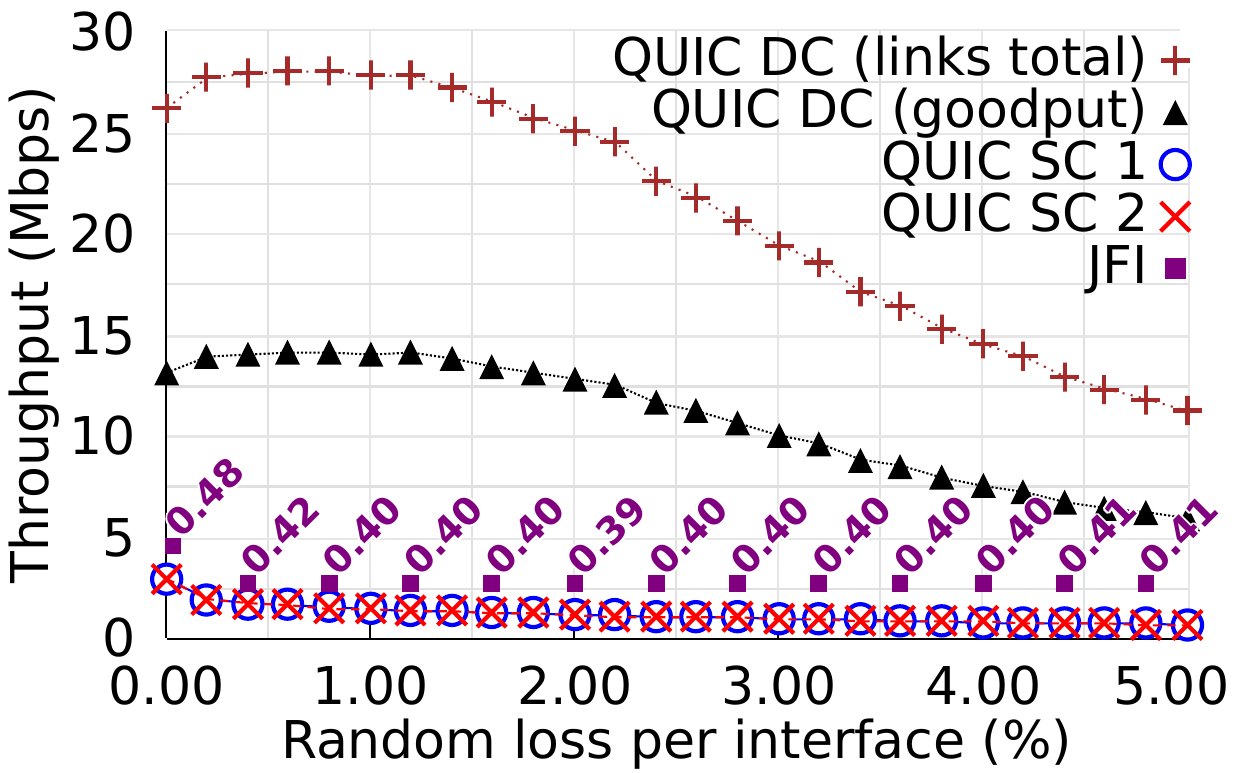}}
    \vspace{-9pt}
    \caption{DC throughput and fairness with duplicate packets}
    \label{fig:H}
    \vspace{-2pt}
\end{figure*}

\begin{figure*}[t]
    \subfigure[Throughput\label{fig:I-1}]{\includegraphics[trim = 0mm 4mm 0mm 0mm, width=\onethirdwidth]{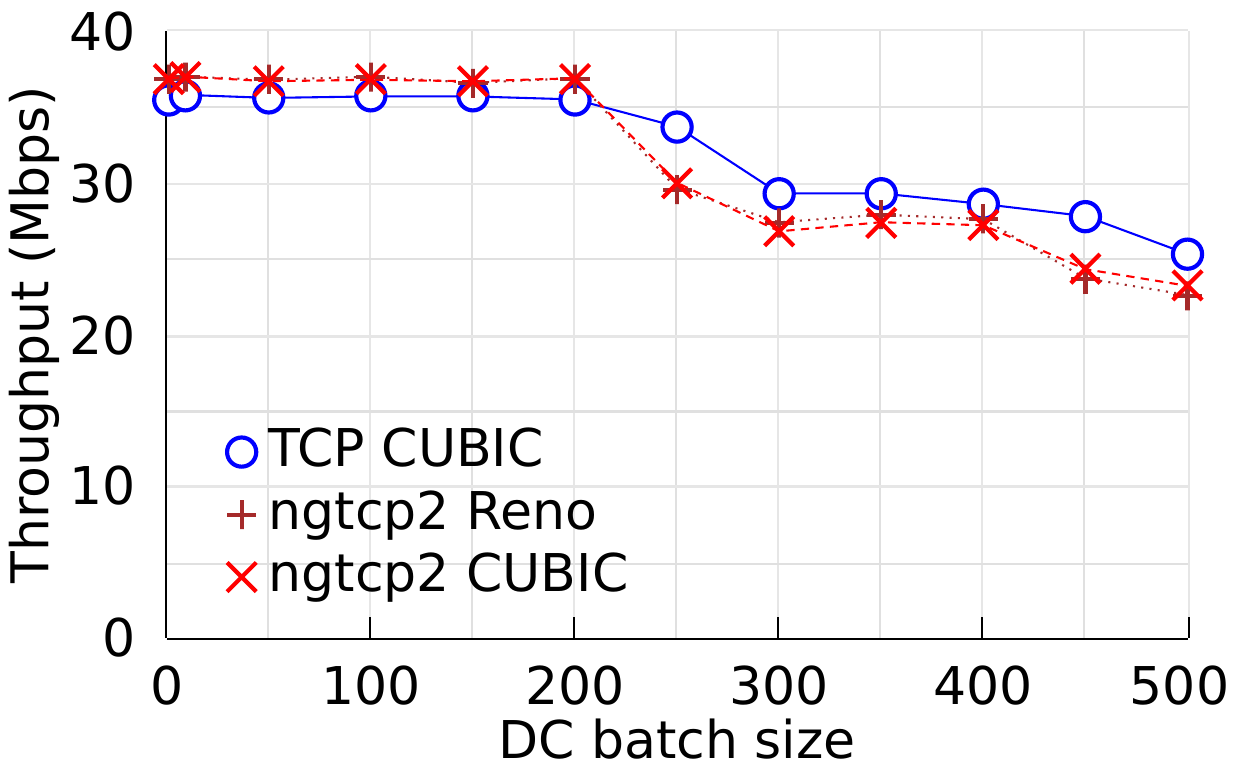}}
    \hspace*{\fill}
    \subfigure[Fairness 
    \revtwo{}{with} 
    NewReno\label{fig:I-2}]{\includegraphics[trim = 0mm 4mm 0mm 0mm, width=\onethirdwidth]{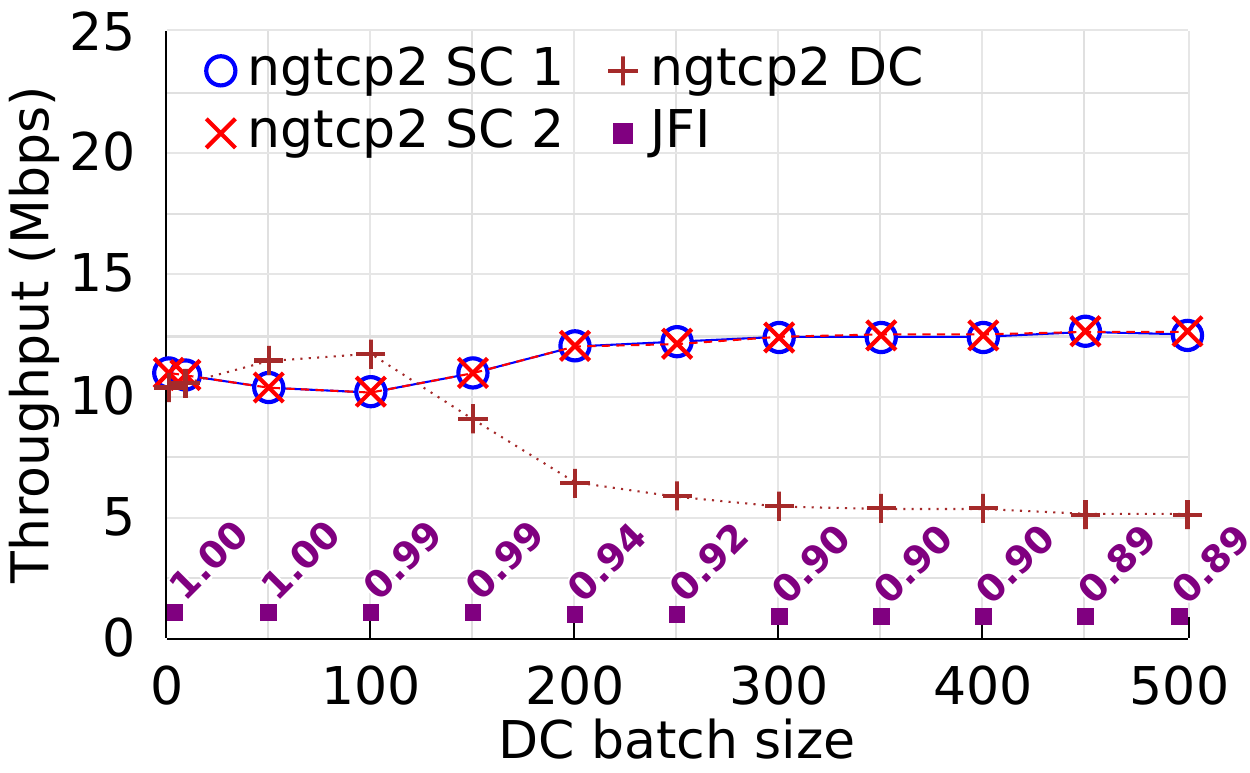}}
    \hspace*{\fill}
    \subfigure[Fairness 
    \revtwo{}{with}
    CUBIC\label{fig:I-3}]{\includegraphics[trim = 0mm 4mm 0mm 0mm, width=\onethirdwidth]{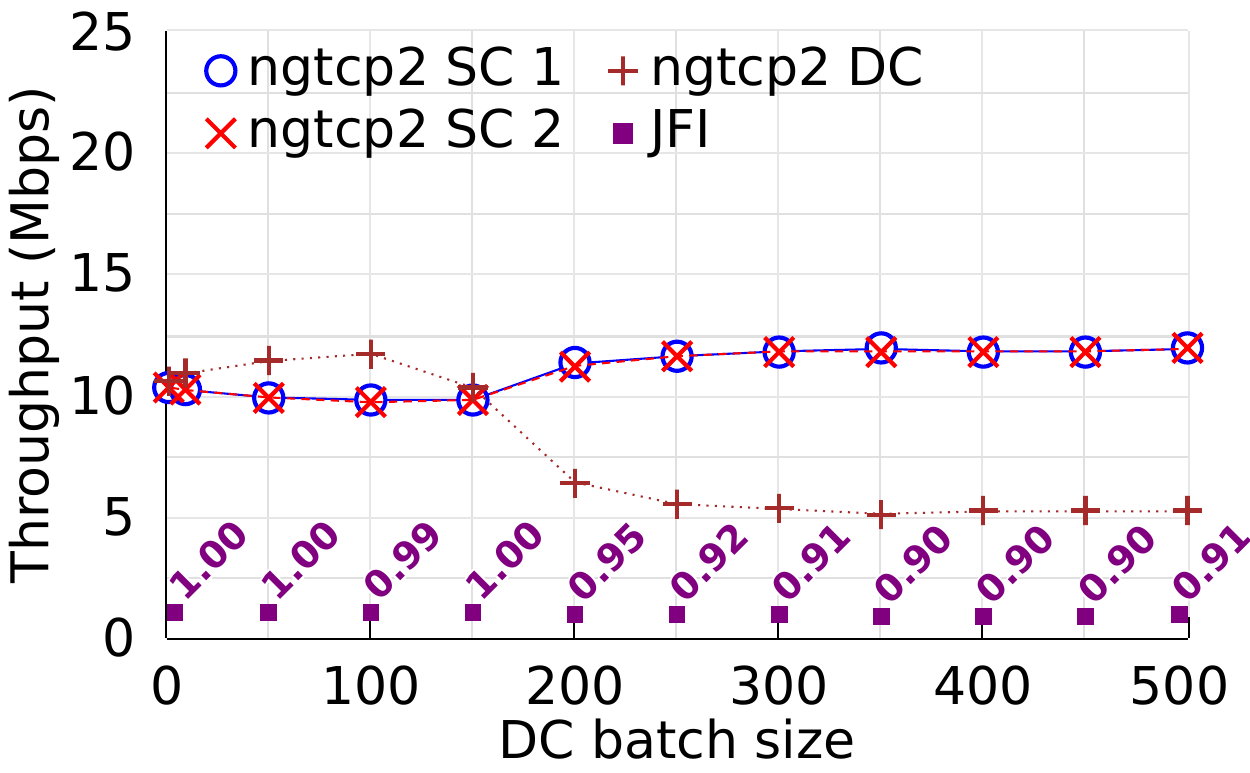}}
    \vspace{-8pt}
    \caption{Impact of congestion control 
    \revtwo{algorithm:}{algorithm (NewReno and CUBIC):} 
    Performance examples with ngtcp2 as QUIC version for different batch sizes}
    \label{fig:I}
\vspace{-4pt}
\end{figure*}

\subsection{Network conditions}\label{ssec:results-network-conditions}

\revone{}{We next look closer at the impact of the network conditions. Again, we consider one parameter at a time, keeping the other parameters fixed as in our default case.}

{\bf Bandwidth ratio:}
Figures~\ref{fig:D-2} and~\ref{fig:D-3} show the throughput for different bandwidth ratios. Figure~\ref{fig:D-2} shows results for the case when the batch split is 50/50, and Figure~\ref{fig:D-3} shows results for when the batch split is selected to match the bandwidth ratio. Figure~\ref{fig:D-2} illustrates the importance of matching ratios, as the highest throughputs are achieved with a ratio of 1:1.  As the ratio increases, a 50/50 batch split underutilizes the link with higher bandwidth. In contrast, when the DC batch split is selected to match the bandwidth ratio (Figure~\ref{fig:D-3}), a much better overall throughput is achieved. With QUIC buffer modified and TCP, the impact is very small. The reason for the worse performance of QUIC with default buffers is the higher burstiness caused by increased reordering. Despite PDCP mitigating reordering, it results in increasing RTTs.

The fairness results for the cases when we vary the bandwidth ratio of the two links are shown in Figures~\ref{fig:E-2} and~\ref{fig:E-3}. Similar to the throughput results, 
\revthree{}{relatively}
higher fairness is achieved when the DC split is selected based on the capacity of the two links. For example, even when the bandwidth ratio is 5:1, the scenario in which the DC split matches the bandwidth ratio achieves a JFI of 0.70, compared to 0.59 in the case a 50/50 split is used. In both cases, the bandwidth usage is dominated by the SC user with higher bandwidth and the DC user relies heavily on the throughput achieved via the weaker link.  However, the fairness improves as DC moves more traffic to the link with the higher bandwidth. 
\revtwo{}{Yet, the highest fairness (JFI=1.00) is achieved only when the ratios are equal.}

{\bf Delay ratio:}
Both the throughput and fairness are negatively affected by increasing delays, and in the case of a high average delay, these metrics are also negatively affected by an increasing delay ratio. This is illustrated by comparing the throughput Figures~\ref{fig:F-1}~and~\ref{fig:F-2} or fairness Figures~\ref{fig:G-1}~and~\ref{fig:G-2}. For both types of experiments, the two figures show results for low-delay and high-delay scenarios, respectively. In the low-delay scenarios, the sum of the delays over the two links is 20~ms, and in the high-delay scenario the sum is 200~ms. 

The throughput decrease is mostly due to increased packet reordering caused by the higher delays. In these cases, the PDCP layer will buffer more packets before performing a batch delivery to the QUIC client, causing packet bursts as well as a higher RTT. Furthermore, after receiving a batch delivery, the clients will send a cumulative ACK for many packets, which will, for a short time, largely decrease the number of packets in flight when received at the server. The draft for QUIC~\cite{QUIC-RFC-Recovery-29} recommends a pacer, which helps the QUIC server recover from an ACK-burst by sending new packets at \revone{}{a} steadier pace. The advantage of more even pacing can be seen by the higher values observed with a delay ratio of 1:1 in Figure~\ref{fig:F-2}.

The increasing delays and delay ratios also negatively impact 
fairness. 
For example, in the low-delay case (Figure~\ref{fig:G-1}), JFI reduces from 0.9996 to 0.9986 
\revtwo{}{(both rounded to 1.00 in the figure)}
as the delay ratio increases from 1:1 to 5:1, whereas JFI drops from 0.9903 to 0.8758 for the high-delay case (Figure~\ref{fig:G-2}). 
\revtwo{}{While the reductions of the fairness index for the high-delay case are significant, these reductions are still much smaller than those observed when increasing the bandwidth rations equally (e.g., Figures~\ref{fig:E-2} and~\ref{fig:E-3}).}
\revthree{}{These relative comparisons suggest that fairness is less affected by network-delay-ratio differences than by  equally large bandwidth-ratio differences even under 
the 
large-delay 
case.}
The higher throughput of SC 
\revtwo{eno2}{2} compared to that of SC 
\revtwo{eno1}{1}
is due to 
its 
lower RTT.

{\bf Loss rates:}
While increased packet losses negatively impact the throughput (Figure~\ref{fig:F-3}), small packet losses have very limited impact on the fairness index (Figure~\ref{fig:G-3}).

\subsection{Use of duplicate packets}\label{ssec:results-duplicate-packets}

\revthree{}{Thus far we have considered a simple use case in which DC primarily is used to improve throughput and no packet is sent over both interfaces.}
Besides improving throughput, DC can also be used to increase connection reliability. 
\revtwo{}{In Figure~\ref{fig:H-1}, we show the throughput when duplicating every packet and sending them on two separate paths. Compared to previous cases where we use DC to improve throughput and obtain an aggregated bandwidth of 40~Mbps, here we are only able to obtain a maximum bandwidth capacity of 20~Mbps, as the two links are used to send redundant data. This clearly shows the tradeoffs and importance of balancing the throughput and reliability.}
However, DC with packet duplication negatively \revone{effects}{affects} fairness. For example, in fairness tests with loss rates of 0-to-5\% (Figure~\ref{fig:H-3}) JFI is in the range from 0.39 to 0.48. For DC in Figure~\ref{fig:H-3}, we show both the combined interface throughput ($B$) and the goodput ($X$), which under an independence model with retransmissions (after simplification) can be related as $X = B (1+p) / 2$, where $p$ is the loss rate. The low fairness stems from DC having a much higher end-to-end packet delivery probability (i.e., $1$$-$$p^2$ \revone{vs}{vs.} $1$$-$$p$ under independence assumptions) and lower end-to-end packet loss probability (i.e., $p^2$ \revone{vs}{vs.} $p$) compared to SC.
This results in DC 
\revone{}{having less end-to-end packet losses, and}
obtaining a 
\revtwo{}{much}
larger share of the link bandwidths. 
\revtwo{}{As SC packets are not duplicated and have no redundancy, packet losses will lead to much performance degradation.}
These results show that 
\revtwo{}{packet}
duplication can provide much higher reliability at the cost of fairness and goodput.
\revtwo{}{We note that with 0\% added random loss per interface, packet loss still occur due to full buffers dropping packets (e.g., from congestion control bandwidth probing). When duplicating packets, QUIC DC is able to better recover and obtain a large bandwidth share, while QUIC with SC struggles to recover, causing high unfairness.}

\subsection{QUIC implementation and congestion control algorithm}\label{ssec:results-cc-versions}

To explore the impact of other QUIC implementations and congestion control algorithms, experiments were repeated using \revtwo{ngtcp2 and CUBIC.}{(1) the QUIC implementation named ngtcp2, and (2) the congestion control algorithm CUBIC. For throughput experiments,}
Figure~\ref{fig:I-1} shows little to no differences in the results between different congestion control algorithms when varying DC batch size. However, when compared to Figure~\ref{fig:A-1}, differences can be observed between the QUIC implementations. While the results follow the same patterns for both implementations, the considerable throughput drop occurs at different batch sizes. Another noticeable difference is that ngtcp2 has a slightly higher throughput than aioquic at smaller batch sizes, exceeding the throughput for TCP.

Figures~\ref{fig:I-2} and~\ref{fig:I-3} show the corresponding fairness results
\revone{}{for different DC batch sizes}.
Again, only small differences between the NewReno and CUBIC results are observed. For example, the JFI differ by at most 0.02 (DC batch size of 500) between the algorithms. DC using CUBIC is initially slightly more resilient to performance drops occurring with a larger batch size. In contrast, NewReno allows the SC connections to achieve slightly higher throughput at larger batch sizes while the DC throughput is similar to CUBIC at higher batches. When comparing Figures~\ref{fig:I-2} and~\ref{fig:I-3} to~\ref{fig:B-1}, some differences can be seen between the QUIC implementations. Ngtcp2 is more aggressive, leading to the DC connection having slightly higher throughput than the SC connections at smaller batch sizes and a drastic reduction in throughput when the batch size increases. Aioquic has a more balanced sharing of the bandwidths at smaller batch sizes and see a smaller reduction in throughput at larger batch sizes.

\begin{figure*}[!t]
    \subfigure[DC batch split\label{fig:J-1}]{\includegraphics[trim = 0mm 4mm 0mm 0mm, width=\onethirdwidth]{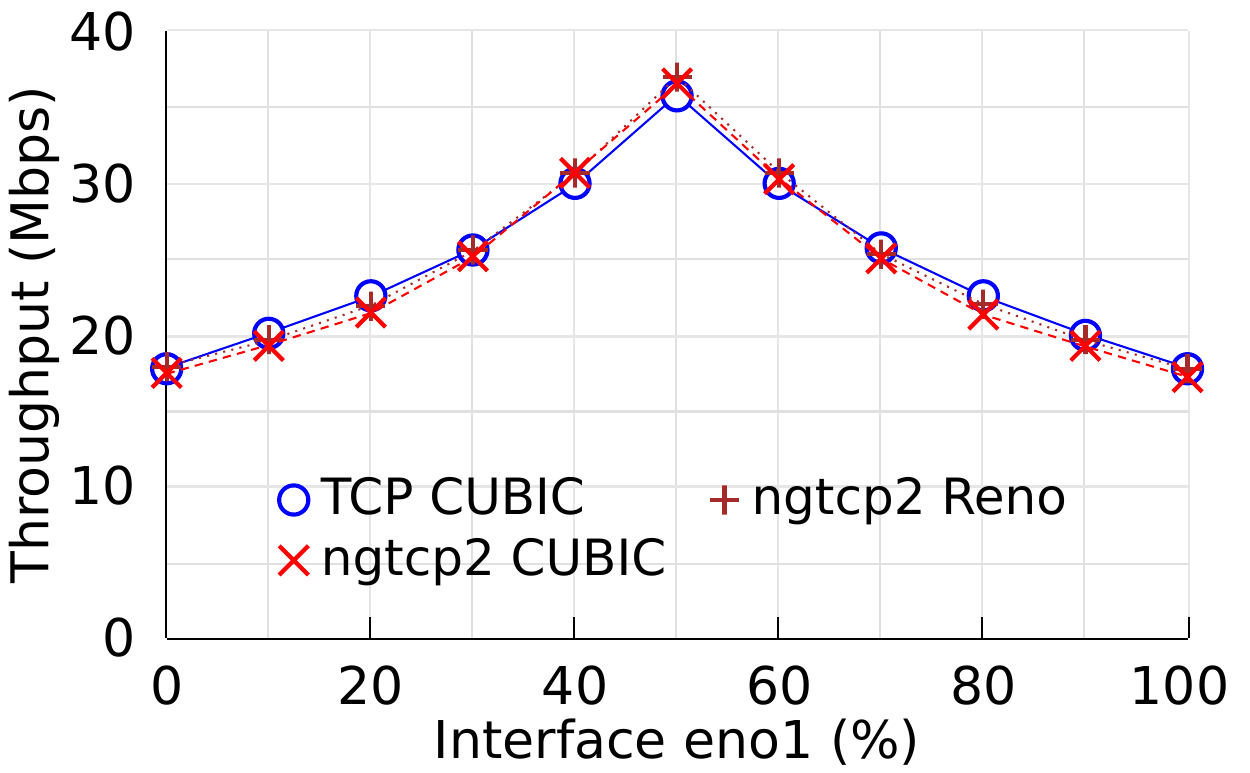}}
    \hspace*{\fill}
    \subfigure[High delay ratio\label{fig:J-2}]{\includegraphics[trim = 0mm 4mm 0mm 0mm, width=\onethirdwidth]{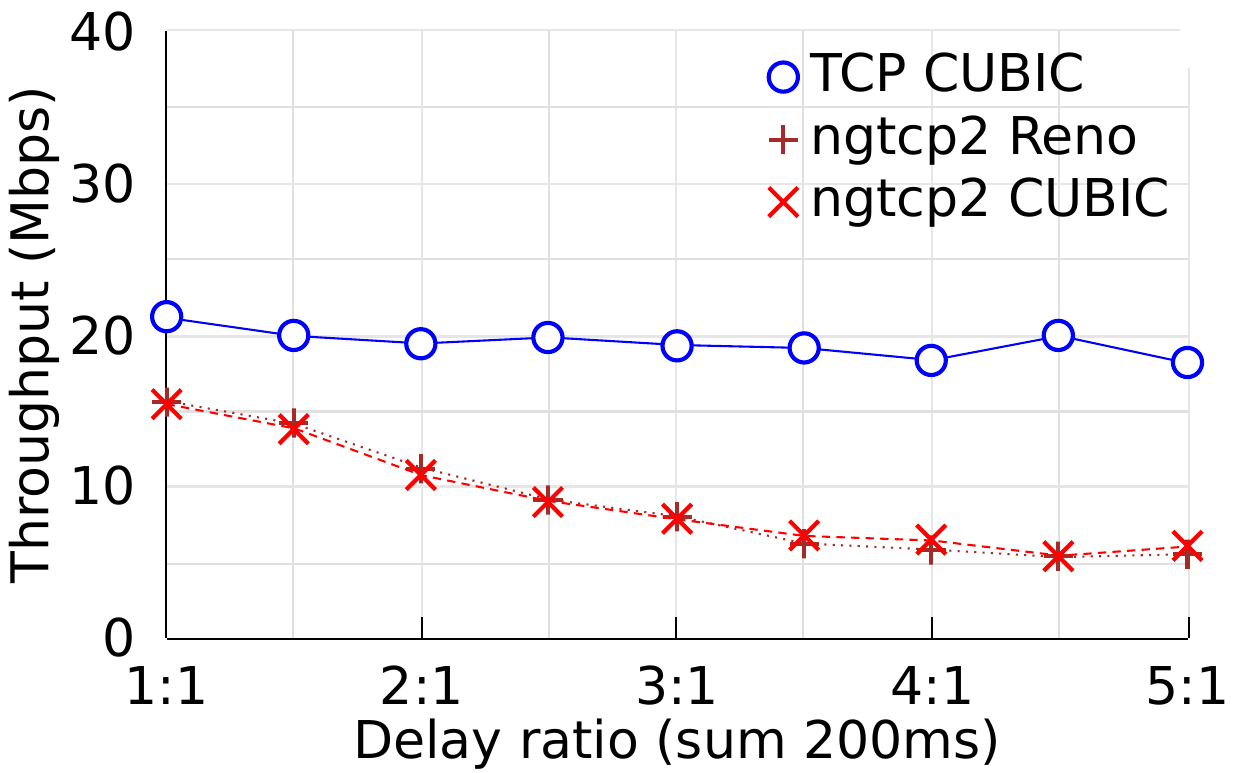}}
    \hspace*{\fill}
    \subfigure[Random loss\label{fig:J-3}]{\includegraphics[trim = 0mm 4mm 0mm 0mm, width=\onethirdwidth]{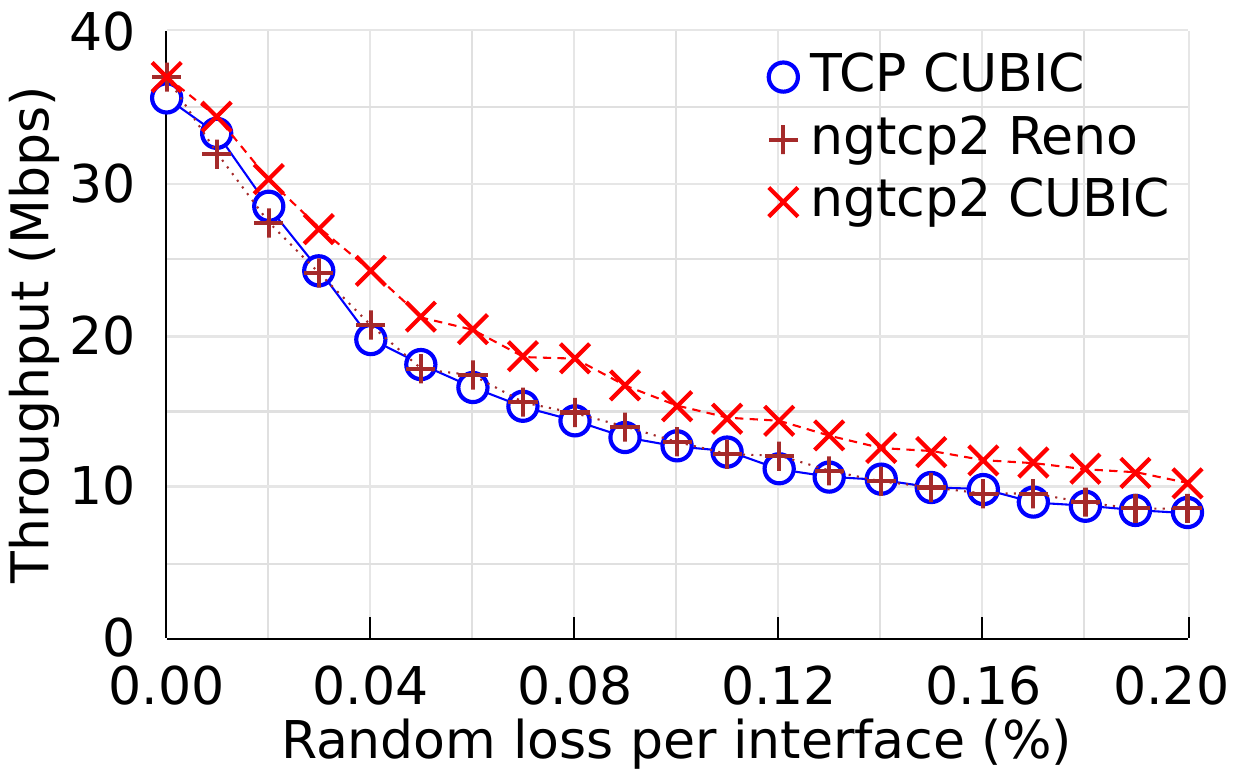}}
    \vspace{-8pt}
    \caption{Throughput when using ngtcp2 (as QUIC implementation) with NewReno and CUBIC 
    \revtwo{}{(as congestion control algorithms)}
    for different parameters}
    \label{fig:J}
\vspace{-5pt}
\end{figure*}

\begin{figure*}[!t]
    \subfigure[DC batch split\label{fig:K-1}]{\includegraphics[trim = 0mm 4mm 0mm 0mm, width=\onethirdwidth]{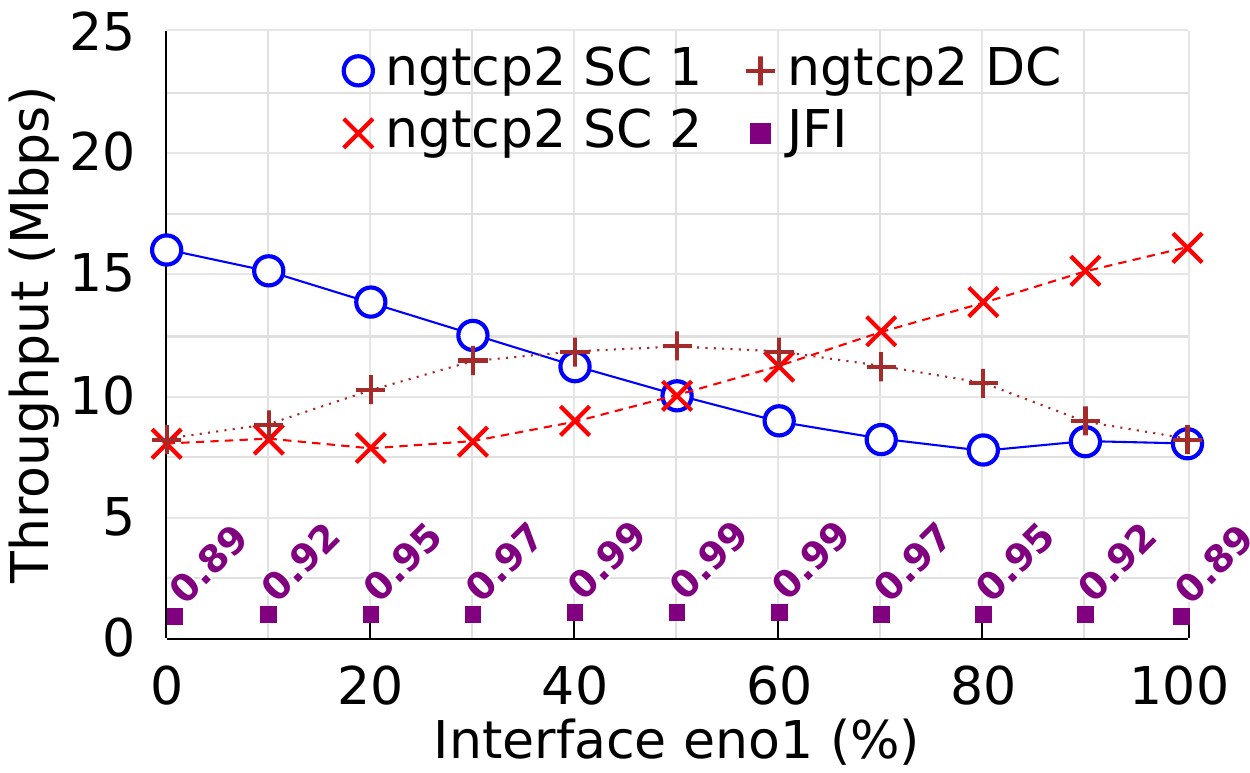}}
    \hspace*{\fill}
    \subfigure[High delay ratio\label{fig:K-2}]{\includegraphics[trim = 0mm 4mm 0mm 0mm, width=\onethirdwidth]{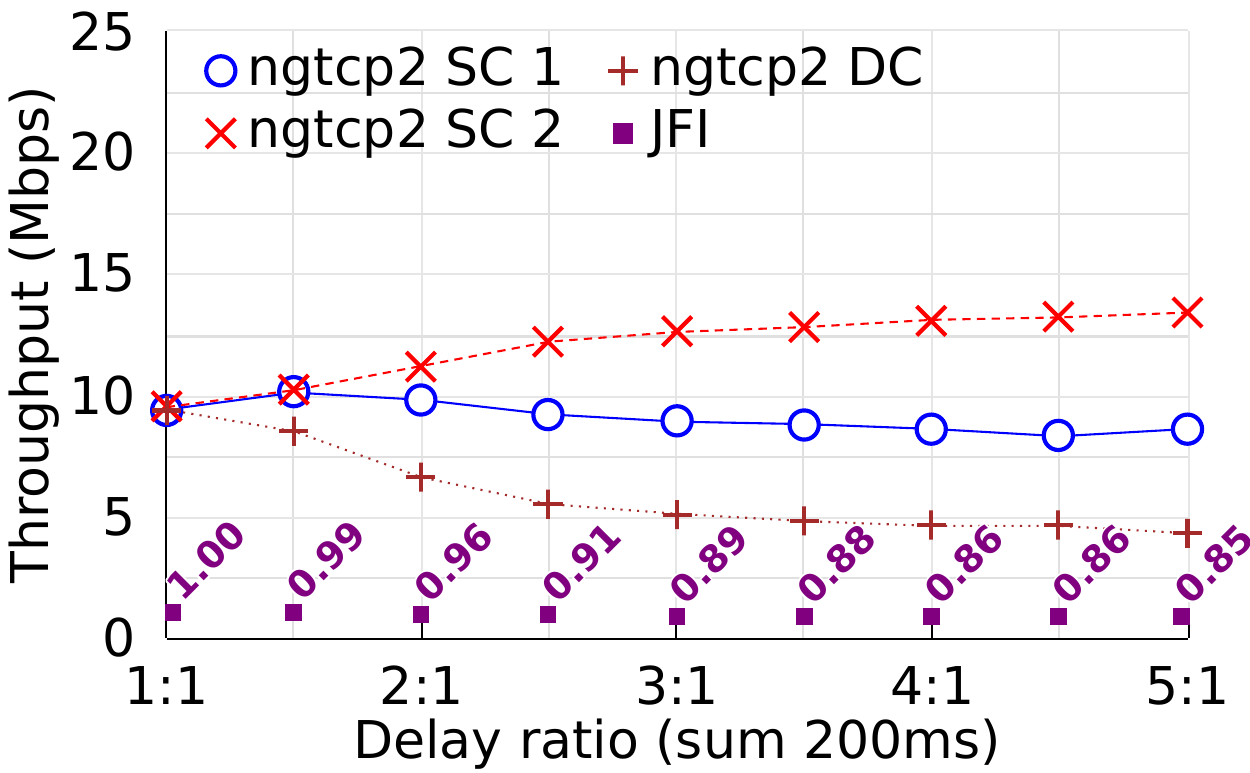}}
    \hspace*{\fill}
    \subfigure[Random loss\label{fig:K-3}]{\includegraphics[trim = 0mm 4mm 0mm 0mm, width=\onethirdwidth]{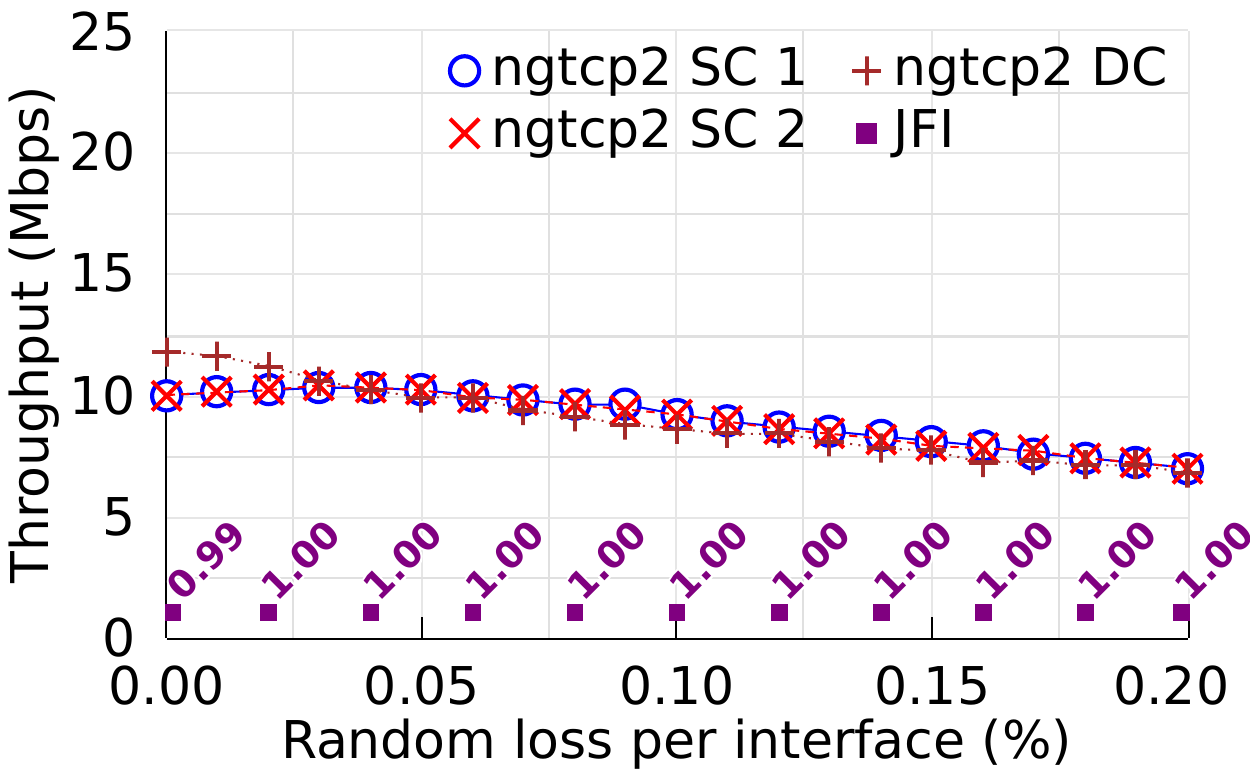}}
    \vspace{-8pt}
    \caption{Fairness when using ngtcp2 
    \revtwo{}{(as QUIC implementation)}
    with NewReno 
    \revtwo{}{(as congestion control algorithm)}
    for different 
    \revone{parameters (CUBIC results very similar)}{parameters}}
    \label{fig:K}
\vspace{-5pt}
\end{figure*}

\begin{figure*}[!t]
    \subfigure[DC batch split\label{fig:KC-1}]{\includegraphics[trim = 0mm 4mm 0mm 0mm, width=\onethirdwidth]{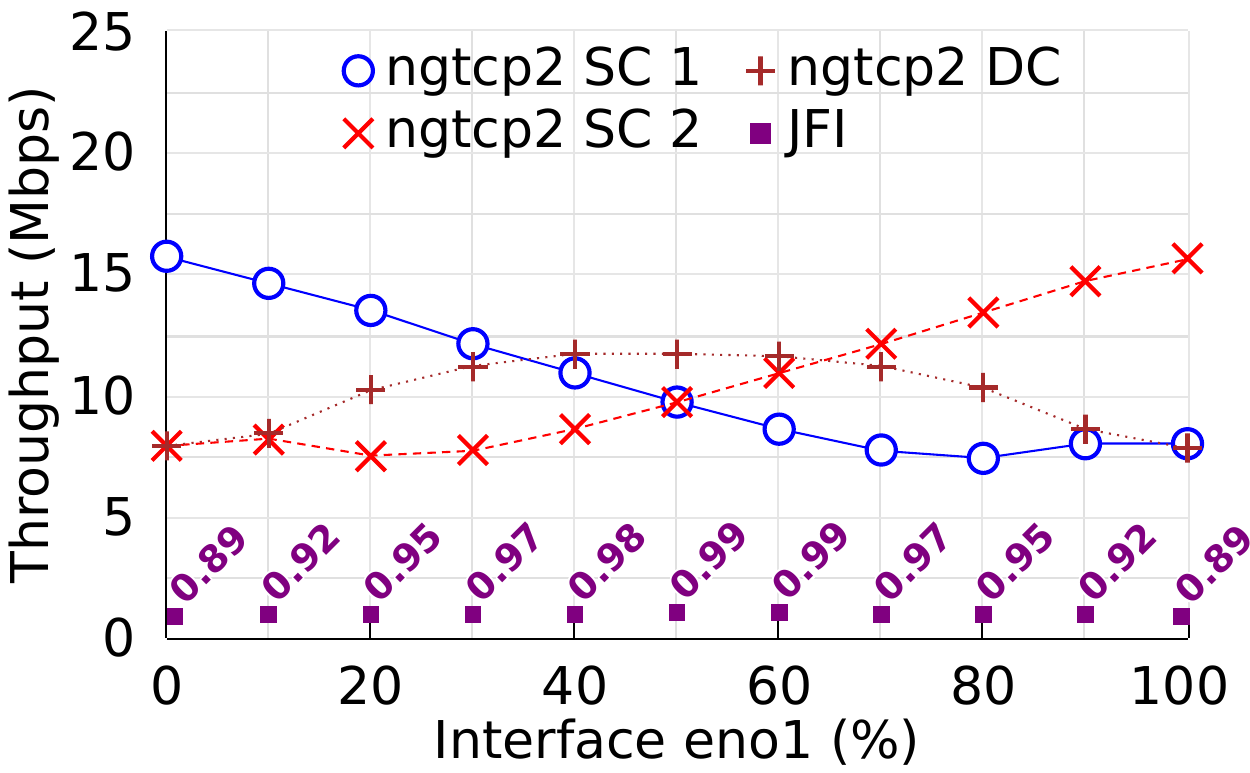}}
    \hspace*{\fill}
    \subfigure[High delay ratio\label{fig:KC-2}]{\includegraphics[trim = 0mm 4mm 0mm 0mm, width=\onethirdwidth]{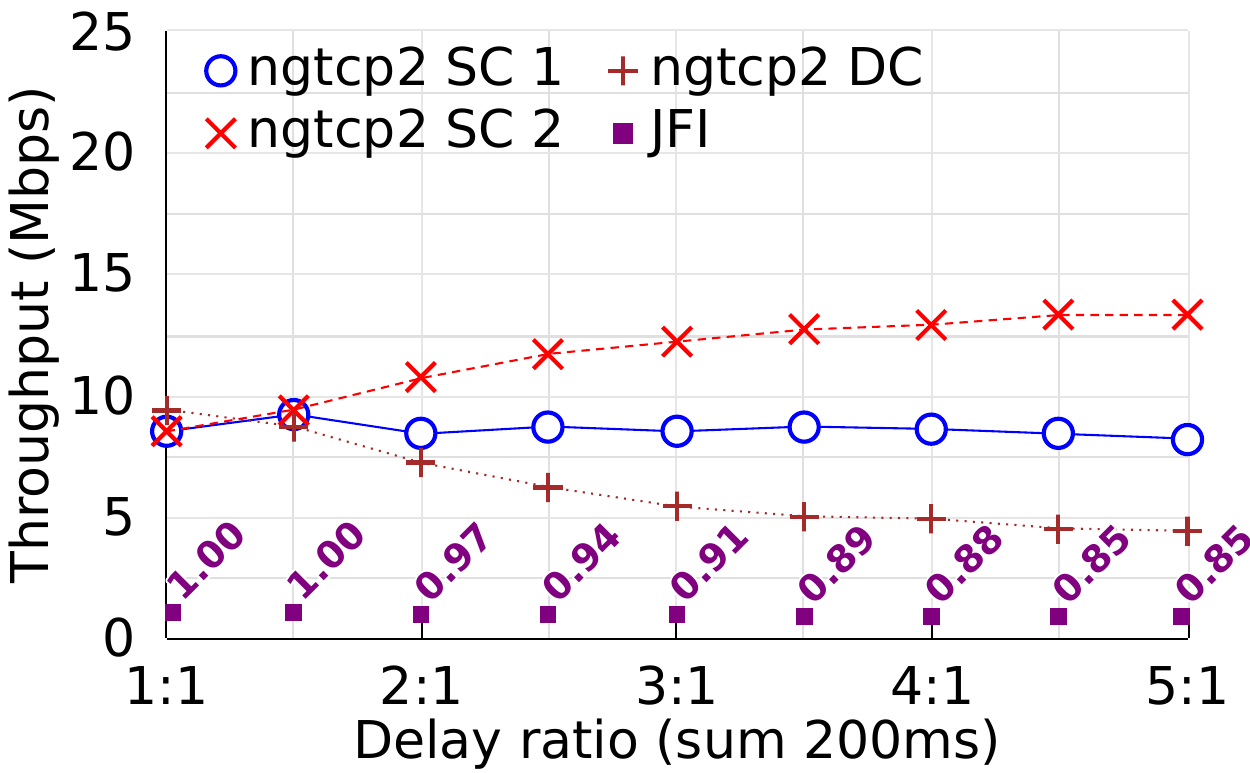}}
    \hspace*{\fill}
    \subfigure[Random loss\label{fig:KC-3}]{\includegraphics[trim = 0mm 4mm 0mm 0mm, width=\onethirdwidth]{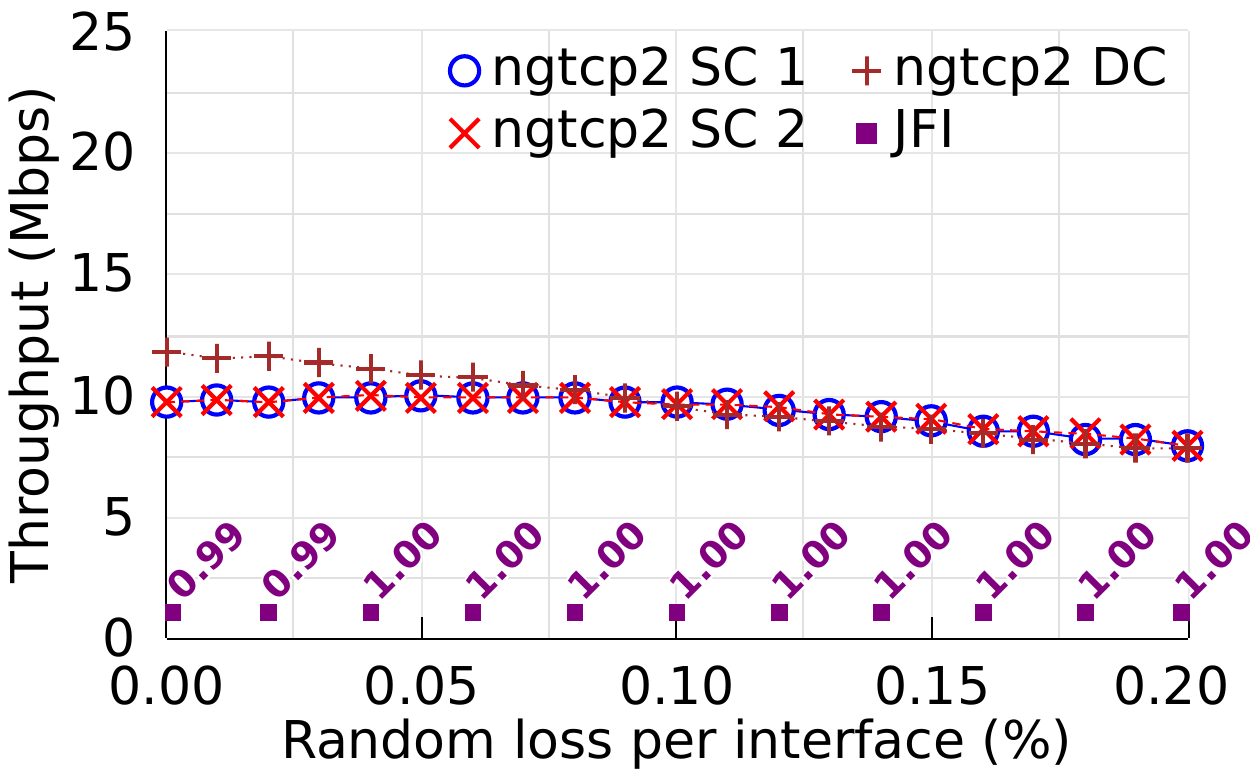}}
    \vspace{-8pt}
    \caption{\revone{}{Fairness when using ngtcp2
    \revtwo{}{(as QUIC implementation)}
    with CUBIC 
    \revtwo{}{(as congestion control algorithm)}
    for different parameters}}
    \label{fig:K-CUBIC}
\vspace{-4pt}
\end{figure*}

When studying the DC batch split using ngtcp2 and different congestion control algorithms (Figure~\ref{fig:J-1}), minimal difference in the overall throughput is observed. 
Compared to aioquic in Figure~\ref{fig:D-1}, little differences are observed at more uneven ratios. Ngtcp2 achieves a higher throughput than aioquic and TCP at more balanced ratios. 
Comparing to
the corresponding fairness results in 
\revone{Figure~\ref{fig:K-1}}{Figures~\ref{fig:K-1} and~\ref{fig:KC-1}} 
to~\ref{fig:E-1}, larger differences 
is 
seen between the QUIC implementations. 
While the two implementations exhibit similar behavior at the most uneven split, the DC connection using ngtcp2 grows more aggressively than the aioquic counterpart when the ratio becomes more balanced. This growth \revone{lead}{leads} to optimal fairness for aioquic, but results in a slightly unfair bandwidth allocation for ngtcp2.

A significant throughput difference between the QUIC implementations can be seen when comparing the high delay ratio experiments in Figures~\ref{fig:J-2} and~\ref{fig:F-2}. Ngtcp2 achieves a significantly lower throughput than aioquic throughout the experiment. This is most likely due to differences in the pacer implementations.
Differences can also be seen when comparing corresponding fairness tests
\revone{(Figures~\ref{fig:K-2} and~\ref{fig:G-2}).}{(Figures~\ref{fig:K-2} and~\ref{fig:KC-2} vs. Figure~\ref{fig:G-2}).} 
Here, the DC connection using ngtcp2 achieves more fair throughput than the aioquic counterpart at balanced ratios and sees a slower drop 
when the ratio gets skewed. However, after the 2:1 ratio point, the DC connections' throughputs become the same for the two implementations. The ngtcp2 SC connection with a higher delay has a much worse performance than the aioquic counterpart at more skewed ratios.

Finally, 
\revtwo{when studying the impact of}{for}
loss rates using ngtcp2 and CUBIC, 
\revtwo{only small differences are observed}{we observe only small differences}
\revtwo{}{compared to our default scenario with aioquic and NewReno}
\revtwo{\revone{(Figures~\ref{fig:J-3} and~\ref{fig:K-3}}{(Figures~\ref{fig:J-3}, \ref{fig:K-3}, and~\ref{fig:KC-3}} compared to Figures~\ref{fig:F-3} and~\ref{fig:G-3}).}{(throughput Figure~\ref{fig:J-3} compared to Figures~\ref{fig:F-3}, and fairness Figures \ref{fig:K-3} and~\ref{fig:KC-3} compared to Figure~\ref{fig:G-3}).}
However, in contrast to the other experiments, ngtcp2 using CUBIC shows a noticeable better performance compared to TCP CUBIC.
Looking closer at the 0.08\% 
\revtwo{loss case,}{loss case in Figure~\ref{fig:J-3},}
we 
\revtwo{have observed}{observe}
that the TCP implementation more often stays in CUBIC's TCP 
\revtwo{mode (used when detecting growth slower than Reno).}{mode (used when detecting growth slower than Reno counterparts, e.g., due to low bandwidth delay products).}
This also explains why TCP CUBIC, TCP NewReno, and ngtcp2 NewReno perform similarly here.

In general, ngtcp2 achieves higher throughput than aioquic, even though both follow the same IETF recommendations. As discussed, differences can occur due to the RFC being 
open for interpretation. The execution speed and resources required by the two implementations also differ. Ngtcp2 is implemented in C and aioquic in Python. With ngtcp2, a larger receive buffer did not impact throughput, as the client buffer was quickly emptied. Ngtcp2 is also noted to be greedier than aioquic over DC, often introducing some unfairness to scenarios that were fair for aioquic.
One potential reason is the difference in pacer implementation, as the IETF only recommends a pacer but does not specify it in detail. The difference in pacer implementation is also clearly shown in high delay ratio tests.


\revone{}{{\bf Additional tests using ngtcp2 with different congestion control algorithms:} To demonstrate the generality of our observations, we next present additional throughput and fairness results when using ngtcp2 with NewReno and CUBIC.}

\begin{figure*}[!t]
    \subfigure[Throughput\label{fig:X-1}]{\includegraphics[trim = 0mm 4mm 0mm 0mm, width=\onethirdwidth]{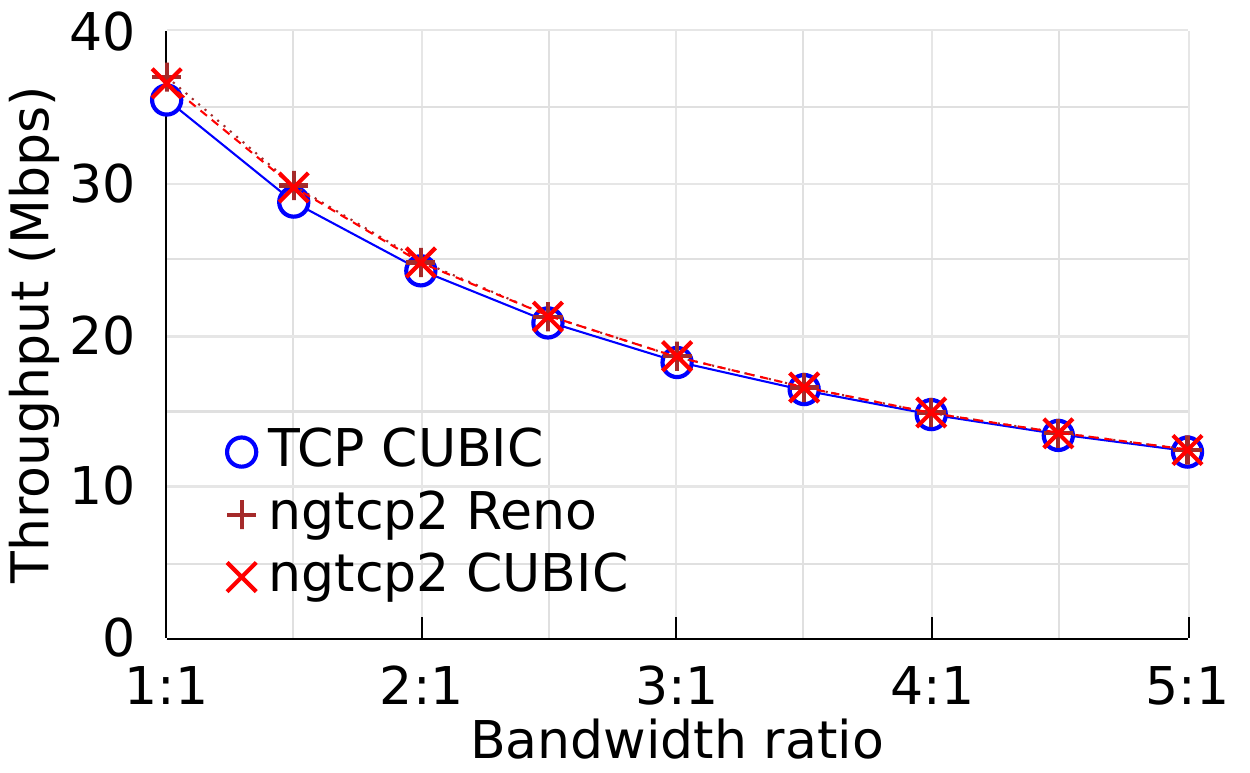}}
    \hspace*{\fill}
    \subfigure[Fairness \revtwo{}{with} NewReno\label{fig:X-2}]{\includegraphics[trim = 0mm 4mm 0mm 0mm, width=\onethirdwidth]{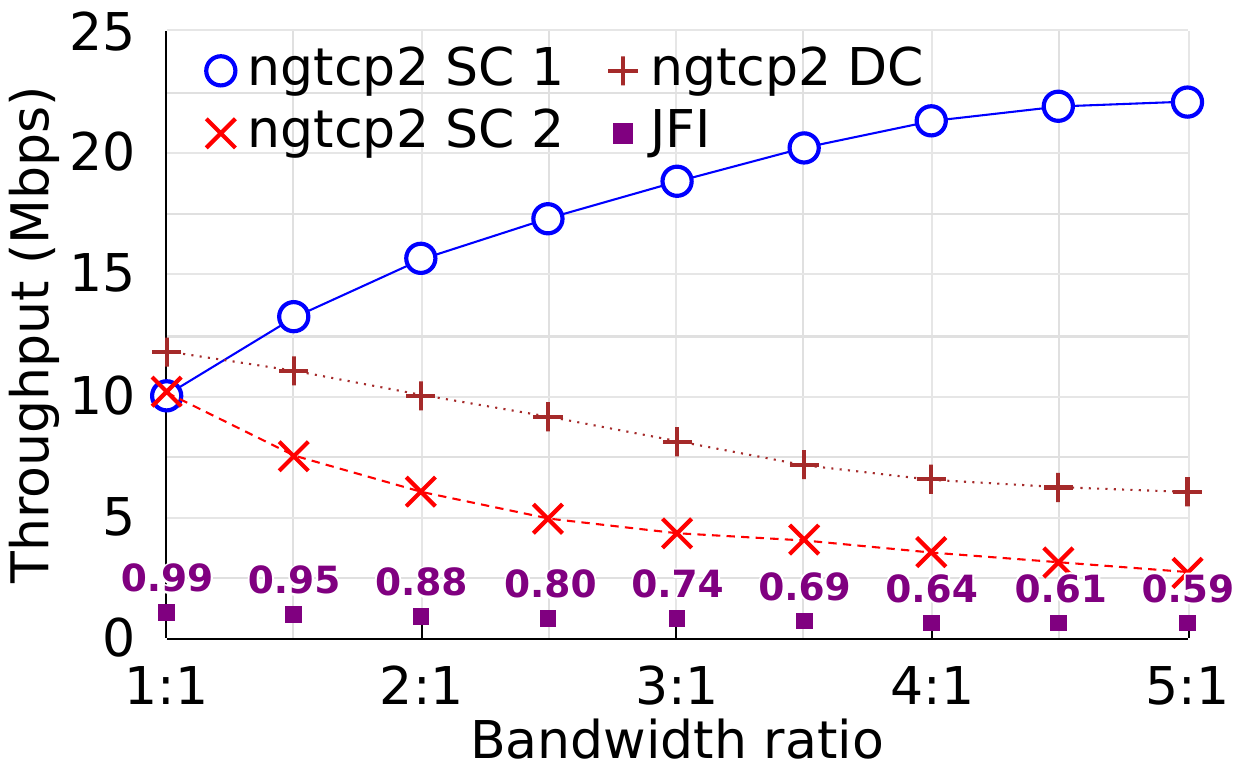}}
    \hspace*{\fill}
    \subfigure[Fairness \revtwo{}{with} CUBIC\label{fig:X-3}]{\includegraphics[trim = 0mm 4mm 0mm 0mm, width=\onethirdwidth]{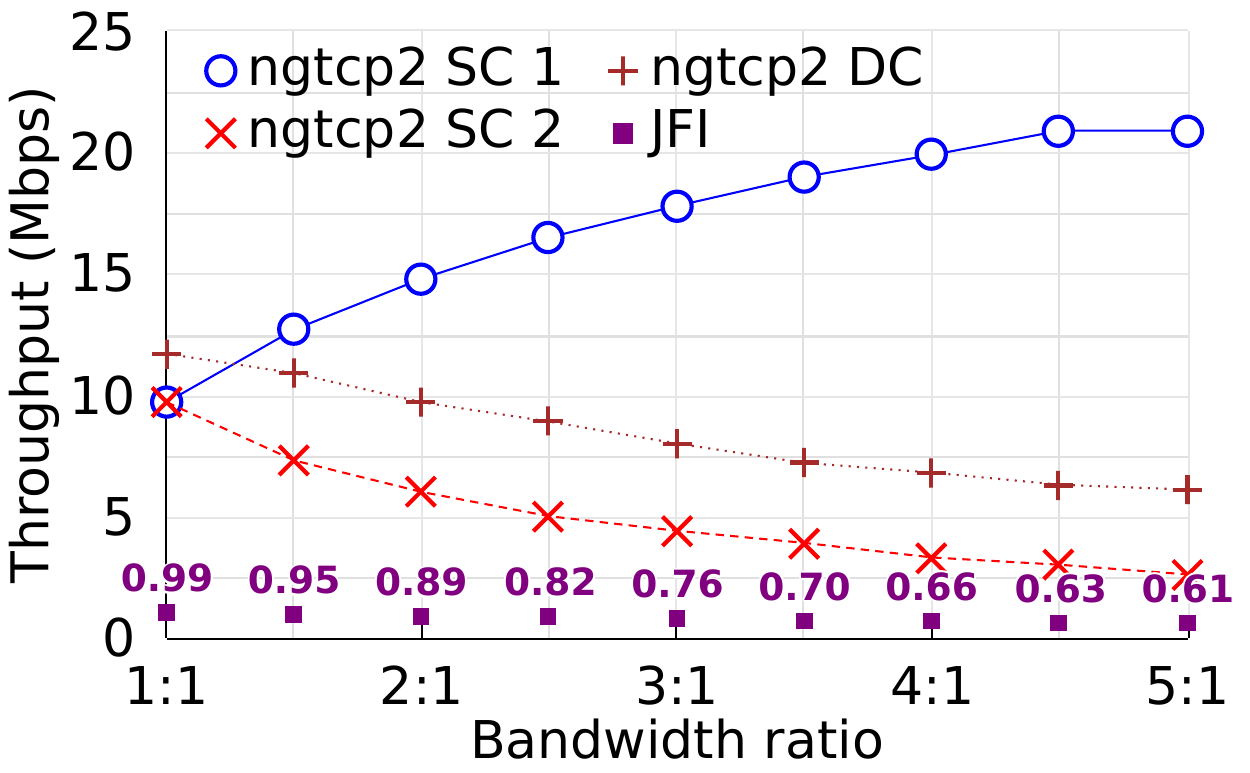}}
    \vspace{-8pt}
    \caption{\revone{}{Impact of the bandwidth ratio when using ngtcp2 as QUIC version with two different congestion control algorithms: NewReno and CUBIC}}
    \label{fig:extra-bw-ratio}
\vspace{-4pt}
\end{figure*}

\begin{figure*}[!t]
     \subfigure[Throughput\label{fig:Y-1}]{\includegraphics[trim = 0mm 4mm 0mm 0mm, width=\onethirdwidth]{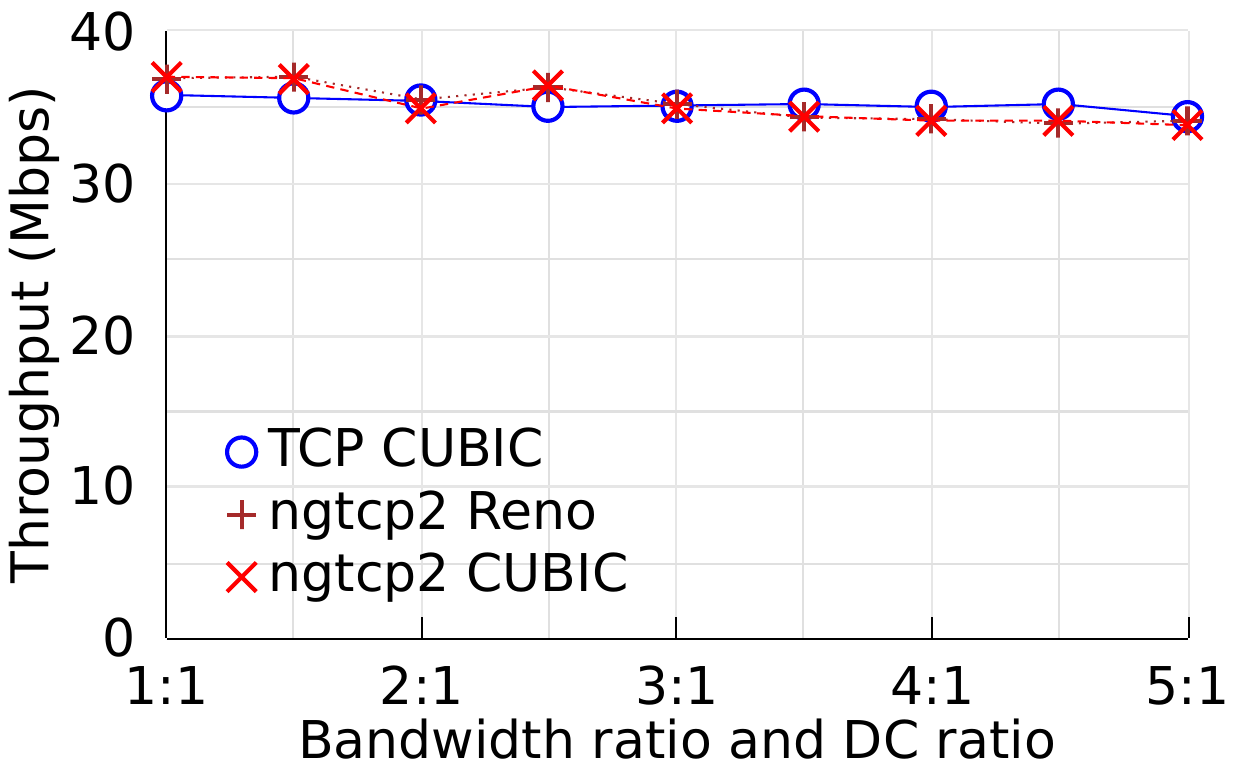}}
     \hspace*{\fill}
     \subfigure[Fairness \revtwo{}{with} NewReno\label{fig:Y-2}]{\includegraphics[trim = 0mm 4mm 0mm 0mm, width=\onethirdwidth]{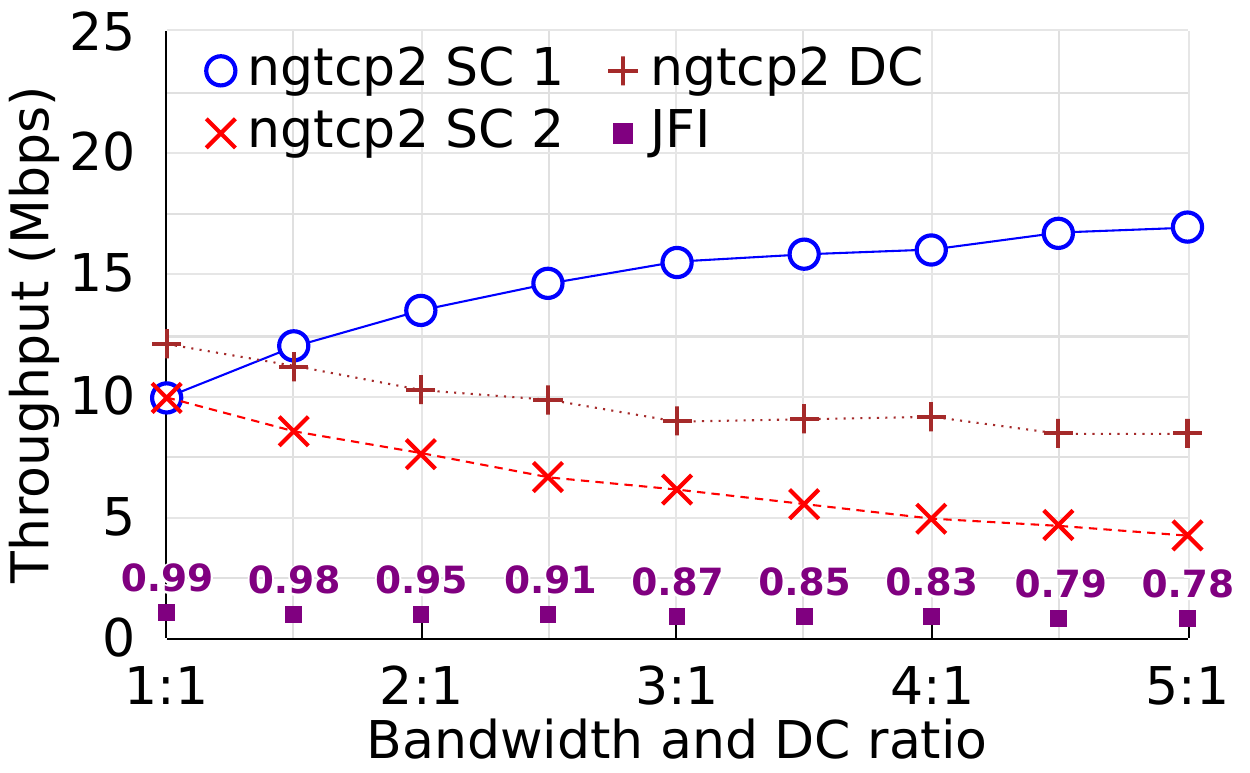}}
     \hspace*{\fill}
     \subfigure[Fairness \revtwo{}{with} CUBIC\label{fig:Y-3}]{\includegraphics[trim = 0mm 4mm 0mm 0mm, width=\onethirdwidth]{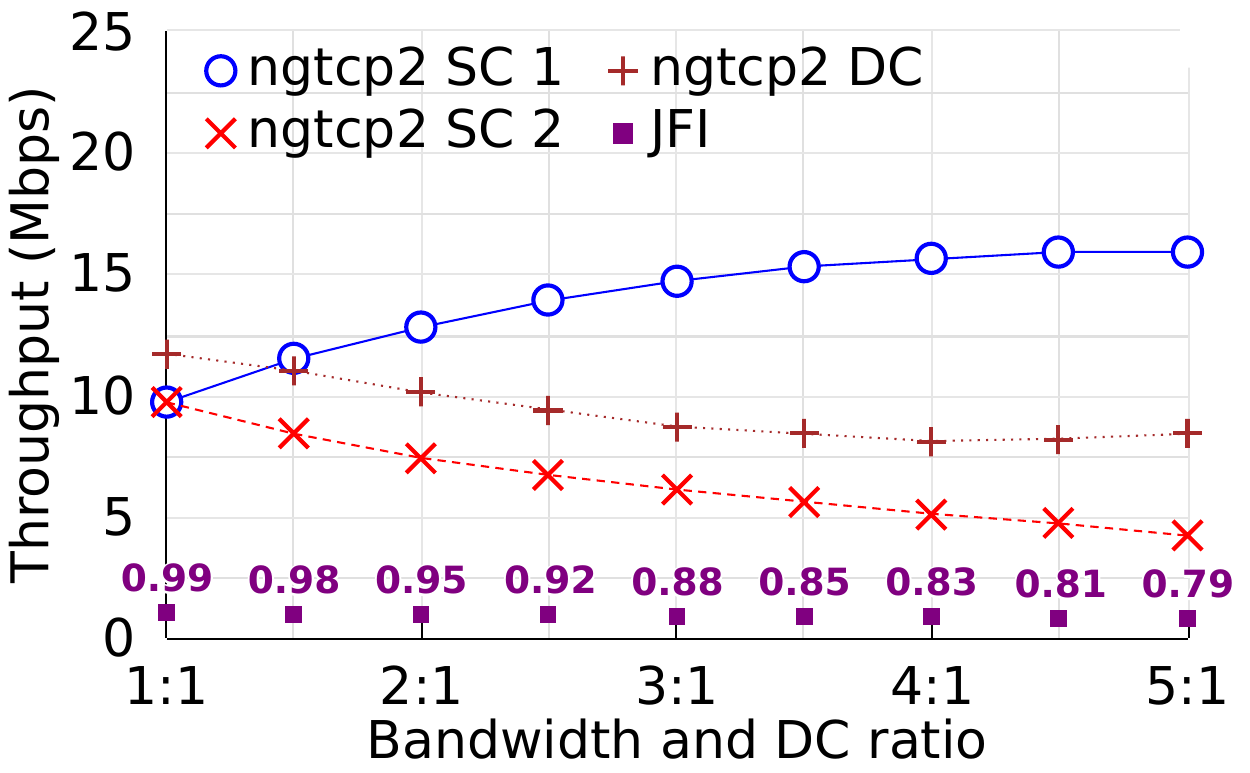}}
    \vspace{-8pt}
    \caption{\revone{}{Impact of the bandwidth ratio when having a matching DC ratio and using ngtcp2 as QUIC version with two different congestion control algorithms: NewReno and CUBIC}}
    \label{fig:extra-bwdc-ratio}
\vspace{-5pt}
\end{figure*}

\begin{figure*}[!t]
    \subfigure[Throughput\label{fig:Z-1}]{\includegraphics[trim = 0mm 4mm 0mm 0mm, width=\onethirdwidth]{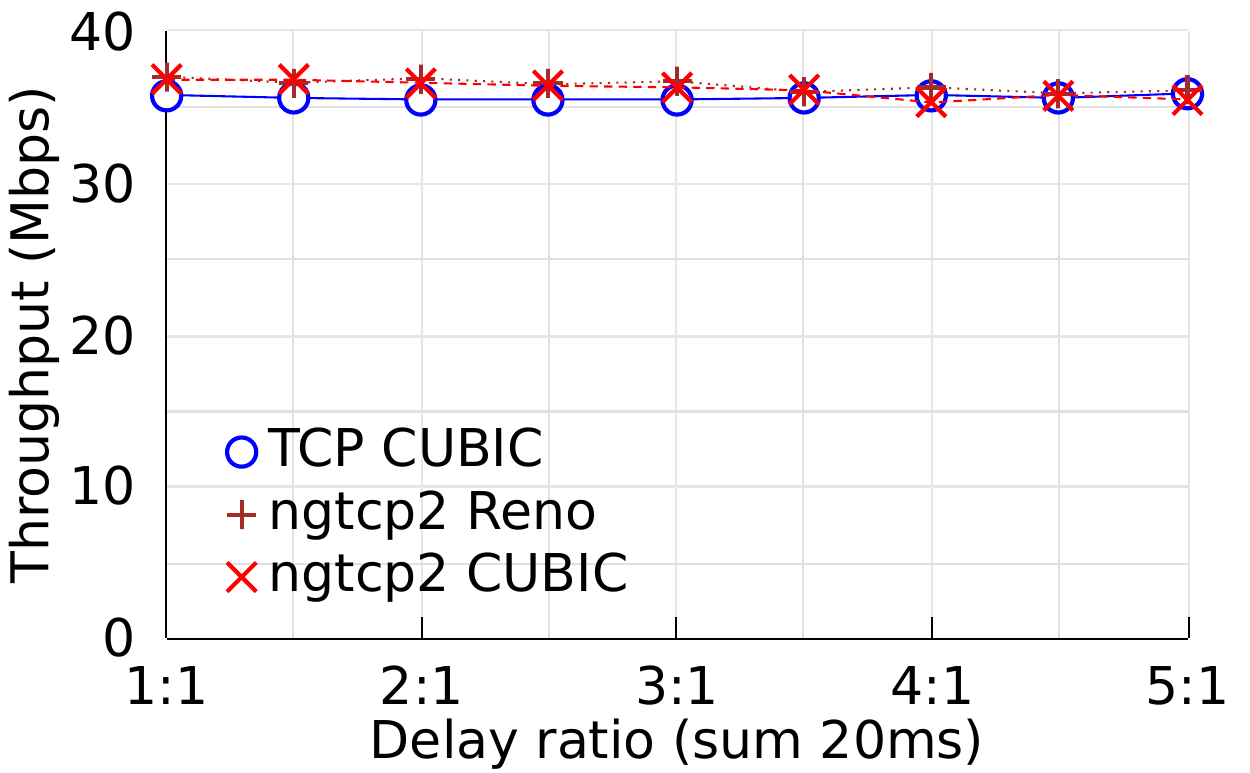}}
    \hspace*{\fill}
    \subfigure[Fairness \revtwo{}{with} NewReno\label{fig:Z-2}]{\includegraphics[trim = 0mm 4mm 0mm 0mm, width=\onethirdwidth]{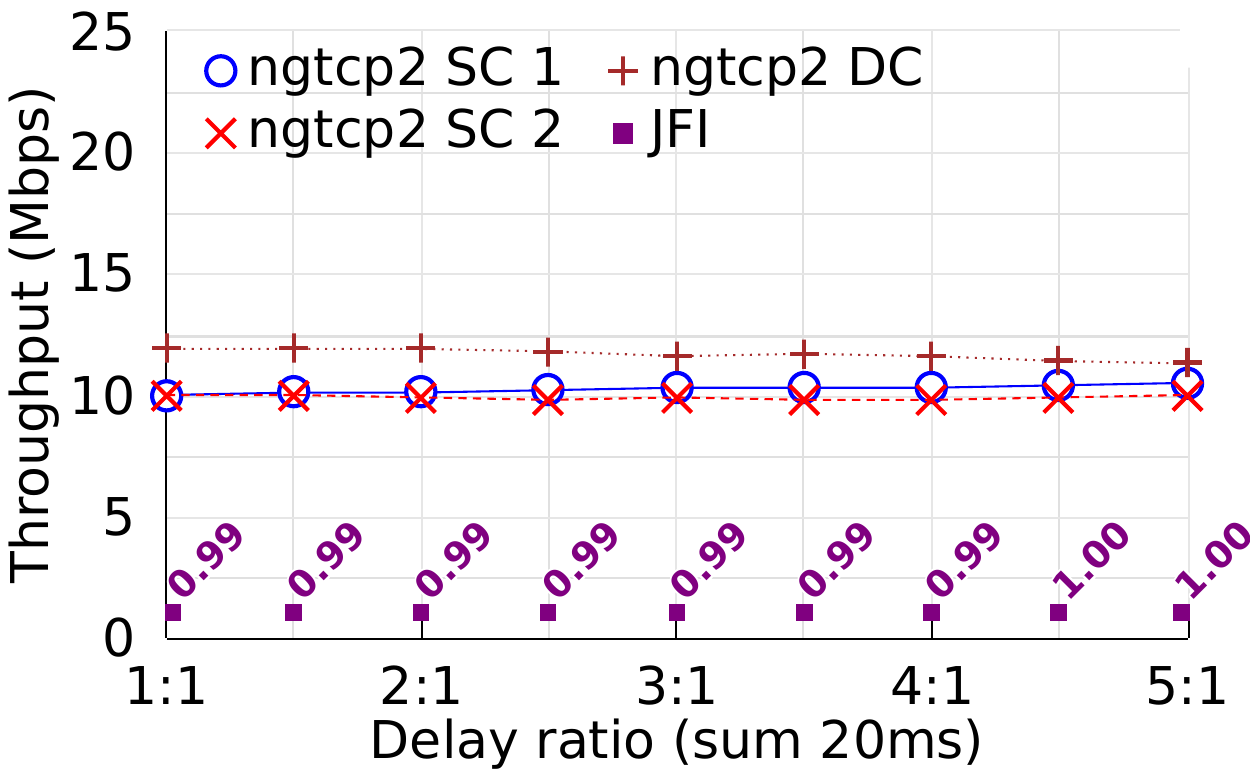}}
    \hspace*{\fill}
    \subfigure[Fairness \revtwo{}{with} CUBIC\label{fig:Z-3}]{\includegraphics[trim = 0mm 4mm 0mm 0mm, width=\onethirdwidth]{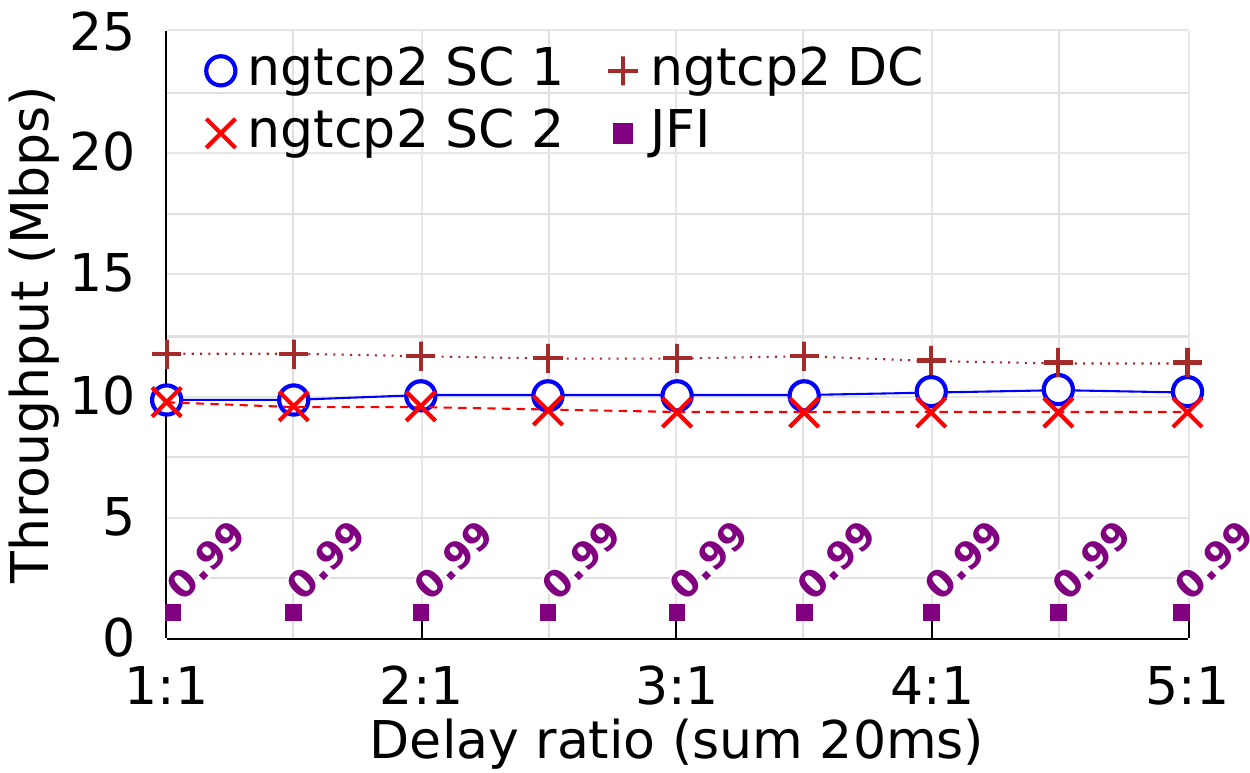}}
    \vspace{-8pt}
    \caption{\revone{}{Impact of the delay ratio in a low-delay scenario, when using ngtcp2 as QUIC version with two different congestion control algorithms: NewReno and CUBIC}}
    \label{fig:extra-delay-ratio}
\vspace{-5pt}
\end{figure*}

\begin{figure*}[!t]
    \subfigure[Throughput\label{fig:Q-1}]{\includegraphics[trim = 0mm 4mm 0mm 0mm, width=\onethirdwidth]{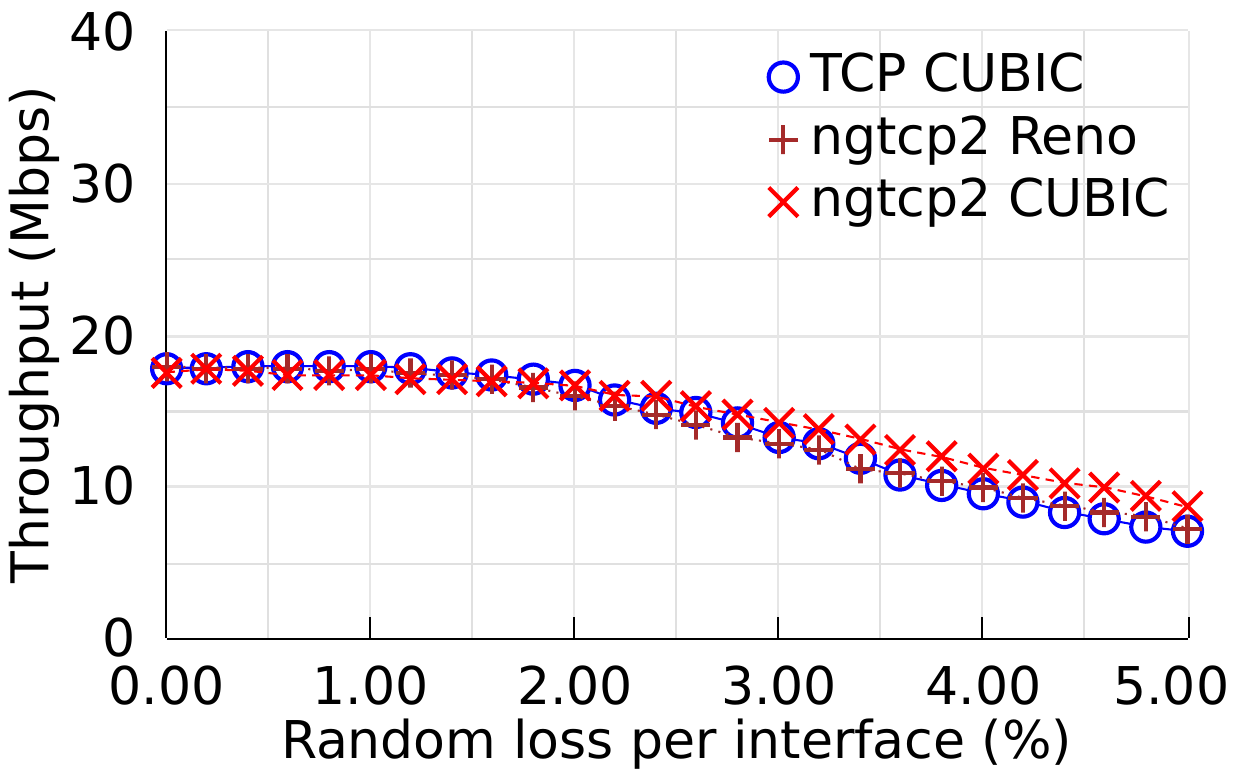}}
    \hspace*{\fill}
    \subfigure[Fairness \revtwo{}{with} NewReno\label{fig:Q-2}]{\includegraphics[trim = 0mm 4mm 0mm 0mm, width=\onethirdwidth]{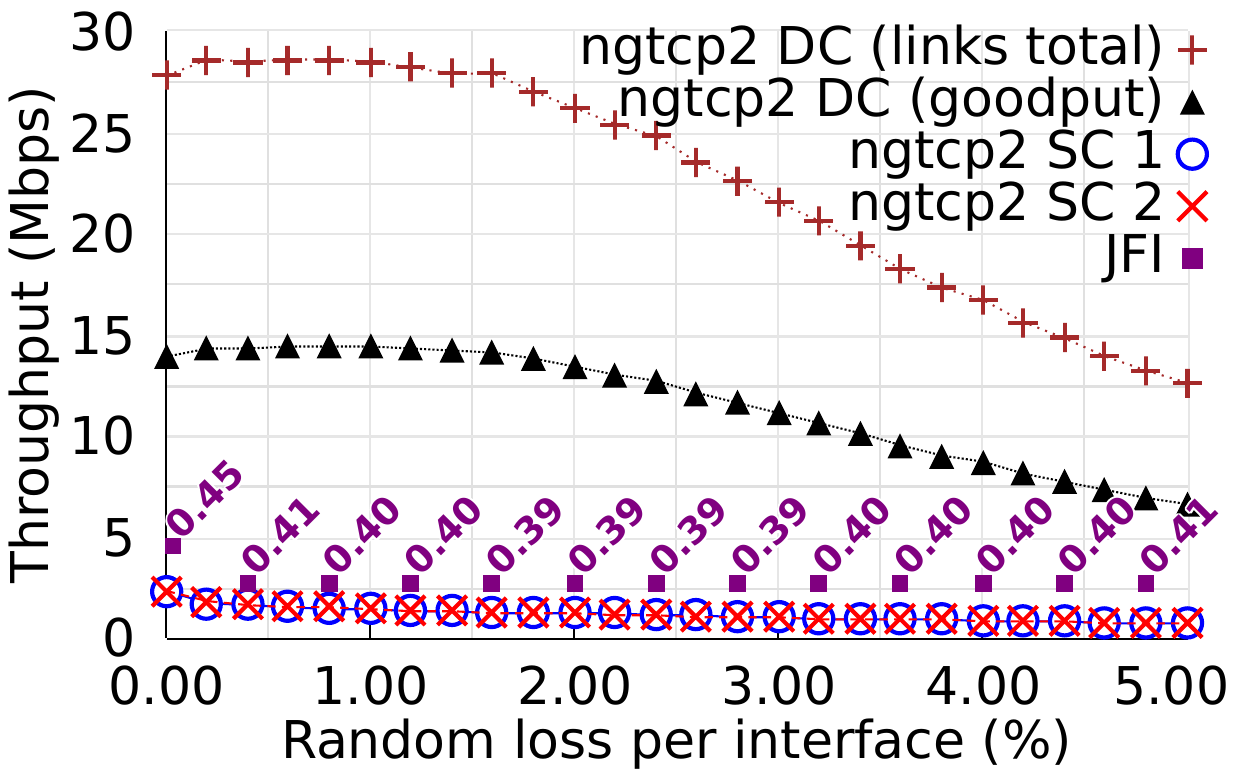}}
    \hspace*{\fill}
    \subfigure[Fairness \revtwo{}{with} CUBIC\label{fig:Q-3}]{\includegraphics[trim = 0mm 4mm 0mm 0mm, width=\onethirdwidth]{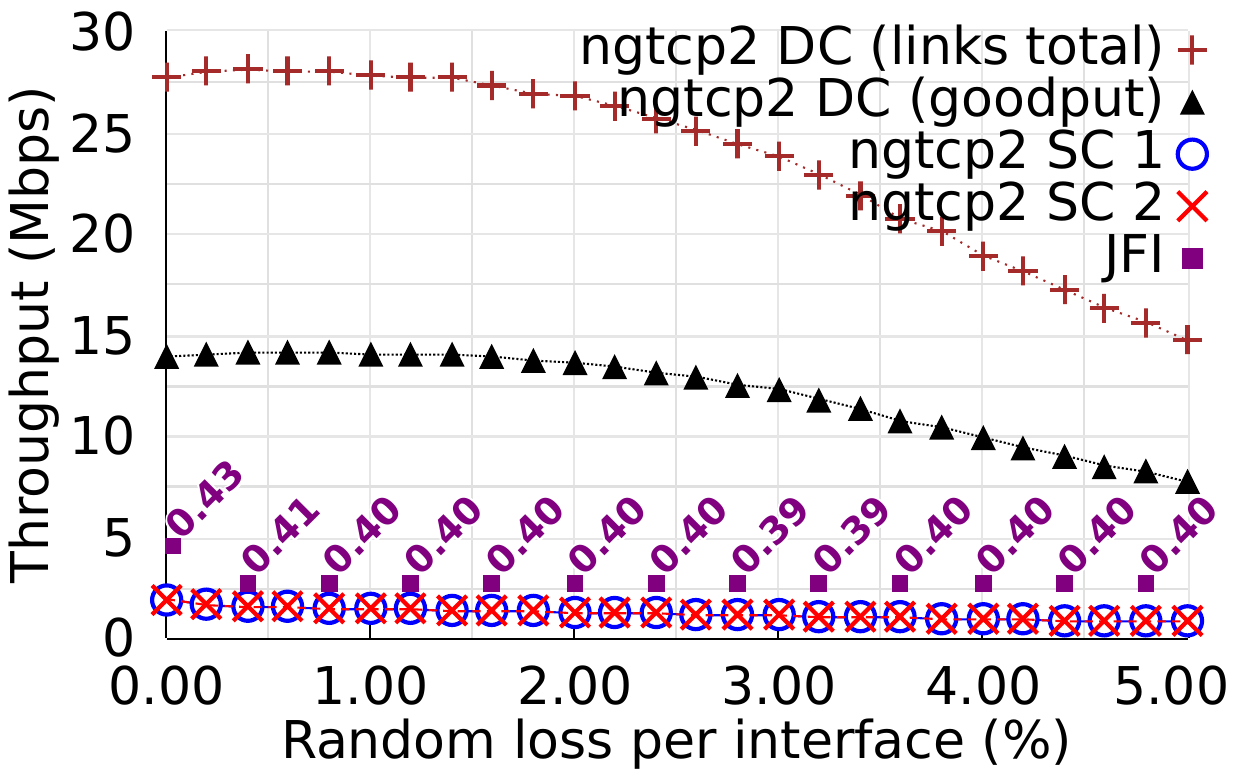}}
    \vspace{-8pt}
    \caption{\revone{}{DC throughput and fairness when using duplicate packets and ngtcp2 as QUIC version with two different congestion control algorithms: NewReno and CUBIC}}
    \label{fig:extra-duplicate}
\vspace{-6pt}
\end{figure*}

\revone{}{Figure~\ref{fig:extra-bw-ratio} shows the 
\revtwo{}{throughput and fairness}
impact of the bandwidth ratio when keeping the DC ratio fixed. (The corresponding sub-figures for aioquic with NewReno are Figures~\ref{fig:D-2} and~\ref{fig:E-2}. Aioquic does not support CUBIC.) Figure~\ref{fig:extra-bwdc-ratio} shows the impact of the bandwidth ratio when using a matching DC ratio. (The corresponding sub-figures for aioquic with NewReno are Figures~\ref{fig:D-3} and ~\ref{fig:E-3}.) Figure~\ref{fig:extra-delay-ratio} shows the impact of the delay ratio for our low-latency scenarios. (The corresponding sub-figures for aioquic with NewReno are Figures~\ref{fig:F-1} and~\ref{fig:G-1}.) Comparing with the corresponding results using aioquic, in all three cases, the observation from Section~\ref{ssec:results-cc-versions} holds true, including that ngtcp2 is more aggressive and achieves higher throughput than aioquic. The results are consistent regardless of whether NewReno or CUBIC is used as congestion control algorithm.}

\begin{figure*}[t!]
    \subfigure[DC batch size, aioquic (repeated Fig.~\ref{fig:A-1})\label{fig:L-1}]{\includegraphics[trim = 0mm 4mm 0mm 0mm, width=\onethirdwidth]{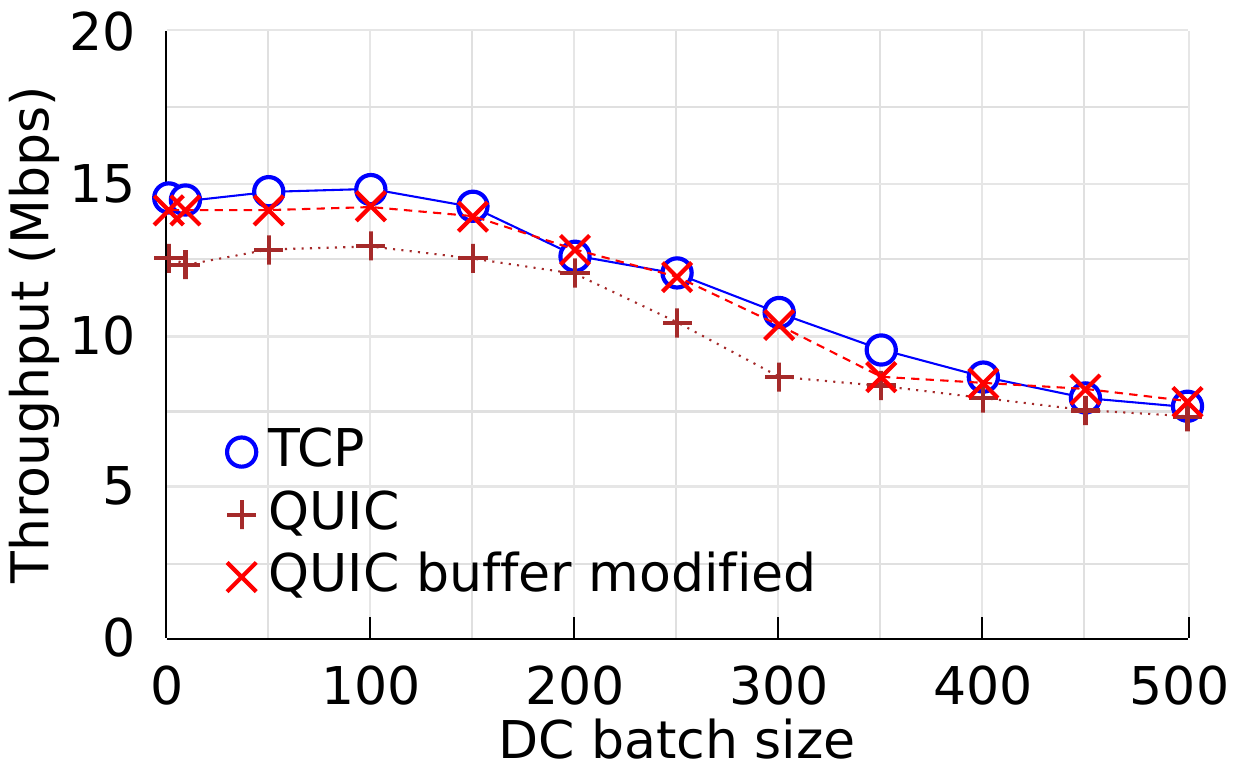}}
    \hspace*{\fill}
    \subfigure[High delay ratio, aioquic (repeated Fig.~\ref{fig:F-2})\label{fig:L-2}]{\includegraphics[trim = 0mm 4mm 0mm 0mm, width=\onethirdwidth]{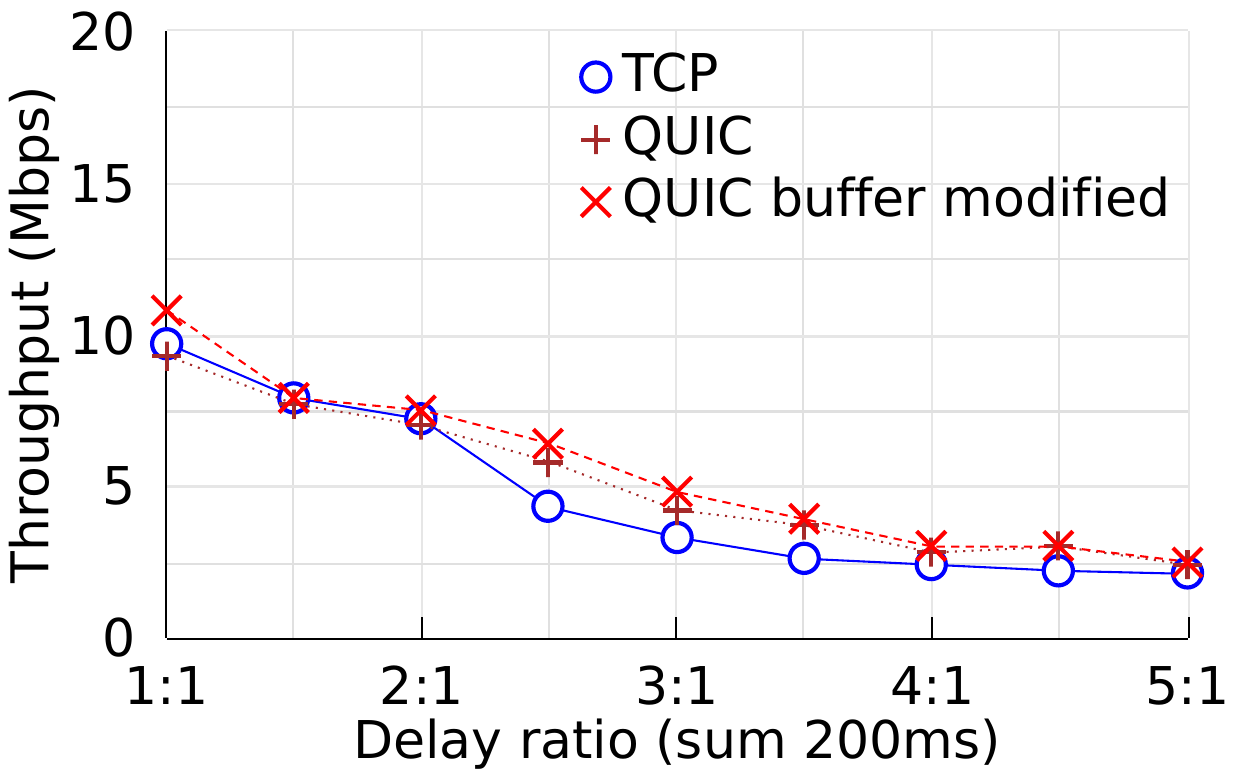}}
    \hspace*{\fill}
    \subfigure[Random loss, aioquic (repeated Fig.~\ref{fig:F-3})\label{fig:L-3}]{\includegraphics[trim = 0mm 4mm 0mm 0mm, width=\onethirdwidth]{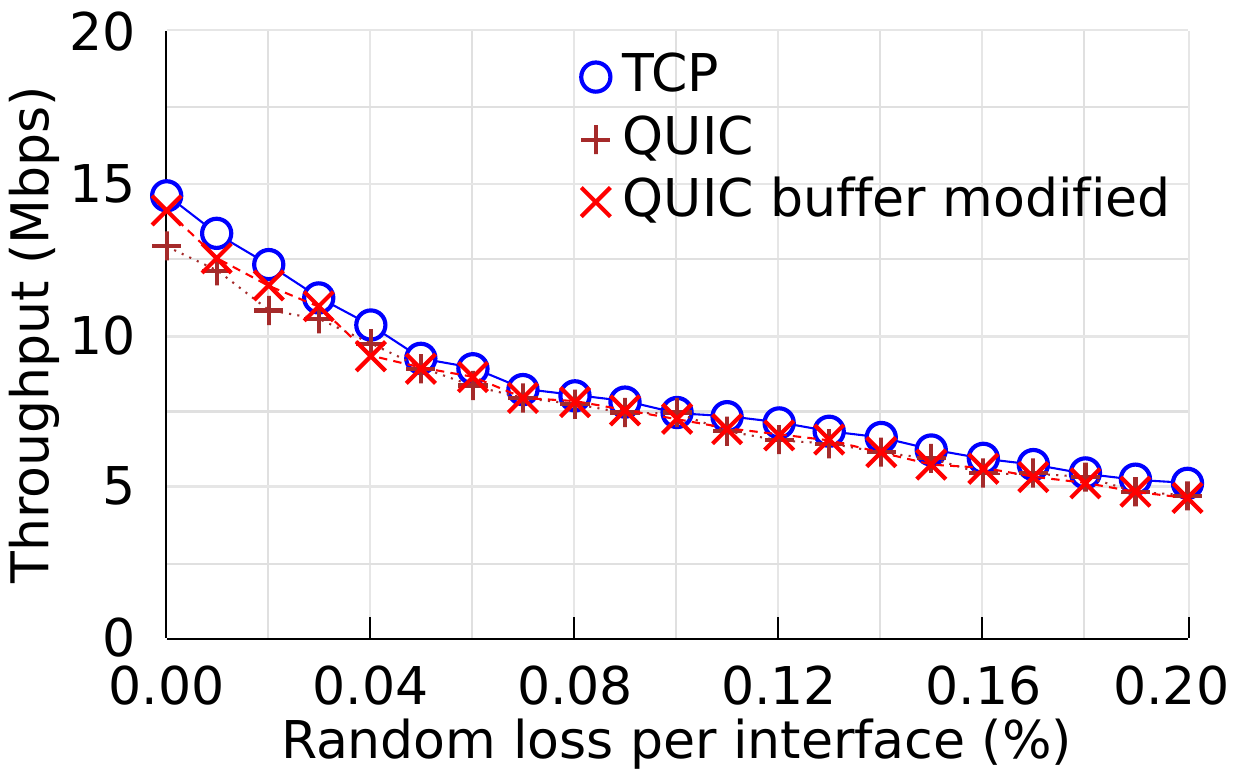}}

    \vspace{-2pt}
    \subfigure[DC batch size, ngtcp2 
    (repeated Fig.~\ref{fig:I-1})\label{fig:L-4}]{\includegraphics[trim = 0mm 4mm 0mm 0mm, width=\onethirdwidth]{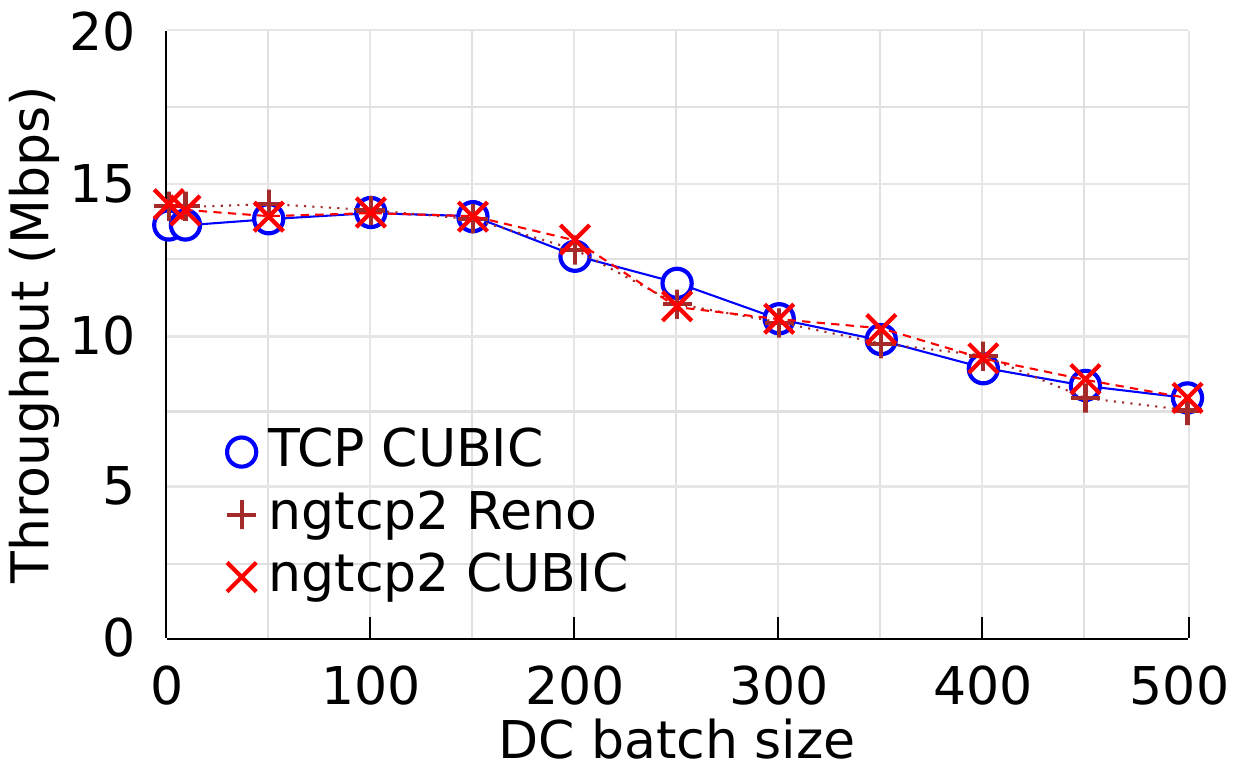}}
    \hspace*{\fill}
    \subfigure[High delay ratio, ngtcp2 (rep. Fig.~\ref{fig:J-2})\label{fig:L-5}]{\includegraphics[trim = 0mm 4mm 0mm 0mm, width=\onethirdwidth]{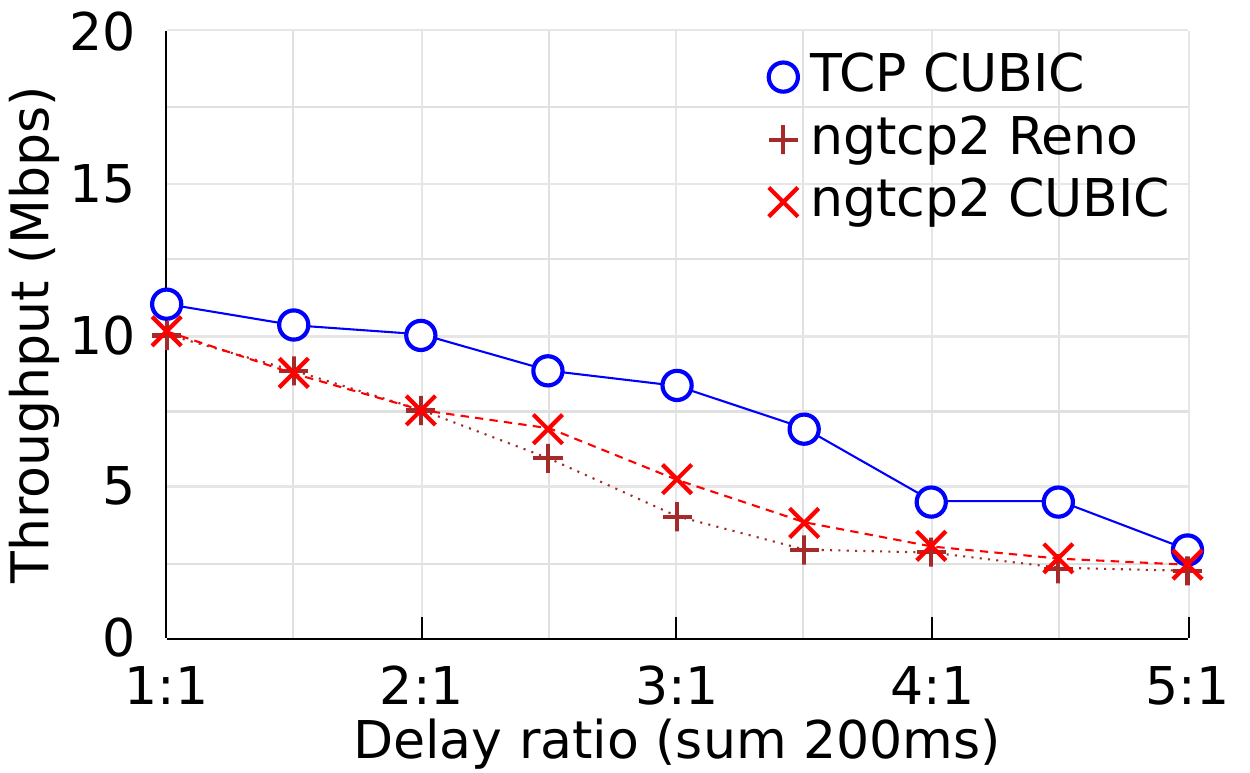}}
    \hspace*{\fill}
    \subfigure[Random loss, ngtcp2 (repeated Fig.~\ref{fig:J-3})\label{fig:L-6}]{\includegraphics[trim = 0mm 4mm 0mm 0mm, width=\onethirdwidth]{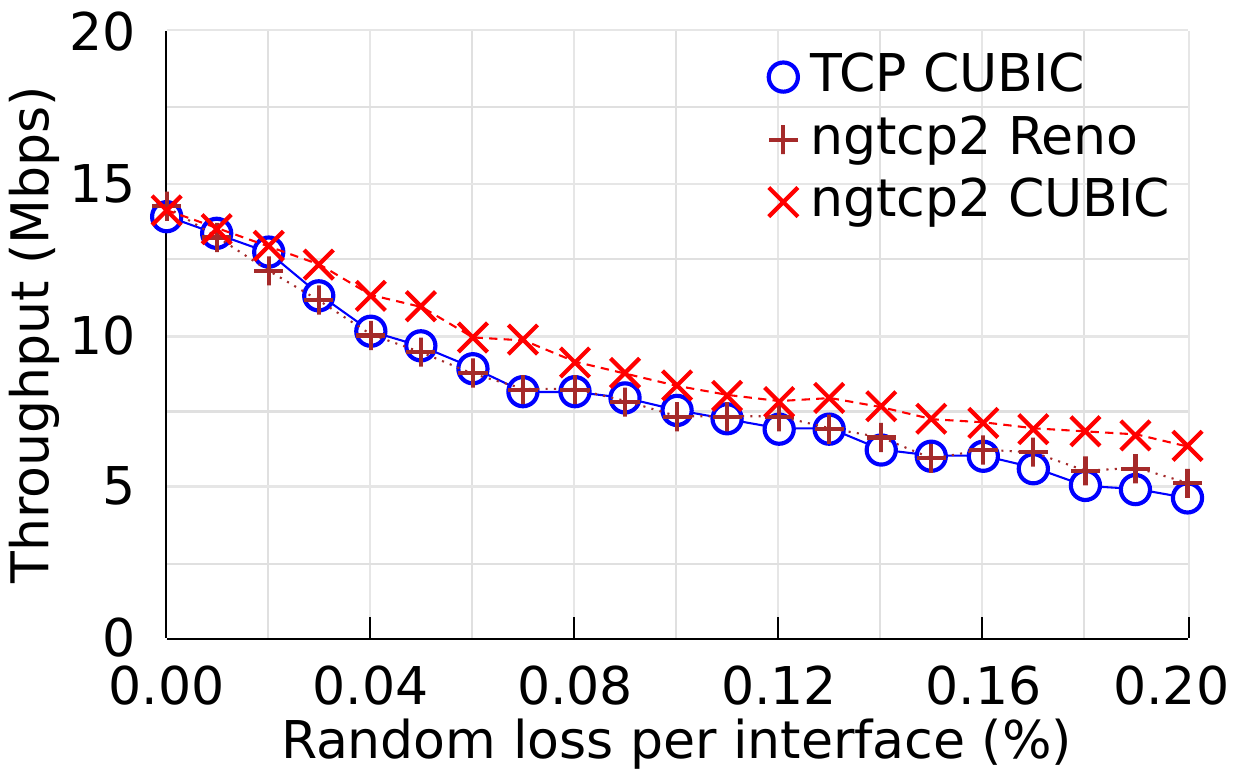}}
    \vspace{-8pt}
    \caption{Trace-driven bandwidth variability tests using aioquic and ngtcp2 as QUIC implementations}
    \label{fig:throughput-trace}
\vspace{-4pt}
\end{figure*}

\begin{figure*}[t!]
    \subfigure[DC \revtwo{ratio}{batch split} (repeated Fig.~\ref{fig:D-1})\label{fig:U-1}]{\includegraphics[trim = 0mm 4mm 0mm 0mm, width=\onethirdwidth]{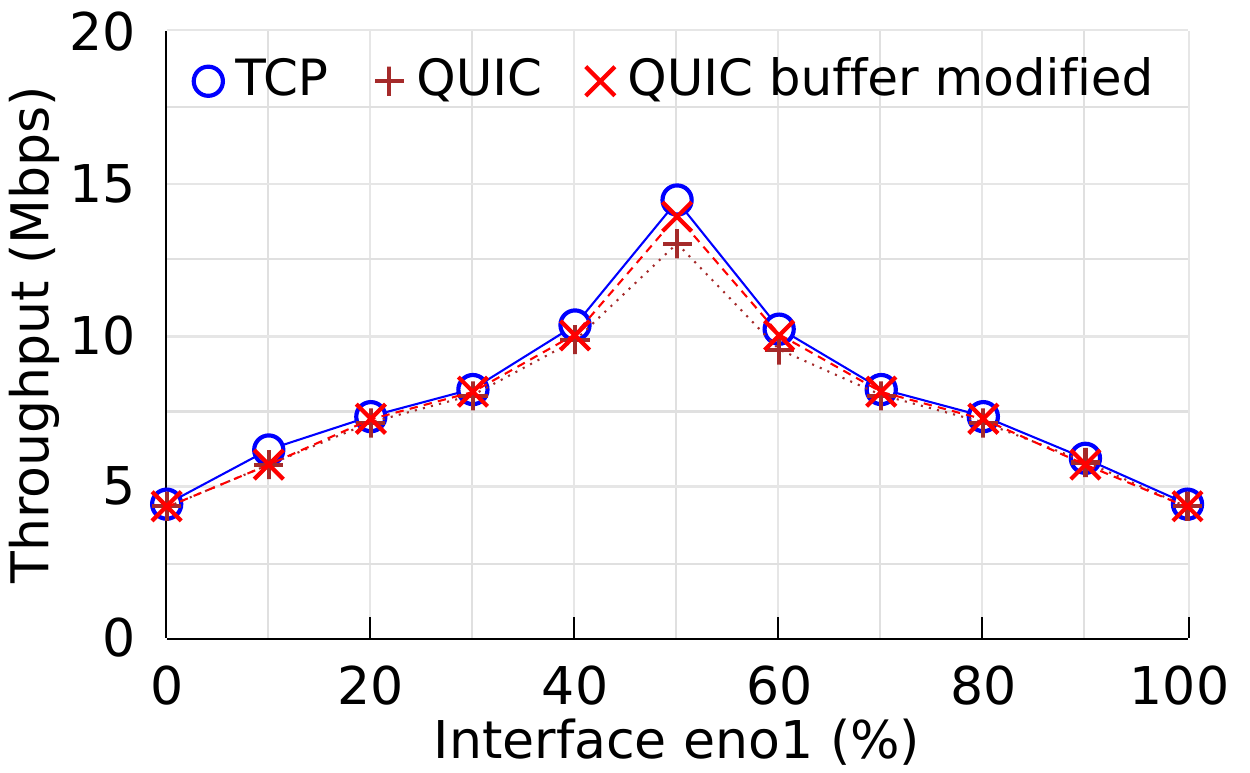}}
    \hspace*{\fill}
    \subfigure[Low delay ratio (repeated Fig.~\ref{fig:F-1})\label{fig:U-2}]{\includegraphics[trim = 0mm 4mm 0mm 0mm, width=\onethirdwidth]{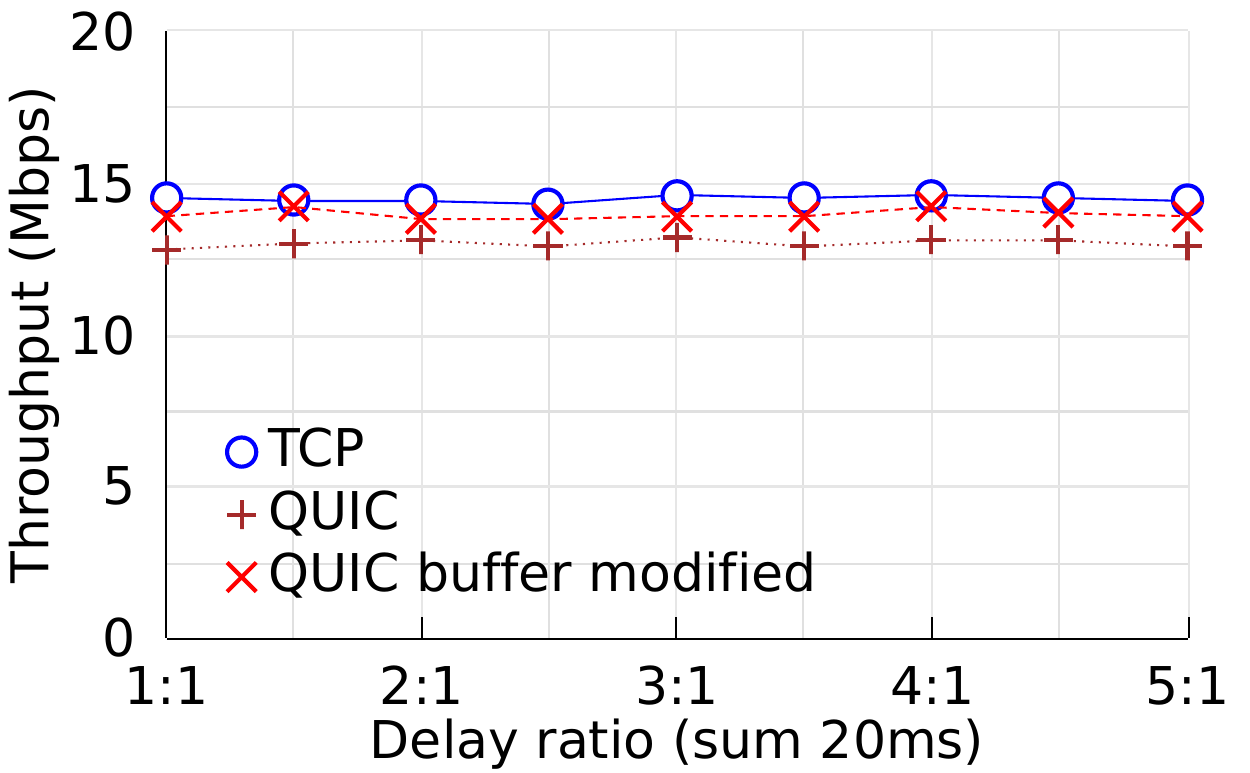}}
    \hspace*{\fill}
    \subfigure[Duplicate packets (repeated Fig.~\ref{fig:H-1})\label{fig:U-3}]{\includegraphics[trim = 0mm 4mm 0mm 0mm, width=\onethirdwidth]{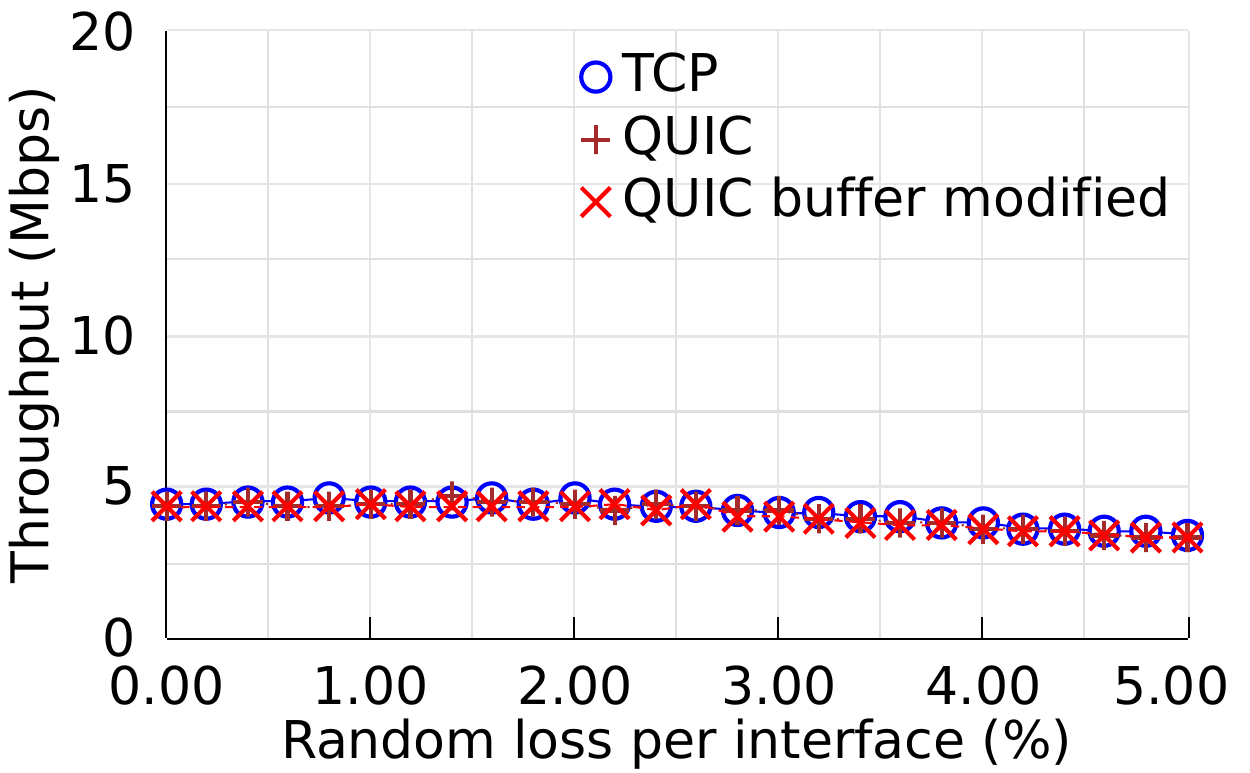}}
    \vspace{-8pt}
    \caption{\revone{}{Additional trace-based throughput validation tests using aioquic as QUIC implementation}}
    \label{fig:extra-trace-aioquic}
\vspace{-4pt}
\end{figure*}

\begin{figure*}[t!]
    \subfigure[DC \revtwo{ratio}{batch split} (repeated Fig.~\ref{fig:J-1})\label{fig:V-1}]{\includegraphics[trim = 0mm 4mm 0mm 0mm, width=\onethirdwidth]{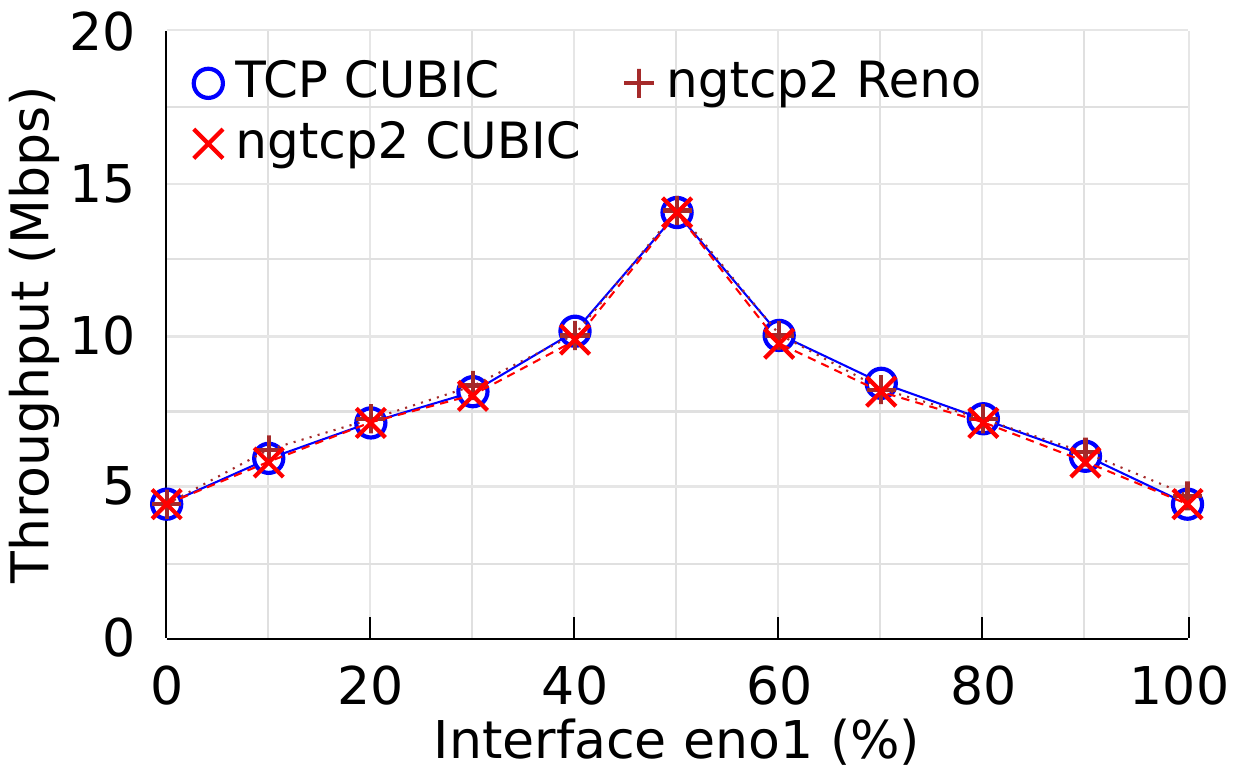}}
    \hspace*{\fill}
    \subfigure[Low delay ratio (repeated Fig.~\ref{fig:Z-1})\label{fig:V-2}]{\includegraphics[trim = 0mm 4mm 0mm 0mm, width=\onethirdwidth]{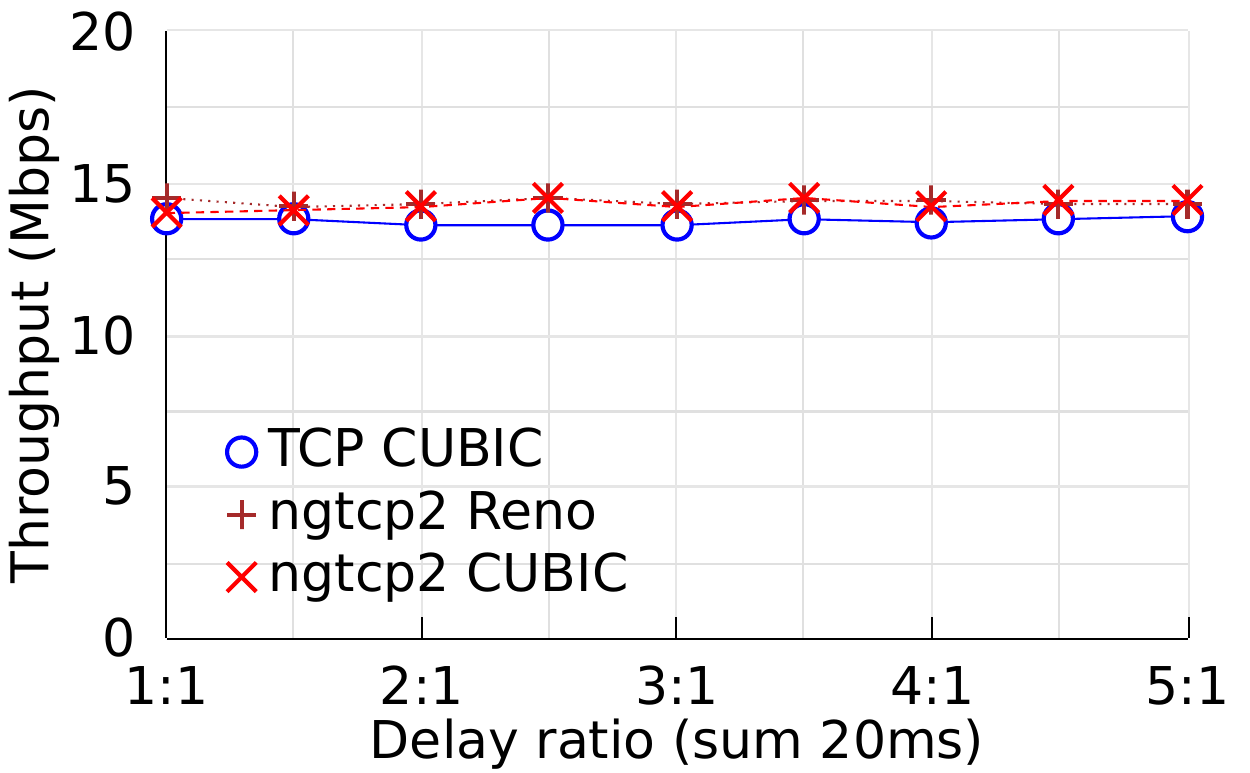}}
    \hspace*{\fill}
    \subfigure[Duplicate packets (repeated Fig.~\ref{fig:Q-1})\label{fig:V-3}]{\includegraphics[trim = 0mm 4mm 0mm 0mm, width=\onethirdwidth]{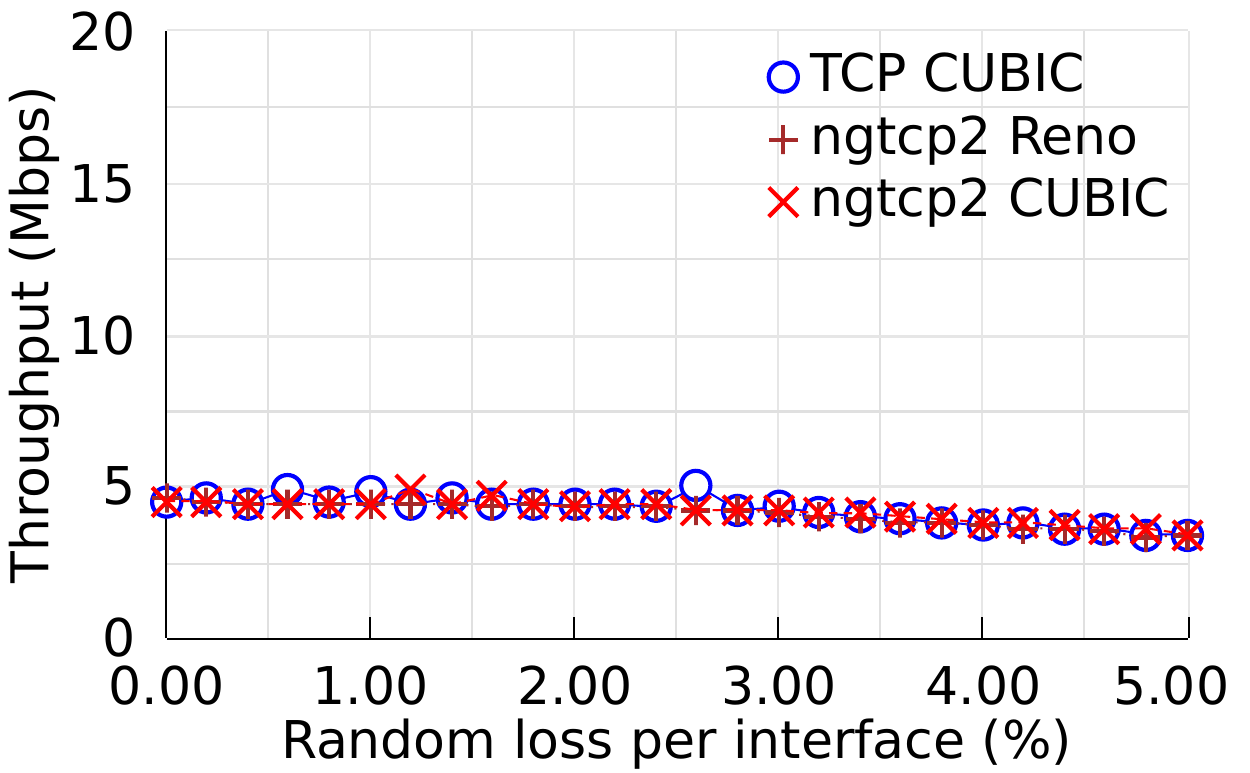}}
    \vspace{-8pt}
    \caption{\revone{}{Additional trace-based throughput validation tests using ngtcp2 as QUIC implementation}}
    \label{fig:extra-trace-ngtcp2}
\vspace{-6pt}
\end{figure*}

\revone{}{Finally, we note that the conclusions regarding the protection that duplicate packets offer during lossy scenarios are consistent 
across QUIC versions and congestion control algorithms.
In all cases, duplication provides better loss protection at the cost of fairness and goodput. Also in this special scenario does ngtcp2 achieve higher throughput than aioquic. These results are shown in Figure~\ref{fig:extra-duplicate} and can be compared 
to
Figure~\ref{fig:H}.}


\subsection{Bandwidth variability scenario}\label{ssec:results-bw-variability}

{\bf Baseline comparisons:}
To capture a more realistic bandwidth user scenario, 
Figures~\ref{fig:L-1} to~\ref{fig:L-3} show repeated experiments with aioquic for DC batch size, high delay ratio and loss rates performed over a LTE sampled bandwidth trace. Figures~\ref{fig:L-4} to~\ref{fig:L-6} show these results but using ngtcp2 with CUBIC. For DC batch size and loss rates, similar trends are observed as when using a fixed bandwidth capability. Similar trends are also observed in the case of delay ratio, but with the effect of the pacer more clearly shown. Lastly, we note that ngtcp2 is more aggressive than aioquic and that CUBIC achieves higher throughput.

\revone{}{{\bf Additional trace-based results:}
Figures~\ref{fig:extra-trace-aioquic} and~\ref{fig:extra-trace-ngtcp2} show additional trace-based results using the two QUIC versions aioquic and ngtcp2, respectively. Here, we show results for (a) the impact of the DC ratio, (b) the impact of the delay ratio when operating in the low-latency range, and (c) the loss protection provided by the use of duplicate packets in loss scenarios. 
While the throughput differences here are smaller between ngtcp2 and aioquic than for the other scenarios (Figures~\ref{fig:extra-trace-aioquic} and~\ref{fig:extra-trace-ngtcp2} compared to Figure~\ref{fig:throughput-trace}), as well as between CUBIC and NewReno (Figure~\ref{fig:extra-trace-ngtcp2}), also for these cases do we observe the same trends as to when using a fixed bandwidth capability.}

\subsection{\revtwo{}{Summary of results}\label{ssec:results-summary}}

\revtwo{}{
\textbf{Throughput:}
In general, QUIC performs better with a larger receive buffer size. With a modified buffer, it reaches similar throughput as the TCP counterparts. For DC parameters batch size and batch split, the throughput is negatively affected by a larger batch size and an uneven split due to reduced link utilization. With a large batch size, we see significant periods where one link is underutilized and idle while packets are forwarded through the second link. We achieve the highest throughput using a batch size near 100-150 and an even batch split of 50/50 to send half of the packets over each interface. When varying the bandwidth ratio, we observe the highest throughput when the two links have a similar capacity and that the throughput decreases with an uneven link capacity. However, we obtain high throughput for all ratios if the DC ratio is selected to match the bandwidth ratio. When studying the delay, we observe the highest throughput achieved with a ratio of 1:1. For other ratios, the throughput is only negatively impacted in the case of having a high delay. Finally, as expected, higher loss results in lower throughput. However, when using DC to duplicate packets, the throughput is much more stable as the packet losses increase.
}

\revtwo{}{
\textbf{Fairness:}
The fairness of competing connections reduces with a large DC batch size as SC clients monopolize the link during DC client's off periods. We observe the highest fairness using a small batch size and an even batch split. High fairness is achieved using a bandwidth ratio of 1:1, but decreases with a more uneven ratio. When the DC ratio is selected to match the bandwidth ratio, the fairness still decreases but at a slower rate. When varying the delay ratio, the fairness is only significantly impacted when having a high delay. For packet losses, the fairness is in general not impacted. However, the fairness is much affected when using DC packet duplication, showing much unfairness regardless of the loss rate.
}

\revtwo{}{
\textbf{Additional experiments:}
We generally observe similar trends when repeating the experiments using another QUIC implementation (ngtcp2) and congestion control algorithm (CUBIC). The main differences observed are (1) the optimal point can vary slightly due to implementation differences (e.g., pacer), (2) ngtcp2 is more aggressive than aioquic, leading to higher throughput and less fairness, and (3) only small differences are observed between CUBIC and NewReno. Our performance experiments over variable bandwidth links show similar trends, validating our results under the more realistic scenario.
}

\section{Conclusions}
\label{sec:conclusion}

In this paper, we present the first performance study of QUIC over DC. Key insights are given for network operators to understand how different DC parameters and network conditions affect QUIC performance.
\revone{}{Regardless if we use aioquic or ngtcp2 as the QUIC implementation or whether we use CUBIC or NewReno as the congestion control algorithm,}
QUIC's throughput is found to be similar to that of TCP in general cases, provided that the UDP receive buffer (when using aioquic) has been increased to \revone{}{a} similar size as the corresponding TCP buffer. We show that QUIC can take advantage of DC when the links share similar properties, and the DC batch size is small. When the properties of the links are too far apart, QUIC performance suffers to the degree that the performance would be better if DC was turned off. Furthermore, we show that QUIC can achieve system-wide fairness, provided that the link properties are similar. 
\revone{}{Otherwise, the DC connection will suffer more than the single connections, showing poor performance and a high degree of unfairness.}
We also show that packet duplication allows QUIC to improve throughput for lossy environments at the cost of substantially increased unfairness.

With aioquic, the QUIC throughput is considerably lower if the UDP receive buffer remains at default values for Linux, as PDCP introduces packet bursts, causing packet drops due to full buffers. This occurs especially often in asymmetric link scenarios with high throughput. With the increased use of QUIC, we emphasize the importance of studying and optimizing the resources provided by the kernel to QUIC.

\revtwo{}{As QUIC is implemented in user-space, interesting future work include studying CPU usage and NIC offloading when used in conjunction with DC. Other 
future work include mathematical modeling and evaluation of the performance of using QUIC over DC.  Here, we take only an experimental approach.
}

\begin{figure*}[t]
     \subfigure[QUIC vs. QUIC (JFI=1.000)]{\includegraphics[trim = 0mm 4mm 0mm 0mm, width=\onethirdwidth]{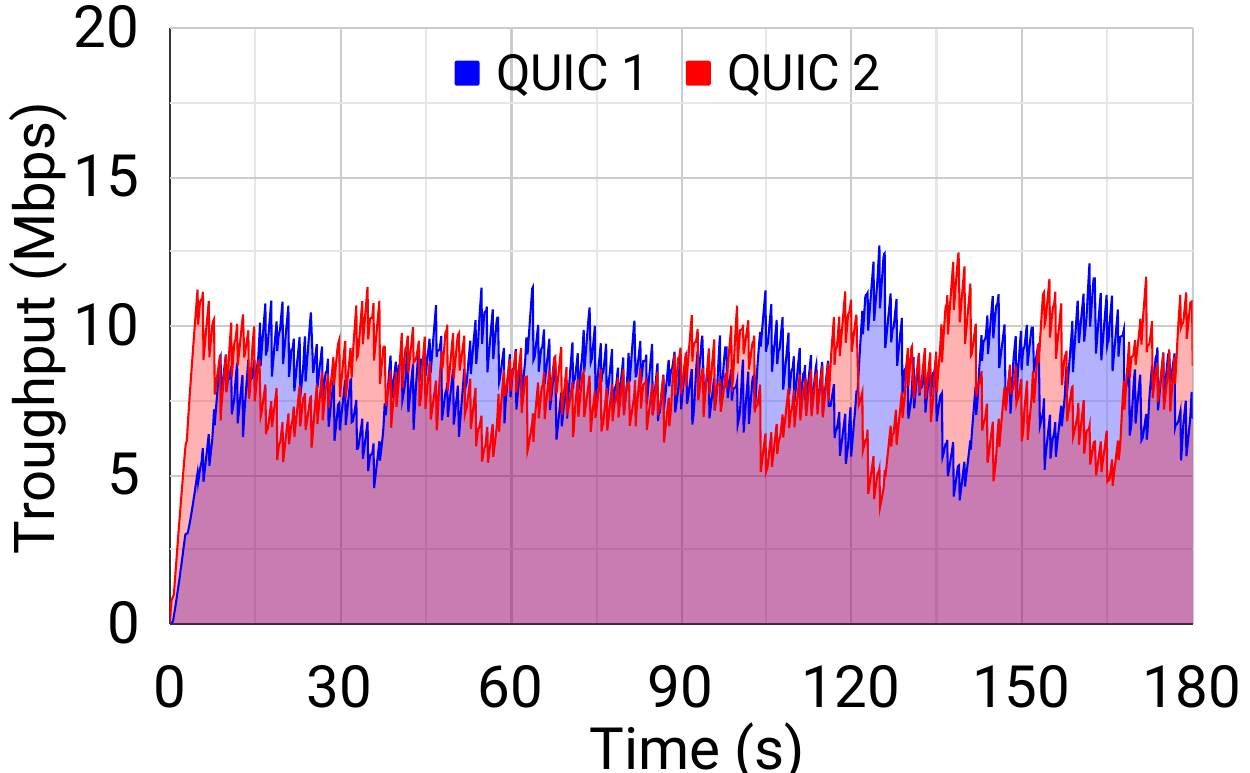}}
     \hspace*{\fill}
     \subfigure[QUIC vs. TCP (JFI=0.995)]{\includegraphics[trim = 0mm 4mm 0mm 0mm, width=\onethirdwidth]{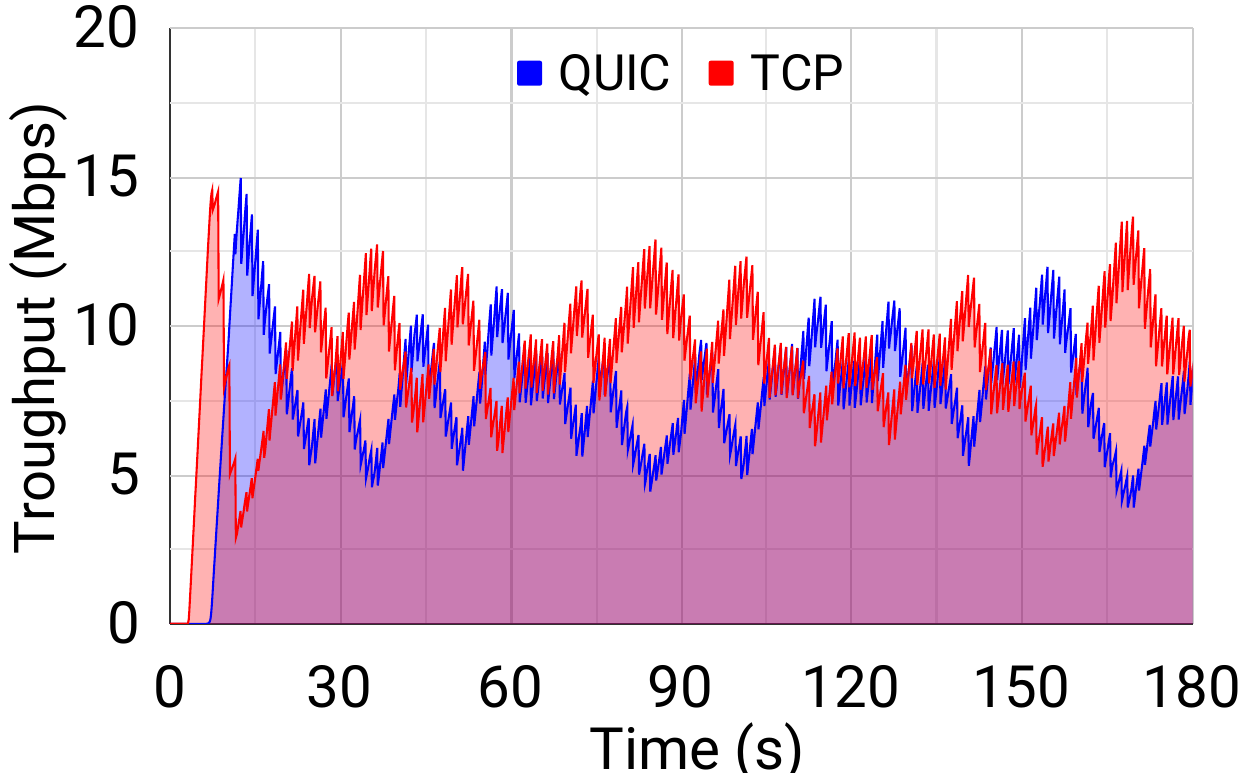}}
     \hspace*{\fill}
     \subfigure[TCP vs. TCP (JFI=0.999)]{\includegraphics[trim = 0mm 4mm 0mm 0mm, width=\onethirdwidth]{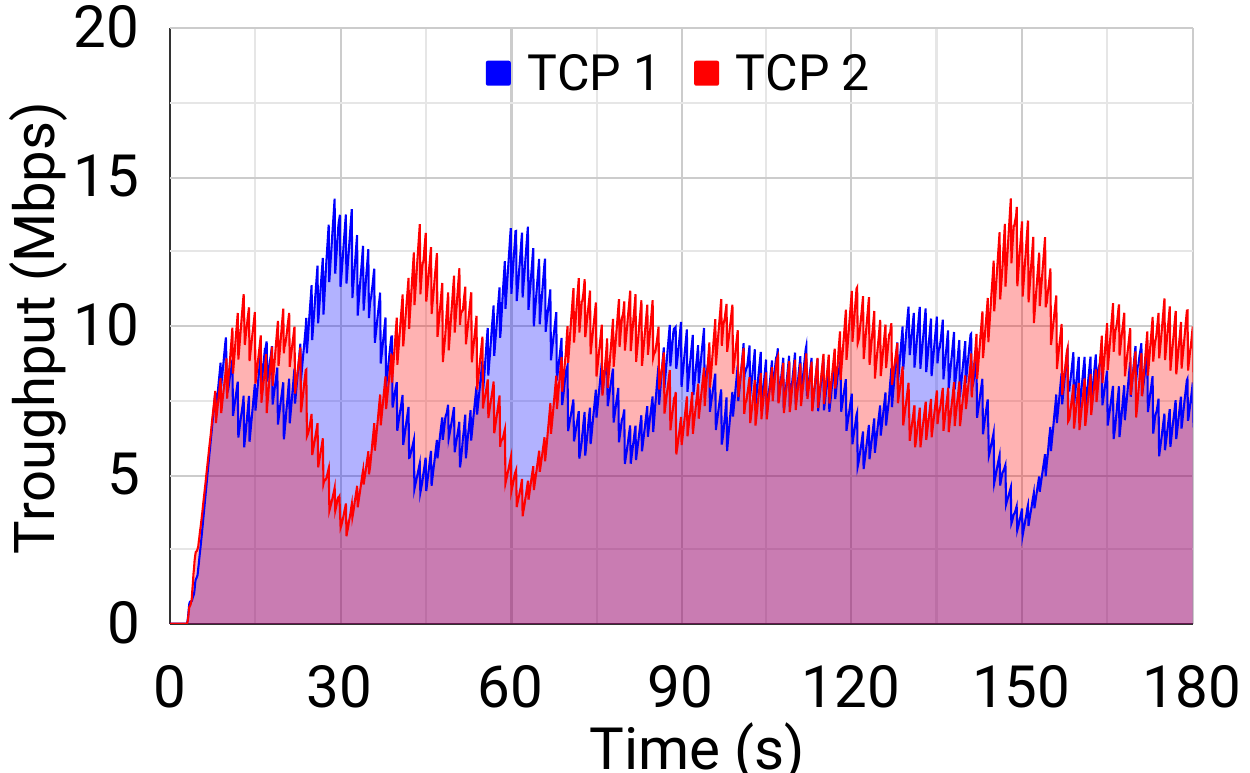}}
    \vspace{-6pt}
    \caption{\revone{}{Pairwise example traces using SC}}
    \label{fig:fairness-example}
\vspace{-8pt}
\end{figure*}

\subsection*{Acknowledgement}
\revone{}{The authors are very thankful to Stefan Sundkvist (at Ericsson) for his feedback and help with this work.}
This work was funded in part by the Swedish Research Council (VR)
and the Wallenberg AI, Autonomous Systems and Software Program (WASP) funded by the Knut and Alice Wallenberg Foundation.


\bibliographystyle{elsarticle-num-names} 
\bibliography{reference}

\begin{thebibliography}{36}
\expandafter\ifx\csname natexlab\endcsname\relax\def\natexlab#1{#1}\fi
\providecommand{\url}[1]{\texttt{#1}}
\providecommand{\href}[2]{#2}
\providecommand{\path}[1]{#1}
\providecommand{\DOIprefix}{doi:}
\providecommand{\ArXivprefix}{arXiv:}
\providecommand{\URLprefix}{URL: }
\providecommand{\Pubmedprefix}{pmid:}
\providecommand{\doi}[1]{\href{http://dx.doi.org/#1}{\path{#1}}}
\providecommand{\Pubmed}[1]{\href{pmid:#1}{\path{#1}}}
\providecommand{\bibinfo}[2]{#2}
\ifx\xfnm\relax \def\xfnm[#1]{\unskip,\space#1}\fi
\bibitem[{Hasselquist et~al.(2020)Hasselquist, Lindstrom, Korzhitskii,
  Carlsson, and Gurtov}]{HLK+20}
\bibinfo{author}{D.~Hasselquist}, \bibinfo{author}{C.~Lindstrom},
  \bibinfo{author}{N.~Korzhitskii}, \bibinfo{author}{N.~Carlsson},
  \bibinfo{author}{A.~Gurtov},
\newblock \bibinfo{title}{{QUIC Throughput and Fairness over Dual
  Connectivity}},
\newblock in: \bibinfo{booktitle}{Proc. IEEE Modelling, Analysis, and
  Simulation of Computer and Telecommunication Systems (MASCOTS) Workshop},
  \bibinfo{year}{2020}.
\bibitem[{{Alfredsson} et~al.(2007){Alfredsson}, {Brunstrom}, and
  {Sternad}}]{ABS2007}
\bibinfo{author}{S.~{Alfredsson}}, \bibinfo{author}{A.~{Brunstrom}},
  \bibinfo{author}{M.~{Sternad}},
\newblock \bibinfo{title}{{Cross-layer analysis of TCP performance in a 4G
  system}},
\newblock in: \bibinfo{booktitle}{Proc. of SoftCOM}, \bibinfo{year}{2007}.
\bibitem[{Da~Silva et~al.(2015)Da~Silva, Mildh, Rune, Wallentin, Vikberg,
  Schliwa-Bertling, and Fan}]{smrwf15}
\bibinfo{author}{I.~Da~Silva}, \bibinfo{author}{G.~Mildh},
  \bibinfo{author}{J.~Rune}, \bibinfo{author}{P.~Wallentin},
  \bibinfo{author}{J.~Vikberg}, \bibinfo{author}{P.~Schliwa-Bertling},
  \bibinfo{author}{R.~Fan},
\newblock \bibinfo{title}{{Tight integration of new 5G air interface and LTE to
  fulfill 5G requirements}},
\newblock in: \bibinfo{booktitle}{Proc. of VTC Spring}, \bibinfo{year}{2015}.
\bibitem[{{Mahmood} et~al.(2018){Mahmood}, {Lopez}, {Laselva}, {Pedersen}, and
  {Berardinelli}}]{MLLPB18}
\bibinfo{author}{N.~H. {Mahmood}}, \bibinfo{author}{M.~{Lopez}},
  \bibinfo{author}{D.~{Laselva}}, \bibinfo{author}{K.~{Pedersen}},
  \bibinfo{author}{G.~{Berardinelli}},
\newblock \bibinfo{title}{{Reliability Oriented Dual Connectivity for URLLC
  services in 5G New Radio}},
\newblock in: \bibinfo{booktitle}{Proc. of ISWCS}, \bibinfo{year}{2018}.
\bibitem[{3GPP(2019)}]{3gpp15}
\bibinfo{author}{3GPP}, \bibinfo{title}{{Summary of Rel-15}},
  \bibinfo{type}{{Tech. Rep. 21.915 Release 15}}, \bibinfo{year}{2019}.
\bibitem[{De~Coninck and Bonaventure(2017)}]{CB17}
\bibinfo{author}{Q.~De~Coninck}, \bibinfo{author}{O.~Bonaventure},
\newblock \bibinfo{title}{{Multipath QUIC: Design and Evaluation}},
\newblock in: \bibinfo{booktitle}{Proc. of ACM CoNEXT}, \bibinfo{year}{2017}.
\bibitem[{Wischik et~al.(2011)Wischik, Raiciu, Greenhalgh, and
  Handley}]{WRGH11}
\bibinfo{author}{D.~Wischik}, \bibinfo{author}{C.~Raiciu},
  \bibinfo{author}{A.~Greenhalgh}, \bibinfo{author}{M.~Handley},
\newblock \bibinfo{title}{{Design, Implementation and Evaluation of Congestion
  Control for Multipath TCP}},
\newblock in: \bibinfo{booktitle}{Proc. of USENIX Symposium on NSDI},
  \bibinfo{year}{2011}.
\bibitem[{Iyengar et~al.(2006)Iyengar, Amer, and Stewart}]{IAS06}
\bibinfo{author}{J.~R. Iyengar}, \bibinfo{author}{P.~D. Amer},
  \bibinfo{author}{R.~Stewart},
\newblock \bibinfo{title}{{Concurrent multipath transfer using SCTP multihoming
  over independent end-to-end paths}},
\newblock \bibinfo{journal}{IEEE/ACM Trans. on Networking}
  (\bibinfo{year}{2006}).
\bibitem[{Langley et~al.(2017)Langley, Riddoch, Wilk, Vicente, Krasic, Zhang,
  Yang, Kouranov, Swett, Iyengar, Bailey, Dorfman, Roskind, Kulik, Westin,
  Tenneti, Shade, Hamilton, Vasiliev, Chang, and Shi}]{LRW+17}
\bibinfo{author}{A.~Langley}, \bibinfo{author}{A.~Riddoch},
  \bibinfo{author}{A.~Wilk}, \bibinfo{author}{A.~Vicente},
  \bibinfo{author}{C.~Krasic}, \bibinfo{author}{D.~Zhang},
  \bibinfo{author}{F.~Yang}, \bibinfo{author}{F.~Kouranov},
  \bibinfo{author}{I.~Swett}, \bibinfo{author}{J.~Iyengar},
  \bibinfo{author}{J.~Bailey}, \bibinfo{author}{J.~Dorfman},
  \bibinfo{author}{J.~Roskind}, \bibinfo{author}{J.~Kulik},
  \bibinfo{author}{P.~Westin}, \bibinfo{author}{R.~Tenneti},
  \bibinfo{author}{R.~Shade}, \bibinfo{author}{R.~Hamilton},
  \bibinfo{author}{V.~Vasiliev}, \bibinfo{author}{W.-T. Chang},
  \bibinfo{author}{Z.~Shi},
\newblock \bibinfo{title}{{The QUIC Transport Protocol: Design and
  Internet-Scale Deployment}},
\newblock in: \bibinfo{booktitle}{Proc. of ACM SIGCOMM}, \bibinfo{year}{2017}.
\bibitem[{{IETF 106 Singapore}(2019)}]{ietf-106-singapore-quic}
\bibinfo{author}{{IETF 106 Singapore}}, \bibinfo{title}{{Some updates on QUIC
  deployment numbers}}, \bibinfo{year}{2019}. \URLprefix
  \url{https://datatracker.ietf.org/meeting/106/
  materials/slides-106-maprg-quic-deployment-update}.
\bibitem[{3GPP(2013)}]{3gpp12}
\bibinfo{author}{3GPP}, \bibinfo{title}{{Study on Small Cell enhancements for
  E-UTRA and E-UTRAN; Higher layer aspects}}, \bibinfo{type}{{Tech. Rep. 36.842
  Release 12}}, \bibinfo{year}{2013}.
\bibitem[{{Ravanshid} et~al.(2016){Ravanshid}, {Rost}, {Michalopoulos}, {Phan},
  {Bakker}, {Aziz}, {Tayade}, {Schotten}, {Wong}, and {Holland}}]{RRM+16}
\bibinfo{author}{A.~{Ravanshid}}, \bibinfo{author}{P.~{Rost}},
  \bibinfo{author}{D.~S. {Michalopoulos}}, \bibinfo{author}{V.~V. {Phan}},
  \bibinfo{author}{H.~{Bakker}}, \bibinfo{author}{D.~{Aziz}},
  \bibinfo{author}{S.~{Tayade}}, \bibinfo{author}{H.~D. {Schotten}},
  \bibinfo{author}{S.~{Wong}}, \bibinfo{author}{O.~{Holland}},
\newblock \bibinfo{title}{{Multi-connectivity functional architectures in 5G}},
\newblock in: \bibinfo{booktitle}{Proc. of IEEE ICC}, \bibinfo{year}{2016}.
\bibitem[{3GPP(2020)}]{3gpp16-etsi5g}
\bibinfo{author}{3GPP}, \bibinfo{title}{{Evolved Universal Terrestrial Radio
  Access; Packet Data Convergence Protocol specification}},
  \bibinfo{type}{{Tech. Rep. 36.323 Release 16}}, \bibinfo{year}{2020}.
\bibitem[{Polese et~al.(2017)Polese, Mezzavilla, Rangan, and Zorzi}]{PMRZ17}
\bibinfo{author}{M.~Polese}, \bibinfo{author}{M.~Mezzavilla},
  \bibinfo{author}{S.~Rangan}, \bibinfo{author}{M.~Zorzi},
\newblock \bibinfo{title}{{Mobility Management for TCP in MmWave Networks}},
\newblock in: \bibinfo{booktitle}{Proc. ACM mmNets}, \bibinfo{year}{2017}.
\bibitem[{{Jin} et~al.(2017){Jin}, {Kim}, {Yun}, {Lee}, {Kim}, and
  {Yi}}]{JKYLKY17}
\bibinfo{author}{B.~{Jin}}, \bibinfo{author}{S.~{Kim}},
  \bibinfo{author}{D.~{Yun}}, \bibinfo{author}{H.~{Lee}},
  \bibinfo{author}{W.~{Kim}}, \bibinfo{author}{Y.~{Yi}},
\newblock \bibinfo{title}{{Aggregating LTE and Wi-Fi: Toward Intra-Cell
  Fairness and High TCP Performance}},
\newblock \bibinfo{journal}{IEEE Trans. on Wireless Communications}
  (\bibinfo{year}{2017}).
\bibitem[{{Khadraoui} et~al.(2016){Khadraoui}, {Lagrange}, and
  {Gravey}}]{KLG16}
\bibinfo{author}{Y.~{Khadraoui}}, \bibinfo{author}{X.~{Lagrange}},
  \bibinfo{author}{A.~{Gravey}},
\newblock \bibinfo{title}{{TCP Performance for Practical Implementation of Very
  Tight Coupling between LTE and WiFi}},
\newblock in: \bibinfo{booktitle}{Proc. of IEEE VTC Fall},
  \bibinfo{year}{2016}.
\bibitem[{Wu et~al.(2018{\natexlab{a}})Wu, He, Qian, Huang, and
  Shen}]{wu2018optimal}
\bibinfo{author}{Y.~Wu}, \bibinfo{author}{Y.~He}, \bibinfo{author}{L.~P. Qian},
  \bibinfo{author}{J.~Huang}, \bibinfo{author}{X.~Shen},
\newblock \bibinfo{title}{Optimal resource allocations for mobile data
  offloading via dual-connectivity},
\newblock \bibinfo{journal}{IEEE Trans. on Mobile Computing}
  (\bibinfo{year}{2018}{\natexlab{a}}).
\bibitem[{Wu et~al.(2018{\natexlab{b}})Wu, Yang, Qian, Zhou, Shen, and
  Awad}]{wu2018optimal-b}
\bibinfo{author}{Y.~Wu}, \bibinfo{author}{X.~Yang}, \bibinfo{author}{L.~P.
  Qian}, \bibinfo{author}{H.~Zhou}, \bibinfo{author}{X.~Shen},
  \bibinfo{author}{M.~K. Awad},
\newblock \bibinfo{title}{Optimal dual-connectivity traffic offloading in
  energy-harvesting small-cell networks},
\newblock \bibinfo{journal}{IEEE Trans. on Green Communications and Networking}
   (\bibinfo{year}{2018}{\natexlab{b}}).
\bibitem[{Sharma et~al.(2019)Sharma, Kumar, and Wu}]{sharma2019performance}
\bibinfo{author}{L.~Sharma}, \bibinfo{author}{B.~B. Kumar},
  \bibinfo{author}{S.-L. Wu},
\newblock \bibinfo{title}{Performance analysis and adaptive {DRX} scheme for
  dual connectivity},
\newblock \bibinfo{journal}{IEEE Internet of Things Journal}
  (\bibinfo{year}{2019}).
\bibitem[{He et~al.(2021)He, Hua, Xu, Gu, and Shen}]{he2021delay}
\bibinfo{author}{M.~He}, \bibinfo{author}{C.~Hua}, \bibinfo{author}{W.~Xu},
  \bibinfo{author}{P.~Gu}, \bibinfo{author}{X.~S. Shen},
\newblock \bibinfo{title}{Delay optimal concurrent transmissions with raptor
  codes in dual connectivity networks},
\newblock \bibinfo{journal}{IEEE Trans. on Network Science and Engineering}
  (\bibinfo{year}{2021}).
\bibitem[{Gurtov and Polishchuk(2009)}]{GuPo09}
\bibinfo{author}{A.~Gurtov}, \bibinfo{author}{T.~Polishchuk},
\newblock \bibinfo{title}{{Secure multipath transport for legacy Internet
  applications}},
\newblock in: \bibinfo{booktitle}{Proc. of IEEE Broadnets},
  \bibinfo{year}{2009}.
\bibitem[{Mogensen et~al.(2019)Mogensen, Markmoller, Madsen, Kolding, Pocovi,
  and Lauridsen}]{MMMKPL19}
\bibinfo{author}{R.~S. Mogensen}, \bibinfo{author}{C.~Markmoller},
  \bibinfo{author}{T.~K. Madsen}, \bibinfo{author}{T.~Kolding},
  \bibinfo{author}{G.~Pocovi}, \bibinfo{author}{M.~Lauridsen},
\newblock \bibinfo{title}{{Selective Redundant MP-QUIC for 5G Mission Critical
  Wireless Applications}},
\newblock in: \bibinfo{booktitle}{Proc. of IEEE VTC Spring},
  \bibinfo{year}{2019}.
\bibitem[{Rabitsch et~al.(2018)Rabitsch, Hurtig, and Brunstrom}]{RHB18}
\bibinfo{author}{A.~Rabitsch}, \bibinfo{author}{P.~Hurtig},
  \bibinfo{author}{A.~Brunstrom},
\newblock \bibinfo{title}{{A Stream-Aware Multipath QUIC Scheduler for
  Heterogeneous Paths}},
\newblock in: \bibinfo{booktitle}{Proc. of ACM SIGCOMM workshop EPIQ},
  \bibinfo{year}{2018}.
\bibitem[{Becke et~al.(2012)Becke, Dreibholz, Adhari, and Rathgeb}]{BDAR12}
\bibinfo{author}{M.~Becke}, \bibinfo{author}{T.~Dreibholz},
  \bibinfo{author}{H.~Adhari}, \bibinfo{author}{E.~P. Rathgeb},
\newblock \bibinfo{title}{{On the fairness of transport protocols in a
  multi-path environment}},
\newblock in: \bibinfo{booktitle}{Proc. of IEEE ICC}, \bibinfo{year}{2012}.
\bibitem[{Raiciu et~al.(2010)Raiciu, Pluntke, Barre, Greenhalgh, Wischik, and
  Handley}]{RPBGWH2010}
\bibinfo{author}{C.~Raiciu}, \bibinfo{author}{C.~Pluntke},
  \bibinfo{author}{S.~Barre}, \bibinfo{author}{A.~Greenhalgh},
  \bibinfo{author}{D.~Wischik}, \bibinfo{author}{M.~Handley},
\newblock \bibinfo{title}{{Data center networking with multipath TCP}},
\newblock in: \bibinfo{booktitle}{Proc. of ACM SIGCOMM workshop HotNets},
  \bibinfo{year}{2010}.
\bibitem[{Zhang and Li(2008)}]{ZL08}
\bibinfo{author}{X.~Zhang}, \bibinfo{author}{B.~Li},
\newblock \bibinfo{title}{{Dice: A Game Theoretic Framework for Wireless
  Multipath Network Coding}},
\newblock in: \bibinfo{booktitle}{Proc. of ACM MobiHoc}, \bibinfo{year}{2008}.
\bibitem[{Jain et~al.(1984)Jain, Chiu, and Hawe}]{JCH84}
\bibinfo{author}{R.~K. Jain}, \bibinfo{author}{D.-M.~W. Chiu},
  \bibinfo{author}{W.~R. Hawe}, \bibinfo{title}{A quantitative measure of
  fairness and discrimination for resource allocation in shared computer
  systems}, \bibinfo{type}{Technical Report} \bibinfo{number}{DEC-TR-301},
  Eastern Research Lab, Digital Equipment Corporation, \bibinfo{year}{1984}.
\bibitem[{aioquic(2020)}]{aioquic}
\bibinfo{author}{aioquic}, \bibinfo{title}{aioquic}, \bibinfo{year}{2020}.
  \URLprefix \url{https://github.com/aiortc/aioquic}.
\bibitem[{Paasch et~al.(2013)Paasch, Khalili, and Bonaventure}]{PKB13}
\bibinfo{author}{C.~Paasch}, \bibinfo{author}{R.~Khalili},
  \bibinfo{author}{O.~Bonaventure},
\newblock \bibinfo{title}{{On the Benefits of Applying Experimental Design to
  Improve Multipath TCP}},
\newblock in: \bibinfo{booktitle}{{Proc. of ACM CoNEXT}}, \bibinfo{year}{2013}.
\bibitem[{Marx et~al.(2018)Marx, Lamotte, Reynders, Pittevils, and
  Quax}]{MaRQ18}
\bibinfo{author}{R.~Marx}, \bibinfo{author}{W.~Lamotte},
  \bibinfo{author}{J.~Reynders}, \bibinfo{author}{K.~Pittevils},
  \bibinfo{author}{P.~Quax},
\newblock \bibinfo{title}{{Towards QUIC Debuggability}},
\newblock in: \bibinfo{booktitle}{Proc. of ACM SIGCOMM workshop EPIQ},
  \bibinfo{year}{2018}.
\bibitem[{Marx et~al.(2020)Marx, Piraux, Quax, and Lamotte}]{MaPQ20}
\bibinfo{author}{R.~Marx}, \bibinfo{author}{M.~Piraux},
  \bibinfo{author}{P.~Quax}, \bibinfo{author}{W.~Lamotte},
\newblock \bibinfo{title}{{Debugging QUIC and HTTP/3 with qlog and qvis}},
\newblock in: \bibinfo{booktitle}{Proc. of Applied Networking Research
  Workshop}, \bibinfo{year}{2020}.
\bibitem[{McMillan and Zuck(2019)}]{McZu19}
\bibinfo{author}{K.~L. McMillan}, \bibinfo{author}{L.~D. Zuck},
\newblock \bibinfo{title}{{Formal Specification and Testing of QUIC}},
\newblock in: \bibinfo{booktitle}{Proc. of ACM SIGCOMM}, \bibinfo{year}{2019}.
\bibitem[{Marx et~al.(2020)Marx, Herbots, Lamotte, and Quax}]{MaHQ20}
\bibinfo{author}{R.~Marx}, \bibinfo{author}{J.~Herbots},
  \bibinfo{author}{W.~Lamotte}, \bibinfo{author}{P.~Quax},
\newblock \bibinfo{title}{{Same Standards, Different Decisions: A Study of QUIC
  and HTTP/3 Implementation Diversity}},
\newblock in: \bibinfo{booktitle}{Proc. of ACM SIGCOMM workshop EPIQ},
  \bibinfo{year}{2020}.
\bibitem[{ngtcp2(2020)}]{ngtcp2}
\bibinfo{author}{ngtcp2}, \bibinfo{title}{ngtcp2}, \bibinfo{year}{2020}.
  \URLprefix \url{https://github.com/ngtcp2/ngtcp2}.
\bibitem[{Raca et~al.(2018)Raca, Quinlan, Zahran, and Sreenan}]{RQZS18}
\bibinfo{author}{D.~Raca}, \bibinfo{author}{J.~J. Quinlan},
  \bibinfo{author}{A.~H. Zahran}, \bibinfo{author}{C.~J. Sreenan},
\newblock \bibinfo{title}{{Beyond Throughput: A 4G LTE Dataset with Channel and
  Context Metrics}},
\newblock in: \bibinfo{booktitle}{Proc. of ACM MMSys}, \bibinfo{year}{2018}.
\bibitem[{Iyengar and Swett(2020)}]{QUIC-RFC-Recovery-29}
\bibinfo{author}{J.~Iyengar}, \bibinfo{author}{I.~Swett}, \bibinfo{title}{{QUIC
  Loss Detection and Congestion Control}}, \bibinfo{type}{Internet-Draft}
  \bibinfo{number}{draft-ietf-quic-recovery-29}, IETF, \bibinfo{year}{2020}.

\end{thebibliography}

\appendix

\section{Pairwise fairness example using SC}\label{appendix:fairness-example}

\revone{}{Figure~\ref{fig:fairness-example} shows throughput example traces with two competing SC clients using the following three configuration pairs: (a) QUIC vs. QUIC, (b) QUIC vs. TCP, and (c) TCP vs. TCP.  While the fairness index in all three cases are close to one (i.e., optimal fairness), the differences in the time-variation between the different connections are smallest for the two competing QUIC flows (i.e., Figure~\ref{fig:fairness-example}(a)). One contributing factor to this difference may be the QUIC pacer smoothing out the saw-tooth behavior of NewReno.}

\end{document}